\documentclass[a4paper,superscriptaddress,twocolumn,aps,prb,showpacs,longbibliography]{revtex4-2}

\usepackage{graphicx}  
\usepackage{physics}   
\usepackage{textgreek} 
\usepackage{hyperref}  
\usepackage{multirow}  
\usepackage{booktabs}  
\usepackage{siunitx}   
\usepackage{soul}

\usepackage{pdfpages} 
\usepackage{pgffor} 
\usepackage{xr} 

\def\supplementfilename{supplemental-material_V12}

\pdfximage{\supplementfilename.pdf}
\def\numbersupplementpages{\the\pdflastximagepages}

\newif\ifarXiv
\arXivtrue 

\makeatletter
\AtBeginDocument{\let\LS@rot\@undefined}
\makeatother

\begin{document}

\title{Efficient Treatment of Relativistic Effects with Periodic Density Functional Methods: Energies, Gradients, and Stress Tensors}

\author{Yannick J. Franzke}
\affiliation{Fachbereich Chemie, Philipps-Universit\"at Marburg,
Hans-Meerwein-Str.\ 4, 35032 Marburg, Germany}
\affiliation{Otto Schott Institute of Materials Research,
Friedrich Schiller University Jena, L\"obdergraben 32,
07743 Jena, Germany}
\thanks{Y.J.F.\ and W.M.S.\ contributed equally to this work}

\author{Werner M. Schosser}
\affiliation{Institute of Physics and Center for Advanced Analytics and Predictive Sciences,
University of Augsburg, Universit\"atsstr.\ 1, 86159 Augsburg, Germany}
\thanks{Y.J.F.\ and W.M.S.\ contributed equally to this work}

\author{Fabian Pauly}
\email{fabian.pauly@uni-a.de}
\affiliation{Institute of Physics and Center for Advanced Analytics and Predictive Sciences,
University of Augsburg, Universit\"atsstr.\ 1, 86159 Augsburg, Germany}

\date{\today}

\begin{abstract}
The implementation of an efficient self-consistent field (SCF) method including both scalar
relativistic effects and spin--orbit interaction in density functional theory (DFT) is presented.
We make use of Gaussian-type orbitals (GTOs) and all integrals are evaluated in real space.
Our implementation supports density functional approximations up to the level of
meta-generalized gradient approximations (mGGAs) for SCF energies and gradients.
The latter can be used to compute the stress tensor and consequently allow us to optimize the
cell structure. Considering spin--orbit interaction requires the extension
of the standard procedures to a two-component (2c) formalism and a non-collinear approach
for open-shell systems. Here, we implemented both the canonical and the Scalmani--Frisch
non-collinear DFT formalisms, with hybrid and range-separated hybrid functionals being
presently restricted to SCF energies.
We demonstrate both efficiency and relevance of spin--orbit effects for the electronic
structure of discrete systems and systems periodic in one to three dimensions.
\end{abstract}

\maketitle

\clearpage
\section{Introduction}
\label{sec:introduction}
Modern physics, chemistry, and materials science rely on the theories of quantum mechanics and
of special relativity. Particularly for systems that include elements with high nuclear
charge, the interplay between both theories is crucial \cite{Dolg.Cao:Relativistic.2012,
Saue:Relativistic.2011, Autschbach:Perspective.2012, Pyykko:Relativistic.2012, Liu:Essentials.2020,
Dyall.Fgri-Jr:2007, Reiher.Wolf:2015, Liu:2017, Pyykko:Physics.2012, Schwerdtfeger:2002,
Schwerdtfeger:2004, Wilson.Grant.ea:1991, InB-Grant:1994}.
The occurring relativistic effects can be divided into two categories, scalar-relativistic
(spin-independent) and spin-dependent relativistic effects. 
The most prominent scalar-relativistic effects are described by the energy--momentum (or $p^4$)
and the Darwin terms. 
The dominating spin-dependent relativistic effect is the interaction between the electron's spin and its
orbital momentum. For lighter elements, this correction is of negligible or small size and can be treated
efficiently with spin--orbit perturbation theory based on the Pauli Hamiltonian \cite{Saue:Relativistic.2011}.
However, for heavy $d$ and $p$ elements, such as platinum, gold, and lead \cite{Pyykko:Theoretical.2008,
Pyykko:Physics.2012, Pyykko.Desclaux:Relativity.1979, Pyykko:Relativistic.1988}, it is of comparable
size as scalar-relativistic effects and thus becomes relevant. Here, more sophisticated spin--orbit
perturbation approaches (e.g.\ Refs.~\cite{Li.Xiao.ea:On.2012, Cheng.Gauss:Perturbative.2014,
Cheng.Wang.ea:Perturbative.2018, Desmarais.Erba.ea:Perturbation.2021})
or a variational two-component ansatz treating both relativistic effects on an
equal footing are needed.

Spin-dependent relativistic effects are important for solids, surface science, and low-dimensional
systems \cite{Ish11,Kue00,Per10}. They are exploited in the field of spin-dependent quantum transport
for magnetoresistive effects \cite{Garcia2009, Hafner2009,Hardrat2012, Jacob2008, Schmaus2011, Kawahara2012,
Hayakawa2016, Strigl2015, Li.Pauly.ea:Giant.2020}, for magnetic excitations such as skyrmions \cite{Kang2016} caused by the
Dzyaloshinskii--Moriya interaction \cite{Hei08,Hei09}, and they may lead to spin-polarized surface states of
topological insulators \cite{Has10,Pau12}.

Both scalar-relativistic and spin--orbit effects are efficiently modeled by effective core potentials (ECPs)
\cite{Dolg.Cao:Relativistic.2012, InC-Dolg:2002, InB-Dolg:2017}. Here, the core electrons are replaced
with pseudopotentials, which are fitted to high-level relativistic theories. This way, only the
one-electron potential operator is directly affected. Therefore, scalar-relativistic effects
can be incorporated at essentially no computational burden, as the electron--electron interaction
operators are unaffected. However, the description of spin--orbit coupling requires a generalization
of the framework. Due to the breakdown of spin symmetry and the complex form of the spin--orbit
pseudopotentials or operators, a so-called 2c formalism employing complex algebra in
real space is needed, which leads to an increase of the computational demands
\cite{Kubler.Hock.ea:Density.1988, Van-Wullen:Relativistic.1999, Van-Wullen:Spin.2002,
Armbruster.Weigend.ea:Self-consistent.2008, Baldes.Weigend:Efficient.2013, Holzer.Franke.ea:Current.2022,
Peralta.Scuseria.ea:Noncollinear.2007, Scalmani.Frisch:New.2012, Bulik.Scalmani.ea:Noncollinear.2013,
Desmarais.Komorovsky.ea:Spin-orbit.2021, Desmarais.Flament.ea:Spin-orbit.2020}.

Offering relatively low computational costs combined with good accuracy, DFT is one of the most important
computational tools concerning electronic structure theory. For periodic systems,
scalar DFT approaches for energies and gradients are widely available in both plane-wave and GTO codes
\cite{Sherrill.Manolopoulos.ea:Electronic.2020, QE-2020,Blaha.Schwarz.ea:WIEN2k.2020,
Garcia.Papior.ea:Siesta.2020, Dovesi.Pascale.ea:CRYSTAL.2020, Kuhne.Iannuzzi.ea:CP2K.2020,
Prentice.Aarons.ea:ONETEP.2020, Sun.Zhang.ea:Recent.2020, g16, Kre96, Balasubramani.Chen.ea:TURBOMOLE.2020}.
\textsc{Turbomole} \cite{TURBOMOLE, Ahlrichs.Bar.ea:Electronic.1989, Furche.Ahlrichs.ea:Turbomole.2014,
Balasubramani.Chen.ea:TURBOMOLE.2020, Franzke.Holzer.ea:TURBOMOLE.2023}, for instance, includes an efficient and stable program to perform
these kinds of calculations for atomic, molecular, and periodic systems, employing GTOs and fast multipole approximations
combined with density fitting to ensure time efficiency. However, the functionalities to include
spin--orbit interaction for periodic systems are still missing in many program suites, preventing the simulation of
promising relativistic systems. Especially open-shell systems require further modifications.
That is, a non-collinear formulation of the exchange-correlation potential is needed
\cite{Kubler.Hock.ea:Density.1988, Van-Wullen:Relativistic.1999, Van-Wullen:Spin.2002,
Armbruster.Weigend.ea:Self-consistent.2008, Baldes.Weigend:Efficient.2013, Holzer.Franke.ea:Current.2022,
Peralta.Scuseria.ea:Noncollinear.2007, Scalmani.Frisch:New.2012, Bulik.Scalmani.ea:Noncollinear.2013,
Desmarais.Komorovsky.ea:Spin-orbit.2021, Desmarais.Flament.ea:Spin-orbit.2020}.
Moreover, for open-shell systems time-reversal symmetry is artificially broken
even in the absence of external fields as a result of the variational optimization in the
single-determinant KS ansatz \cite{Kasper.Jenkins.ea:Perspective.2020}.
In contrast, time-reversal symmetry is well preserved even with a single-determinant KS ansatz
for closed-shell systems. The Kramers' theorem can therefore be exploited in these closed-shell cases,
considerably simplifying the evaluation of the semilocal exchange-correlation potential.

Recently, a GTO-based non-collinear 2c DFT implementation for periodic systems and SCF energies was
presented by the group of Erba in the \textsc{Crystal23} package \cite{Desmarais.Flament.ea:Spin-orbit.2020,
Desmarais.Flament.ea:Adiabatic.2020, Bodo.Desmarais.ea:Spin.2022, Erba.Desmarais.ea:CRYSTAL23.2022}.
Density functional approximations up to the level of (hybrid) generalized gradient approximations (GGAs)
are considered in their work.
However, a corresponding implementation of 2c energy gradients has not been presented prior to the
original submission of our work, c.f.\ Refs.~\cite{Franzke.Schosser.ea:Self-consistent.2023}
and \cite{Desmarais.Erba.ea:2023, Desmarais.Erba.ea:Structural.2023}.
Additionally, meta-generalized gradient approximations (mGGAs) may be superior to GGAs in terms of accuracy
\cite{Mardirossian.Head-Gordon:Thirty.2017, Hao.Sun.ea:Performance.2013, Mo.Car.ea:Assessment.2017,
Goerigk.Hansen.ea:look.2017, Aschebrock.Kummel:Ultranonlocality.2019, Holzer.Franzke.ea:Assessing.2021,
Franzke.Yu:Hyperfine.2022, Franzke.Yu:Quasi-Relativistic.2022, Becke:Perspective.2014,
Becke:Density-Functional.2022, Borlido.Aull.ea:Large-Scale.2019, Ghosh.Jana.ea:Efficient.2022,
Kovacs.Tran.ea:What.2022, Franzke.Holzer:Exact.2023}.
Formally, mGGAs require the inclusion of the current density in the
kinetic energy density \cite{Dobson:Alternative.1993, Becke:Current.2002, Tao:Explicit.2005}
for spin--orbit coupling \cite{Holzer.Franke.ea:Current.2022} and electromagnetic
properties in general \cite{Bates.Furche:Harnessing.2012, Holzer.Franzke.ea:Assessing.2021,
Franzke.Holzer.ea:NMR.2022, Franzke.Holzer:Impact.2022, Bruder.Franzke.ea:Paramagnetic.2022,
Pausch.Holzer:Linear.2022, Schattenberg.Kaupp:Effect.2021, Bruder.Franzke.ea:Zero-Field.2023,
Franzke.Bruder.ea:Paramagnetic.2023, Tellgren.Teale.ea:Non-perturbative.2014,
Furness.Verbeke.ea:Current.2015, Reimann.Ekstrom.ea:importance.2015,
Irons.Spence.ea:Analyzing.2020, Sen.Tellgren:Benchmarking.2021, Irons.Spence.ea:Analyzing.2020}.
Note that this also applies to the introduction of spin--orbit effects within
perturbation theory and linear response methods \cite{Bruder.Franzke.ea:Paramagnetic.2022,
Bruder.Franzke.ea:Zero-Field.2023}.
The impact of the current density in spin--orbit coupling
calculations depends on the specific functional and the respective enhancement factors.
Here, tests for molecular systems have shown that TASK
\cite{Aschebrock.Kummel:Ultranonlocality.2019} and the Minnesota functionals
such as M06-L\cite{Zhao.Truhlar:new.2006}, M06 \cite{Zhao.Truhlar:M06.2008}, or M06-2X
\cite{Zhao.Truhlar:M06.2008} are very sensitive to the inclusion of the current density,
while TPSS \cite{Tao.Perdew.ea:Climbing.2003} and related functionals are rather insensitive
\cite{Holzer.Franke.ea:Current.2022, Holzer.Franzke.ea:Assessing.2021,
Franzke.Holzer.ea:NMR.2022, Franzke.Holzer:Impact.2022, Bruder.Franzke.ea:Paramagnetic.2022,
Pausch.Holzer:Linear.2022, Schattenberg.Kaupp:Effect.2021, Bruder.Franzke.ea:Zero-Field.2023,
Franzke.Bruder.ea:Paramagnetic.2023, Grotjahn.Furche.ea:Importance.2022}.

Our work aims to further fill this gap by introducing a 2c formalism to enable simulations of
systems with periodic boundary conditions in DFT up to the level of meta-generalized gradient
approximations including relativistic effects for energies, band structures, energy gradients,
and the stress tensor to allow for structure optimizations.
For this purpose, existing functionalities for the relativistic treatment of discrete
systems with a 2c formalism \cite{Armbruster.Weigend.ea:Self-consistent.2008, Baldes.Weigend:Efficient.2013,
Holzer.Franke.ea:Current.2022} will be introduced into \textsc{Turbomole}'s module \texttt{RIPER} 
\cite{Burow.Sierka.ea:Resolution.2009, Burow.Sierka:Linear.2011, Lazarski.Burow.ea:Density.2015,
Lazarski.Burow.ea:Density.2016, Grajciar:Low.2015, Irmler.Burow.ea:Robust.2018, Irmler.Pauly:Multipole-based.2019}
for periodic electronic structure calculations.

\section{Theory and Implementation}
\label{sec:theoretical-methods}
In this section, we extend the 2c DFT formalism of Armbruster \textit{et al.}\
\cite{Armbruster.Weigend.ea:Self-consistent.2008, Baldes.Weigend:Efficient.2013}
to periodic systems \cite{Burow.Sierka.ea:Resolution.2009, Burow.Sierka:Linear.2011,
Lazarski.Burow.ea:Density.2015, Lazarski.Burow.ea:Density.2016, Grajciar:Low.2015}
and also consider the non-collinear approach of Scalmani and Frisch
\cite{Scalmani.Frisch:New.2012, Bulik.Scalmani.ea:Noncollinear.2013,
Egidi.Sun.ea:Two-Component.2017} as an alternative to the canonical DFT
formalism. Furthermore, the computational demands are assessed and compared to standard
one-component (1c) procedures.

\subsection{Two-Component SCF Formalism}
\label{subsec:scf-theory}
The 2c DFT formalism for periodic systems is based on the quasi-relativistic Hamiltonian in the Born--Oppenheimer
approximation
\begin{eqnarray}
H^{\vec{L}}= \sum_{i=1}^n \left[ \boldsymbol{\sigma}_0 h^{0,\vec{L}}_i +
\vec{\boldsymbol{\sigma}}\cdot \vec{h}^{\,\textrm{SO},\vec{L}}_{i}\right] +
V^{\vec{L}}_{\textrm{ee}} + V^{\vec{L}}_{\textrm{NN}} \label{Ham}
\end{eqnarray}
for a unit cell (UC) $\vec{L}$ and $n$ electrons. The terms in the Hamiltonian include an effective spin--orbit (SO) operator,
$\vec{\boldsymbol{\sigma}}\cdot \vec{h}^{\,\textrm{SO},\vec{L}}_i$, that consists of the standard $(2 \crossproduct 2)$ Pauli
matrices $\boldsymbol{\sigma}_u ~ (u=x,y,z)$ and the vector operator $\vec{h}^{\,\textrm{SO},\vec{L}}_i$, which depends
on the orbital momentum and parameters fitted to results of, e.g., a relativistic four-component (4c) treatment
\cite{Dolg.Cao:Relativistic.2012, InC-Dolg:2002, InB-Dolg:2017}.
Hence, the operators $\vec{h}^{\,\textrm{SO},\vec{L}}_i$ lead to antisymmetric integrals in real space.
Further, a spin-independent one-particle Hamiltonian $h^{0, \vec{L}}_i$ that describes the kinetic energy of the
electrons and their potential energy in the field of the nuclei is included. $\boldsymbol{\sigma}_0$ is
the identity matrix. $V^{\vec{L}}_{\textrm{ee}}$ and $V^{\vec{L}}_{\textrm{NN}}$ represent the electron-electron
interaction and the repulsion of the nuclei. Additionally, Grimme's DFT dispersion corrections can
be handled like the nuclear repulsion operator \cite{Grimme.Antony.ea:consistent.2010, Grimme.Ehrlich.ea:Effect.2011}.
We stress that the presence of the spin--orbit operator
$\vec{\boldsymbol{\sigma}}\cdot \vec{h}_i^{\,\textrm{SO},\vec{L}}$ causes the spin not to
be a conserved quantity anymore, i.e.\ the spin is not a `good' quantum number.

The orbital wave functions $\vec{\psi}_i{^{\vec{k}}}$ thus have to be adjusted. Due to the translational symmetry of solids,
they are a linear combination of Bloch functions 
\begin{eqnarray}
\phi_\mu^{\vec{k}}(\vec{r})=\frac{1}{\sqrt{N_{\text{UC}}}}\sum_{\vec{L}}e^{\text{i}\vec{k}\cdot\vec{L}}\xi_\mu^{\vec{L}}(\vec{r})
\label{eq:bloch}
\end{eqnarray}
that are expanded in terms of real-valued GTO basis functions
\begin{eqnarray}
\xi_{\mu}^{\vec{L}}\left(\vec{r}\right) = \xi_\mu \left(\vec{r}-\vec{R}_\mu - \vec{L}\right), \label{eq:xi}
\end{eqnarray}
centered at the atomic position $\vec{R}_\mu$ in direct lattice cell $\vec{L}$ over all $N_{\text{UC}}$ unit cells.
The wave functions are so-called 2c spinors
\begin{eqnarray}
\vec{\psi}_i{^{\vec{k}}}(\vec{r}) = \begin{pmatrix} \psi_i^{\alpha, \vec{k}}(\vec{r}) \\ \psi_i^{\beta, \vec{k}}(\vec{r})\end{pmatrix}
= \sum_{\mu} \begin{pmatrix} c_{\mu i}^{\alpha, \vec{k}} \\ c_{\mu i}^{ \beta, \vec{k}} \end{pmatrix}
\phi_\mu^{\vec{k}}(\vec{r}) 
\end{eqnarray}
with the spin indices $\alpha$ and $\beta$, and complex coefficients $c_{\mu i}^{\sigma, \vec{k}}$.
The spinors depend on the wave vector $\vec{k}$ within the unit cell in reciprocal space, called the
first Brillouin zone (FBZ), and the band index $i$.

Using this basis set expansion, the generalized Kohn--Sham (GKS) equations become
\begin{eqnarray}
\begin{pmatrix}
\boldsymbol{F}^{\alpha \alpha, \vec{k}} & \boldsymbol{F}^{\alpha \beta, \vec{k}} \\
\boldsymbol{F}^{\beta \alpha, \vec{k}} & \boldsymbol{F}^{\beta \beta, \vec{k}} \\
\end{pmatrix}
\begin{pmatrix}
\boldsymbol{C}^{\alpha, \vec{k}} \\ \boldsymbol{C}^{ \beta, \vec{k}}
\end{pmatrix}
=
\begin{pmatrix}
\boldsymbol{S}^{\vec{k}} & \boldsymbol{0} \\
\boldsymbol{0} & \boldsymbol{S}^{\vec{k}} \\
\end{pmatrix}
\begin{pmatrix}
\boldsymbol{C}^{\alpha, \vec{k}} \\ \boldsymbol{C}^{ \beta, \vec{k}}
\end{pmatrix}
\boldsymbol{\epsilon}^{\vec{k}}
\quad
\label{eq:RH}
\end{eqnarray}
with doubled dimension compared to the standard Kohn--Sham formulation.
Here, $\boldsymbol{F}$ is the Kohn--Sham--Fock matrix, $\boldsymbol{C}^{\sigma, \vec{k}}$ is the
matrix of the spinor coefficients $c_{\mu i}^{\sigma, \vec{k}}$,
and $\boldsymbol{\epsilon}^{\vec{k}}$ is the diagonal matrix of the spinor eigenvalues $\epsilon_i^{\sigma, \vec{k}}$.
These Roothaan--Hall-like equations can be solved for each $\vec{k}$ point separately in the FBZ.
Both the overlap matrix between the GTOs
\begin{eqnarray}
S_{\mu \nu}^{\vec{k}} = \sum_{\vec{L}} e^{\text{i}\vec{k}\cdot\vec{L}} S_{\mu \nu}^{ \vec{L}}
\end{eqnarray}
with
\begin{eqnarray}
S_{\mu \nu}^{\vec{L}} & = & S_{\mu \nu}^{\vec{0} \vec{L}} =
\int \xi_\mu^{\vec{0}}\left(\vec{r}\right) \xi_\nu^{\vec{L}}\left(\vec{r}\right) \textrm{d}^3r \label{S}
\end{eqnarray}
and the Kohn-Sham--Fock matrix
\begin{eqnarray}
F_{\mu \nu}^{\vec{k}} = \sum_{\vec{L}} e^{\text{i}\vec{k} \cdot \vec{L}} F_{\mu \nu}^{\vec{L}}
\end{eqnarray}
are obtained via a Fourier transformation of the real space matrices.
Following Eq.~(\ref{Ham}), the Kohn--Sham--Fock matrix
components are \cite{Armbruster.Weigend.ea:Self-consistent.2008}
\allowdisplaybreaks
\begin{eqnarray}
F_{\mu \nu}^{\alpha \alpha, \vec{L}} & = & T_{\mu \nu}^{\vec{L}} + J_{\mu \nu}^{\vec{L}} +
h^{\textrm{SO},z, \vec{L}}_{\mu \nu} + X_{ \mu \nu}^{\alpha \alpha, \vec{L}},  \label{F1}\\ 
F_{\mu \nu}^{\alpha \beta, \vec{L}} & = & h^{\textrm{SO}, x, \vec{L}}_{\mu \nu} - \text{i} h^{\textrm{SO}, y, \vec{L}}_{\mu \nu} +  X_{\mu \nu}^{\alpha \beta, \vec{L}}, \label{F2} \\
F_{\mu \nu}^{\beta \beta, \vec{L}} & = & T_{\mu \nu}^{\vec{L}} + J_{\mu \nu}^{\vec{L}} - 
h^{\textrm{SO}, z, \vec{L}}_{\mu \nu} + X_{ \mu \nu}^{\beta \beta, \vec{L}}, \label{F3}
\end{eqnarray}
consisting of the kinetic energy matrix elements $T_{\mu \nu}^{\vec{L}}$, the Coulomb matrix elements $J_{\mu \nu}^{\vec{L}}$,
the exchange-correlation (XC) matrix elements $X_{ \mu \nu}^{\sigma \sigma', \vec{L}}$
(with $\sigma$ and $\sigma'$ being the spin indices),
and the spin--orbit matrix elements $h^{\textrm{SO}, u, \vec{L}}_{\mu \nu}$ ($u = x,y,z$).
Thus, the total energy reads
\begin{equation}
E_{\text{SCF}}= E_{\textrm{T}}
+ E_{\textrm{J}}+ E_{\textrm{SO}}
+ E_{\textrm{XC}} + E_{\textrm{NN}}
\end{equation}
with the nuclear repulsion term $E_{\textrm{NN}}$. Note that the scalar ECP contribution $h_{\mu \nu}^{0,\vec{L}}$
is included in the kinetic energy matrix (see below).
For open-shell systems without space-inversion symmetry, the Kohn--Sham equations generally need to
be solved for $+\vec{k}$ and $-\vec{k}$ explicitly \cite{Desmarais.Flament.ea:Spin-orbit.2020}.

In this 2c formalism, the general density matrix reads
\begin{equation}
\begin{split}
\boldsymbol{\rho}(\vec{r},\vec{r}\,') = & \frac{1}{V_{\textrm{FBZ}}} \sum_{i=1}^n \int_{\textrm{FBZ}}^{\epsilon_i^{\vec{k}} < \epsilon_{\textrm{F}}} \vec{\psi}_i{^{\vec{k}}}(\vec{r})\left(\vec{\psi}_i{^{\vec{k}}}(\vec{r}\,')\right)^\dagger \textrm{d}^3k \\
= & \left(\begin{array}{cc} \rho^{\alpha \alpha}(\vec{r},\vec{r}\,') & \rho^{\alpha \beta}(\vec{r},\vec{r}\,') \\
\rho^{\beta \alpha}(\vec{r},\vec{r}\,') & \rho^{\beta \beta}(\vec{r},\vec{r}\,') \end{array}\right).
\label{Density}
\end{split}
\end{equation}
It involves an integral over the FBZ, $V_\text{FBZ}$ denotes the volume of the FBZ, and 
$\epsilon_{\textrm{F}}$ is the Fermi level.
The density matrix not only contains the particle contribution
\begin{eqnarray}
\rho_{\textrm{p}}(\vec{r},\vec{r}\,') = \textrm{Tr}\left[\boldsymbol{\rho}(\vec{r},\vec{r}\,')\right] =
\sum_{\vec{L}} \textrm{Tr}\left[\boldsymbol{\rho}^{\vec{L}}(\vec{r},\vec{r}\,')\right],
\end{eqnarray}
but also the spin-vector contribution
\begin{eqnarray}
\vec{\rho}_{\textrm{m}}(\vec{r}, \vec{r}\,') = \textrm{Tr}\left[\vec{\boldsymbol{\sigma}} \boldsymbol{\rho}(\vec{r}, \vec{r}\,')\right]
= \sum_{\vec{L}} \textrm{Tr}\left[\vec{\boldsymbol{\sigma}} \boldsymbol{\rho}^{\vec{L}}(\vec{r}, \vec{r}\,')\right]
\end{eqnarray}
in terms of $N_{\text{UC}}$ unit-cell contributions associated with the lattice vectors $\vec{L}$, c.f.~Eqs.~(\ref{eq:bloch})
and (\ref{eq:xi}). For easy handling within the program structure, the density matrix $\rho(\vec{r},\vec{r}\,')$
is divided into the four spin contributions according to
\begin{eqnarray}
\rho^{\sigma \sigma'}(\vec{r}, \vec{r}\,') & = &  \sum_{\vec{L}}\rho^{\sigma \sigma', \vec{L}}(\vec{r}, \vec{r}\,'), \\
\rho^{\sigma \sigma',\vec{L}}(\vec{r}, \vec{r}\,') & = & \sum_{\mu \nu} \sum_{\vec{L}\,'} \xi_\mu^{\vec{L}} \left(\vec{r}\right)
D_{\mu \nu}^{\sigma \sigma', \vec{L} \vec{L}\,'}
\xi_\nu^{\vec{L}\,'}\left(\vec{r}\,'\right). \quad
\end{eqnarray}
Thus, the electron density $\rho^{\sigma \sigma'}(\vec{r})$ can be obtained from the atomic-orbital (AO) real space density matrix
\begin{equation}
{\small
\begin{split} D_{\mu \nu}^{\sigma \sigma', \vec{L} \vec{L}\,'}
 = \frac{1}{V_{\textrm{FBZ}}} \sum_{i=1}^n \int_{\textrm{FBZ}}^{\epsilon_i^{\vec{k}} < \epsilon_{\textrm{F}}} 
e^{\text{i} \vec{k}\cdot \left[\vec{L} - \vec{L}\,'\right]}
\left(c_{\mu i}^{\sigma,\vec{k}} c_{\nu i}^{*\sigma',\vec{k}} \right) \textrm{d}^3k. \label{D} \quad
\end{split}
}
\end{equation}
We also define the shorthand notation $D_{\mu \nu}^{\sigma \sigma',\vec{L}} = D_{\mu \nu}^{\sigma \sigma',\vec{L} \vec{0}}$
with
\begin{equation}
D_{\mu \nu}^{\sigma \sigma',\vec{L}} = \frac{1}{V_{\textrm{FBZ}}} \sum_{i=1}^n
\int_{\textrm{FBZ}}^{\epsilon_i^{\vec{k}} < \epsilon_{\textrm{F}}} 
e^{\text{i} \vec{k}\cdot \vec{L}}
\left(c_{\mu i}^{\sigma,\vec{k}} ~ c_{\nu i}^{*\sigma',\vec{k}} \right) ~ \textrm{d}^3k. \label{D0} \quad
\end{equation}
Note that the AO density matrix is a complex quantity in the 2c formalism and
consists of eight blocks, i.e.\ four real and four imaginary spin blocks.
The particle density, $\rho_{\textrm{p}}(\vec{r})=\rho_{\textrm{p}}(\vec{r},\vec{r})$,
relates to the electron number $n$ via
\begin{eqnarray}
\int \rho_{\textrm{p}}(\vec{r}) ~ \textrm{d}^3r = n,
\end{eqnarray}
while the spin-vector density or spin-magnetization vector, $\vec{\rho}_{\textrm{m}}(\vec{r}) = \vec{\rho}_{\textrm{m}}(\vec{r}, \vec{r})$,
is related to the expectation value of the system's spin according to
\begin{eqnarray}
\left\langle \vec{S} \right\rangle =
\frac{1}{2}\int\vec{\rho}_{\textrm{m}}(\vec{r}) ~ \textrm{d}^3r.
\end{eqnarray}
In the absence of spin--orbit coupling, this is simply half the number of unpaired
or spin-polarized electrons. In terms of AO density matrices,
the electron number and the spin expectation value make use of
$\Re(D_{\mu \nu}^{\alpha \alpha,\vec{L}}) + \Re(D_{\mu \nu}^{\beta \beta,\vec{L}})$, and
$\Re(D_{\mu \nu}^{\alpha \beta,\vec{L}}) + \Re(D_{\mu \nu}^{\beta \alpha,\vec{L}})$,
$\Im(D_{\mu \nu}^{\alpha \beta,\vec{L}}) - \Im(D_{\mu \nu}^{\beta \alpha,\vec{L}})$,
$\Re(D_{\mu \nu}^{\alpha \alpha,\vec{L}}) - \Re(D_{\mu \nu}^{\beta \beta,\vec{L}})$,
respectively.

Evaluation of the kinetic term 
\begin{eqnarray}
T_{\mu \nu}^{\vec{L}} = \int \xi_\mu^{\vec{0}}\left(\vec{r}\right) \left[-\frac{\nabla^2}{2} + V^{\textrm{ECP}}_{\textrm{sr}}\left(\vec{r}\right) \right] \xi_\nu^{\vec{L}}\left(\vec{r}\right) ~ \textrm{d}^3r, \label{Tmat}
\end{eqnarray}
that also includes scalar relativistic energy corrections, employing the ECPs $V^{\textrm{ECP}}_{\textrm{sr}}$,
needs no further modification compared to the 1c case, i.e.\ the elements $T_{\mu \nu}^{\vec{L}}$ form
a 1c matrix. Therefore, the respective energy is obtained as
{\begin{equation}
\begin{split}
E_{\text{T}} = & \sum_{\mu \nu} \sum_{\vec{L}} \Tr \left[ T^{\vec{L}}_{\mu \nu} \boldsymbol{\sigma}_0
\boldsymbol{D}^{\vec{L}}_{\nu \mu} \right] \\
= & \, \sum_{\mu \nu} \sum_{\vec{L}} T_{\mu \nu}^{\vec{L}} \left[\Re(D_{\nu \mu}^{\alpha \alpha, \vec{L}}) +
\Re(D_{\nu \mu}^{\beta \beta, \vec{L}}) \right],
\end{split}
\end{equation}
where $\boldsymbol{D}^{\vec{L}}_{\nu \mu}$ denotes the $\left(2 \crossproduct 2\right)$ AO density matrix made up of the
four complex spin contributions $\boldsymbol{D}^{\sigma \sigma', \vec{L}}_{\nu \mu}$, c.f.\ Eq.~(\ref{D0}).

Evaluation of the Coulomb interaction 
\begin{eqnarray}
J_{\mu \nu}^{\vec{L}} = \int \xi_\mu^{\vec{0}} \left(\vec{r}\right) \frac{\rho_{\textrm{p}}(\vec{r}\,')-\rho_{\textrm{N}}(\vec{r}\,')}{\left|\vec{r}-\vec{r}\,'\right|} \xi_\nu^{\vec{L}}\left(\vec{r}\right) ~ \textrm{d}^3r \; \textrm{d}^3r' \label{J}
\end{eqnarray}
utilizes a hierarchical scheme and density fitting \cite{Burow.Sierka.ea:Resolution.2009}
with the particle density $\rho_{\textrm{p}}(\vec{r})$ and the charge distribution $\rho_{\textrm{N}}(\vec{r})$.
The latter is also written as a sum of UC contributions according to
\begin{eqnarray}
\rho_{\textrm{N}}(\vec{r}) & = & \sum_{\vec{L}} \rho_{\textrm{N}}^{\vec{L}}(\vec{r}), \\
\rho_{\textrm{N}}^{\vec{L}}(\vec{r}) & = & \sum_{I} Z^{\textrm{eff}}_I \delta(\vec{r} -\vec{R}_I - \vec{L}),
\end{eqnarray}
with the effective charges $Z^{\textrm{eff}}_I$ and positions $\vec{R}_I$ of the nuclei $I$ in the reference cell.
The 2c implementation of $J_{\mu \nu}^{\vec{L}}$ is straightforward based on the 1c Kohn--Sham scheme,
as the Coulomb interaction is independent of the spin and only depends on the particle density.
The elements $J_{\mu \nu}^{\vec{L}}$ hence form a 1c matrix.
We refer to Refs.~\cite{Burow.Sierka.ea:Resolution.2009,Lazarski.Burow.ea:Density.2015}
for details on the construction of the Coulomb matrix and the corresponding energy.
Here, we just state that the Coulomb integrals are split into a crystal near field (CNF) and
a crystal far field (CFF), yielding the following two contributions to the Coulomb energy
\begin{equation}
E_{\text{J}} = E_{\text{J}, \text{CNF}} + E_{\text{J}, \text{CFF}}.
\end{equation}

As the first new terms compared to the 1c formalism, the spin--orbit ECP (SO-ECP)
terms $h^{\textrm{SO}, x, \vec{L}}_{\mu \nu}$, $h^{\textrm{SO}, y,\vec{L}}_{\mu \nu}$, and
$h^{\textrm{SO}, z, \vec{L}}_{\mu \nu}$ are taken into account. They read
\begin{equation}
\begin{split}
\boldsymbol{h}^{\textrm{SO},\vec{L}}_{\mu \nu} = & \begin{pmatrix}
h^{\textrm{SO}, z, \vec{L}}_{\mu \nu} & h^{\textrm{SO}, x, \vec{L}}_{\mu \nu} - \text{i} h^{\textrm{SO}, y, \vec{L}}_{\mu \nu} \\
h^{\textrm{SO}, x, \vec{L}}_{\mu \nu} + \text{i} h^{\textrm{SO}, y, \vec{L}}_{\mu \nu} &  - h^{\textrm{SO}, z, \vec{L}}_{\mu \nu}\\ \end{pmatrix} \\
 = &  \int \xi_\mu^{\vec{0}}\left(\vec{r}\right) \boldsymbol{V}^{\textrm{ECP}}_{\textrm{SO}}\left(\vec{r}\right)
\xi_\nu^{\vec{L}}\left(\vec{r}\right) ~ \textrm{d}^3r \label{eq:H_SO}
\end{split}
\end{equation}
with the effective spin--orbit potential $\boldsymbol{V}^{\textrm{ECP}}_{\textrm{SO}}\left(\vec{r}\right)$.
The associated spin--orbit energy is obtained as
\begin{equation}
E_{\textrm{SO}} =  \sum_{\mu \nu} \sum_{\vec{L}} \Tr \left[ \boldsymbol{h}^{\textrm{SO},\vec{L}}_{\mu \nu}
\boldsymbol{D}^{\vec{L}}_{\nu \mu} \right].
\end{equation}
Here, the antisymmetric linear combinations of the AO density matrices are needed,
i.e.\ $\Im(D_{\nu \mu}^{\alpha \beta,\vec{L}}) + \Im(D_{\nu \mu}^{\beta \alpha,\vec{L}})$,
$\Re(D_{\nu \mu}^{\alpha \beta,\vec{L}}) - \Re(D_{\nu \mu}^{\beta \alpha,\vec{L}})$, and
$\Im(D_{\nu \mu}^{\alpha \alpha,\vec{L}}) - \Im(D_{\nu \mu}^{\beta \beta,\vec{L}})$
for the spin--orbit $x$, $y$, and $z$ components, respectively. These density matrices
also give rise to a spin-current density contribution \cite{Holzer.Franke.ea:Current.2022}.
In contrast, $E_{\textrm{SO}}$ does not depend explicitly on the particle current density.

The final term, that has to be considered, involves the exchange-correlation
matrices $\boldsymbol{X}^{\sigma \sigma'}$. We will first discuss the semilocal
or ``pure'' (i.e.\ non-hybrid) contributions and then describe the generalization
to include Fock exchange. The semilocal XC matrices not only depend on the
particle-number density $\rho_{\textrm{p}}$ but also on the spin-vector density
or spin-magnetization vector $\vec{\rho}_{\textrm{m}}$. In the non-collinear
Kramers-unrestricted (KU) formalism, the XC potential operator of the local
spin density approximation is given by
\begin{equation}
\label{eq:vxc1}
\begin{split}
\boldsymbol{V}^{\textrm{XC}}\left[\rho_{\textrm{p}}(\vec{r}), \vec{\rho}_{\textrm{m}}(\vec{r}) \right] =
&  \frac{\delta E_{\text{XC}}}{\delta \rho_{\textrm{p}}(\vec{r})} \boldsymbol{\sigma}_0
+ \sum_{u = \{x,y,z\}} \frac{\delta E_{\text{XC}}}{\delta \rho_{\textrm{m},u}(\vec{r})} \boldsymbol{\sigma}_u \\
= &  \frac{\delta E_{\text{XC}}}{\delta \rho_{\textrm{p}}(\vec{r})} \boldsymbol{\sigma}_0
+ \frac{\delta E_{\text{XC}}}{\delta |\vec{\rho}_{\textrm{m}}(\vec{r})|} \frac{\vec{\rho}_{\textrm{m}}(\vec{r})
\cdot \vec{\boldsymbol{\sigma}}}{|\vec{\rho}_{\textrm{m}}(\vec{r})|}.
\end{split}
\end{equation}
It is based on the exchange-correlation energy as a functional of the particle density
and the norm of the spin-vector density
\begin{eqnarray}
E_{\text{XC}} & = &  E_{\text{XC}}\left[ \rho_{\textrm{p}}(\vec{r}), |\vec{\rho}_{\textrm{m}}(\vec{r})| \right]
= E_{\text{XC}}\left[  \rho_{\uparrow,\downarrow} (\vec{r}) \right], \quad \\
 \rho_{\uparrow,\downarrow} (\vec{r}) & = & \frac{1}{2} \left[ \rho_{\textrm{p}}(\vec{r}) \pm |\vec{\rho}_{\textrm{m}}(\vec{r})| \right].
\label{eq:spin-up-down-density}
\end{eqnarray}
More generally, the semilocal XC energy for a given density functional approximation $f^{\text{XC}}$ reads
\begin{equation}
 E_{\text{XC}} = \int_{\text{UC}} f^{\text{XC}} \left[
\rho_{\uparrow,\downarrow}(\vec{r}), 
\gamma_{\uparrow \uparrow,\uparrow \downarrow,\downarrow \downarrow}(\vec{r}),
\tau_{\uparrow,\downarrow}(\vec{r}) \right] ~ \textrm{d}^3 r,
\end{equation}
using the auxiliary variable $\gamma_{\sigma \sigma'} = \frac{1}{4} (\vec{\nabla} \rho_{\sigma}) \cdot (\vec{\nabla} \rho_{\sigma'})$
with $\sigma, \sigma' \in \{\uparrow, \downarrow \}$ 
and the kinetic energy density $\tau$ for meta-generalized gradient approximations.
The kinetic energy density may be defined analogously to the electron density as
\begin{equation}
{\small
\boldsymbol{\tau}(\vec{r},\vec{r}\,') = \frac{1}{V_{\textrm{FBZ}}} \sum_{i=1}^n
\int_{\textrm{FBZ}}^{\epsilon_i^{\vec{k}} < \epsilon_{\textrm{F}}} \left( \vec{\nabla} \vec{\psi}_i{^{\vec{k}}}(\vec{r}) \right)
\left(\vec{\nabla} \vec{\psi}_i{^{\vec{k}}}(\vec{r}\,') \right)^\dagger \mathrm{d}^3 k.
\label{tau}
}
\end{equation}
The particle kinetic energy density and the spin contributions then follow as
\begin{eqnarray}
\tau_{\textrm{p}}(\vec{r}) & = & \textrm{Tr}\left[\boldsymbol{\tau}(\vec{r})\right] = \textrm{Tr}\left[\boldsymbol{\tau}(\vec{r},\vec{r})\right], \\
\vec{\tau}_{\textrm{m}}(\vec{r}) & = & \textrm{Tr}\left[\vec{\boldsymbol{\sigma}} \boldsymbol{\tau}(\vec{r})\right]
= \textrm{Tr}\left[\vec{\boldsymbol{\sigma}} \boldsymbol{\tau}(\vec{r}, \vec{r})\right],
\end{eqnarray}
which allow to construct $\tau_{\uparrow,\downarrow} (\vec{r})$ analogously to Eq.~(\ref{eq:spin-up-down-density}).
We note that the current density correction for $\boldsymbol{\tau}(\vec{r},\vec{r}\,')$
is presently neglected in \texttt{RIPER}, while we have implemented it for molecular systems \cite{Holzer.Franke.ea:Current.2022}.
Therefore, only the real and symmetric part of the spin magnetization in direct space is used for
the semilocal XC contribution in \texttt{RIPER}, i.e.
\begin{eqnarray}
\rho_{\textrm{m}}^x (\vec{r}) & = & \rho^{\alpha \beta}(\vec{r}) + \rho^{\beta \alpha}(\vec{r}), \\
\rho_{\textrm{m}}^y (\vec{r}) & = & \text{i} \left( \rho^{\alpha \beta}(\vec{r}) - \rho^{\beta \alpha}(\vec{r}) \right), \\
\rho_{\textrm{m}}^z (\vec{r}) & = & \rho^{\alpha \alpha}(\vec{r}) - \rho^{\beta \beta}(\vec{r}).
\end{eqnarray}
In terms of AO density matrices, we need the symmetric linear combinations
$\Re(D_{\nu \mu}^{\alpha \beta,\vec{L}}) + \Re(D_{\nu \mu}^{\beta \alpha,\vec{L}})$,
$\Im(D_{\nu \mu}^{\alpha \beta,\vec{L}}) - \Im(D_{\nu \mu}^{\beta \alpha,\vec{L}})$, and
$\Re(D_{\nu \mu}^{\alpha \alpha,\vec{L}}) - \Re(D_{\nu \mu}^{\beta \beta,\vec{L}})$
for the non-collinear $x$, $y$, and $z$ components.

For an implementation with real AO basis functions, it is advantageous to directly calculate the
semilocal exchange-correlation potential in a matrix representation as
\begin{equation}
\boldsymbol{X}_{\mu\nu}^{\vec{L}}=\int
\xi_\mu^{\vec{0}}\left(\vec{r}\right) \boldsymbol{V}^{\textrm{XC}}(\vec{r}) \xi_\nu^{\vec{L}}
\left(\vec{r}\right)\textrm{d}^3r
\end{equation}
by differentiation of the exchange-correlation energy with respect to the corresponding AO density matrices
\cite{Gill.Johnson.ea:standard.1993}.
Thus, the quantities for the 2c formalism follow as
\begin{equation}
\label{eq:Vxc0}
{\small
\begin{split}
X^{0, \vec{L}}_{\mu \nu} = & \phantom{+} \int \frac{1}{2} \left[
\frac{\partial f^{\text{XC}}}{\partial \rho_{\uparrow}} +
\frac{\partial f^{\text{XC}}}{\partial \rho_{\downarrow}} \right]
\xi_{\mu}^{\vec{0}} (\vec{r}) ~ \xi_{\nu}^{\vec{L}} (\vec{r}) ~ \textrm{d}^3 r \\
 - \int & \frac{1}{2}
\left[ 
2\frac{\partial f^{\text{XC}}}{\partial \gamma_{\uparrow \uparrow}} \vec{\nabla} \rho_{\uparrow} +
2\frac{\partial f^{\text{XC}}}{\partial \gamma_{\downarrow \downarrow}} \vec{\nabla} \rho_{\downarrow}  +
\frac{\partial f^{\text{XC}}}{\partial \gamma_{\uparrow \downarrow}} (\vec{\nabla} \rho_{\uparrow} + \vec{\nabla} \rho_{\downarrow}) \right] \cdot \\
 & \left[ \left\{ \vec{\nabla} \xi_{\mu}^{\vec{0}} (\vec{r}) \right\} \xi_{\nu}^{\vec{L}} (\vec{r})
+ \xi_{\mu}^{\vec{0}} (\vec{r}) \left\{ \vec{\nabla} \xi_{\nu}^{\vec{L}} (\vec{r}) \right\}
\right]  \textrm{d}^3 r \\
 + \int & \frac{1}{2} \left[ 
\frac{\partial f^{\text{XC}}}{\partial \tau_{\uparrow}} + 
\frac{\partial f^{\text{XC}}}{\partial \tau_{\downarrow}}
\right]
\left[ \vec{\nabla} \xi_{\mu}^{\vec{0}} (\vec{r})  \right] \cdot
\left[ \vec{\nabla} \xi_{\nu}^{\vec{L}} (\vec{r})  \right] \textrm{d}^3 r
\end{split}
}
\end{equation}
for the scalar contribution, which is multiplied with $\boldsymbol{\sigma}_0$, and
\begin{equation}
\label{eq:Vxcm}
{\small
\begin{split}
X^{u, \vec{L}}_{\mu \nu} = & \frac{\rho_{\textrm{m}}^u}{|\vec{\rho}_{\textrm{m}}|} \int \frac{1}{2} \left[
\frac{\partial f^{\text{XC}}}{\partial \rho_{\uparrow}} -
\frac{\partial f^{\text{XC}}}{\partial \rho_{\downarrow}}
\right] \xi_{\mu}^{\vec{0}} (\vec{r}) ~ \xi_{\nu}^{\vec{L}}  (\vec{r}) ~ \textrm{d}^3 r \\
- \int & \frac{1}{2}
\left[ 
2\frac{\partial f^{\text{XC}}}{\partial \gamma_{\uparrow \uparrow}} \vec{\nabla} \rho_{\uparrow} -
2\frac{\partial f^{\text{XC}}}{\partial \gamma_{\downarrow \downarrow}} \vec{\nabla} \rho_{\downarrow}  -
\frac{\partial f^{\text{XC}}}{\partial \gamma_{\uparrow \downarrow}} (\vec{\nabla} \rho_{\uparrow} - \vec{\nabla} \rho_{\downarrow}) \right] \cdot \\
 & \left[ \left\{ \vec{\nabla} \xi_{\mu}^{\vec{0}} (\vec{r}) \right\} \xi_{\nu}^{\vec{L}} (\vec{r})
+ \xi_{\mu}^{\vec{0}} (\vec{r}) \left\{ \vec{\nabla} \xi_{\nu}^{\vec{L}}  (\vec{r}) \right\}
\right]  \textrm{d}^3 r \\
 + \int & \frac{1}{2} \left[ 
\frac{\partial f^{\text{XC}}}{\partial \tau_{\uparrow}} - 
\frac{\partial f^{\text{XC}}}{\partial \tau_{\downarrow}}
\right]
\left[ \vec{\nabla} \xi_{\mu}^{\vec{0}} (\vec{r})  \right] \cdot
\left[ \vec{\nabla} \xi_{\nu}^{\vec{L}} (\vec{r})  \right]  \textrm{d}^3 r
\end{split}
}
\end{equation}
for the spin contributions with $u \in \{x,y,z\}$. The four spin blocks for the Kohn--Sham--Fock
equations, c.f.\ Eqs.~(\ref{F1})--(\ref{F3}), can be evaluated from Eqs.~(\ref{eq:Vxc0}) and (\ref{eq:Vxcm})
according to
\begin{align}
X^{\alpha \alpha, \vec{L}}_{\mu \nu} = & \frac{1}{2}
(X^{0, \vec{L}}_{\mu \nu} + X^{z, \vec{L}}_{\mu \nu}), \label{eq:xaa} \\
X^{\beta \beta, \vec{L}}_{\mu \nu} = & \frac{1}{2}
(X^{0, \vec{L}}_{\mu \nu} - X^{z, \vec{L}}_{\mu \nu}), \label{eq:xbb} \\
X^{\alpha \beta, \vec{L}}_{\mu \nu} = & \frac{1}{2}
(X^{x, \vec{L}}_{\mu \nu} - \text{i} X^{y, \vec{L}}_{\mu \nu}), \label{eq:xab} \\
X^{\beta \alpha, \vec{L}}_{\mu \nu} = & \frac{1}{2}
(X^{x, \vec{L}}_{\mu \nu} - \text{i} X^{y, \vec{L}}_{\mu \nu})^{*} \label{eq:xba}.
\end{align}
We note that this form is a well established approximation in relativistic quantum mechanics
\cite{Desmarais.Komorovsky.ea:Spin-orbit.2021}.
The evaluation of the exchange-correlation matrix blocks uses numerical integration on a grid
and the hierarchical scheme presented in Ref.~\cite{Burow.Sierka:Linear.2011}.
Details on the underlying molecular 2c implementation can be found in
Refs.~\cite{Armbruster.Weigend.ea:Self-consistent.2008, Baldes.Weigend:Efficient.2013,
Holzer.Franke.ea:Current.2022}.

Equations~(\ref{eq:Vxc0}) and (\ref{eq:Vxcm}) correspond to the canonical non-collinear
formalism \cite{Kubler.Hock.ea:Density.1988, Van-Wullen:Spin.2002}.
An alternative approach with a local magnetic torque was presented by
Scalmani and Frisch (SF) \cite{Scalmani.Frisch:New.2012} for molecules and later extended
to periodic systems in Ref.~\cite{Bulik.Scalmani.ea:Noncollinear.2013}.
This approach is numerically slightly more stable than the canonical ansatz
\cite{Desmarais.Komorovsky.ea:Spin-orbit.2021}.
The following generalizations are used for the extension of the UKS
framework to the GKS picture
\begin{eqnarray}
\gamma_{\uparrow \uparrow, \downarrow \downarrow}^{\textrm{SF}} & = &
\frac{1}{4} \left[ \vec{\nabla} \rho_{\textrm{p}} \cdot \vec{\nabla} \rho_{\textrm{p}}
+ \vec{\nabla} \vec{\rho}_{\textrm{m}} \odot \vec{\nabla} \vec{\rho}_{\textrm{m}} \right]
\pm \frac{f_{\nabla}}{2} \Gamma, \\
\Gamma & = & \sqrt{ \left( \vec{\nabla} \rho_{\textrm{p}} \cdot \vec{\nabla} \vec{\rho}_{\textrm{m}} \right)
\circ \left( \vec{\nabla} \rho_{\textrm{p}} \cdot \vec{\nabla} \vec{\rho}_{\textrm{m}} \right) }, \\
\gamma_{\uparrow \downarrow}^{\textrm{SF}} & = & \frac{1}{4} \left[ \vec{\nabla} \rho_{\textrm{p}} \cdot \vec{\nabla} \rho_{\textrm{p}}
- \vec{\nabla} \vec{\rho}_{\textrm{m}} \odot \vec{\nabla} \vec{\rho}_{\textrm{m}} \right], \\
f_{\nabla} & = & \textrm{sgn} \left( \left[\vec{\nabla} \rho_{\textrm{p}} \cdot \vec{\nabla} \vec{\rho}_{\textrm{m}}  \right]
\circ \vec{\rho}_{\textrm{m}} \right), \\
\tau_{\uparrow,\downarrow}^{\textrm{SF}} & = & \frac{1}{2} \tau_{\textrm{p}} \pm \frac{f_{\tau}}{2} | \vec{\tau}_{\textrm{m}} |, \\
f_{\tau} & = & \textrm{sgn} \left( \vec{\tau}_{\textrm{m}} \circ \vec{\rho}_{\textrm{m}} \right).
\end{eqnarray}
Here, sgn is the signum function and $\circ$ indicates a scalar product of spin components,
while $\cdot$ refers to the scalar product in real space. Simultaneous scalar products are
denoted with $\odot$ according to
\begin{equation}
\vec{\nabla} \vec{\rho}_{\textrm{m}} \odot \vec{\nabla} \vec{\rho}_{\textrm{m}}
= \sum_{u \in \{x,y,z\}} \left( \vec{\nabla} \rho_{\textrm{m}}^u \right) \cdot \left( \vec{\nabla} \rho_{\textrm{m}}^u \right).
\end{equation}
The densities $\rho_{\uparrow}^{\textrm{SF}}$ and $\rho_{\downarrow}^{\textrm{SF}}$ are the same as
for the canonical ansatz \cite{Scalmani.Frisch:New.2012}, i.e., $\rho_{\uparrow}$ and $\rho_{\downarrow}$
of Eq.~(\ref{eq:spin-up-down-density}). The semilocal XC potential is again obtained via derivatives with respect
to the AO density matrices \cite{Egidi.Sun.ea:Two-Component.2017}, and this results in the scalar XC contribution
\begin{equation}
\label{eq:Vxc0-sf}
{\small
\begin{split}
X^{\textrm{SF}, 0, \vec{L}}_{\mu \nu} = & \phantom{+} \int \frac{1}{2}
\left[ \frac{\partial f^{\text{XC}}}{\partial \rho_{\uparrow}^{\textrm{SF}}} +
\frac{\partial f^{\text{XC}}}{\partial \rho_{\downarrow}^{\textrm{SF}}} \right]
\xi_{\mu}^{\vec{0}} (\vec{r}) ~ \xi_{\nu}^{\vec{L}} (\vec{r}) ~ \textrm{d}^3 r \\
 - \int & \frac{1}{2}
\left[ 
\frac{\partial f^{\text{XC}}}{\partial \gamma_{\uparrow \uparrow}^{\textrm{SF}}} +
\frac{\partial f^{\text{XC}}}{\partial \gamma_{\downarrow \downarrow}^{\textrm{SF}}} +
\frac{\partial f^{\text{XC}}}{\partial \gamma_{\uparrow \downarrow}^{\textrm{SF}}} \right] \vec{\nabla} \rho_{\textrm{p}} \cdot \\
 & \left[ \left\{ \vec{\nabla} \xi_{\mu}^{\vec{0}} (\vec{r}) \right\} \xi_{\nu}^{\vec{L}} (\vec{r})
+ \xi_{\mu}^{\vec{0}} (\vec{r}) \left\{ \vec{\nabla} \xi_{\nu}^{\vec{L}} (\vec{r}) \right\}
\right]  \textrm{d}^3 r \\
 - \int &
\left[ 
\frac{\partial f^{\text{XC}}}{\partial \gamma_{\uparrow \uparrow}^{\textrm{SF}}} -
\frac{\partial f^{\text{XC}}}{\partial \gamma_{\downarrow \downarrow}^{\textrm{SF}}} \right]
\frac{f_{\nabla}}{2} \frac{ \left( \vec{\nabla} \rho_{\textrm{p}} \cdot \vec{\nabla} \vec{\rho}_{\textrm{m}} \right)
\circ \vec{\nabla} \vec{\rho}_{\textrm{m}}}{\Gamma} ~ \cdot \\
 & \left[ \left\{ \vec{\nabla} \xi_{\mu}^{\vec{0}} (\vec{r}) \right\} \xi_{\nu}^{\vec{L}} (\vec{r})
+ \xi_{\mu}^{\vec{0}} (\vec{r}) \left\{ \vec{\nabla} \xi_{\nu}^{\vec{L}} (\vec{r}) \right\}
\right]  \textrm{d}^3 r \\
 + \int & \frac{1}{2} \left[ 
\frac{\partial f^{\text{XC}}}{\partial \tau_{\uparrow}^{\textrm{SF}}} + 
\frac{\partial f^{\text{XC}}}{\partial \tau_{\downarrow}^{\textrm{SF}}}
\right]
\left[ \vec{\nabla} \xi_{\mu}^{\vec{0}} (\vec{r})  \right] \cdot
\left[ \vec{\nabla} \xi_{\nu}^{\vec{L}} (\vec{r})  \right] \textrm{d}^3 r
\end{split}
}
\end{equation}
and the three spin--orbit contributions
\begin{equation}
\label{eq:Vxcu-sf}
{\small
\begin{split}
X^{\textrm{SF}, u, \vec{L}}_{\mu \nu} = & \phantom{+} \frac{\rho_{\textrm{m}}^u}{|\vec{\rho}_{\textrm{m}}|} \int \frac{1}{2} \left[
\frac{\partial f^{\text{XC}}}{\partial \rho_{\uparrow}^{\textrm{SF}}} - 
\frac{\partial f^{\text{XC}}}{\partial \rho_{\downarrow}^{\textrm{SF}}} \right]
\xi_{\mu}^{\vec{0}} (\vec{r}) ~ \xi_{\nu}^{\vec{L}} (\vec{r}) ~ \textrm{d}^3 r \\
 - \int & \frac{1}{2}
\left[ 
\frac{\partial f^{\text{XC}}}{\partial \gamma_{\uparrow \uparrow}^{\textrm{SF}}} +
\frac{\partial f^{\text{XC}}}{\partial \gamma_{\downarrow \downarrow}^{\textrm{SF}}} -
\frac{\partial f^{\text{XC}}}{\partial \gamma_{\uparrow \downarrow}^{\textrm{SF}}} \right] \vec{\nabla} \rho_{\textrm{m}}^u \cdot \\
 & \left[ \left\{ \vec{\nabla} \xi_{\mu}^{\vec{0}} (\vec{r}) \right\} \xi_{\nu}^{\vec{L}} (\vec{r})
+ \xi_{\mu}^{\vec{0}} (\vec{r}) \left\{ \vec{\nabla} \xi_{\nu}^{\vec{L}} (\vec{r}) \right\}
\right]  \textrm{d}^3 r \\
 - \int &
\left[ 
\frac{\partial f^{\text{XC}}}{\partial \gamma_{\uparrow \uparrow}^{\textrm{SF}}} -
\frac{\partial f^{\text{XC}}}{\partial \gamma_{\downarrow \downarrow}^{\textrm{SF}}} \right]
\frac{f_{\nabla}}{2} \frac{ \left( \vec{\nabla} \rho_{\textrm{p}} \cdot \vec{\nabla} \rho_{\textrm{m}}^u \right)
\vec{\nabla} \rho_{\textrm{p}}}{\Gamma} ~ \cdot \\
 & \left[ \left\{ \vec{\nabla} \xi_{\mu}^{\vec{0}} (\vec{r}) \right\} \xi_{\nu}^{\vec{L}} (\vec{r})
+ \xi_{\mu}^{\vec{0}} (\vec{r}) \left\{ \vec{\nabla} \xi_{\nu}^{\vec{L}} (\vec{r}) \right\}
\right]  \textrm{d}^3 r \\
+ \int & \left[ 
\frac{\partial f^{\text{XC}}}{\partial \tau_{\uparrow}^{\textrm{SF}}} -
\frac{\partial f^{\text{XC}}}{\partial \tau_{\downarrow}^{\textrm{SF}}}
\right] \frac{f_{\tau}}{2} \frac{\tau_{\textrm{m}}^u}{|\vec{\tau}_{\textrm{m}}|}
\left[ \vec{\nabla} \xi_{\mu}^{\vec{0}} (\vec{r})  \right] \cdot
\left[ \vec{\nabla} \xi_{\nu}^{\vec{L}} (\vec{r})  \right] \textrm{d}^3 r.
\end{split}
}
\end{equation}
The denominator including $\Gamma$ is detrimental for some applications as noted
in Ref.~\cite{Egidi.Sun.ea:Two-Component.2017}.
That is, an another transformation is used for regions with a small magnetization
length, i.e.\ spatial regions with $| \vec{\rho}_{\textrm{m}} (\vec{r})| < 10^{-12}$ in atomic units.
The expressions for this modified Scalmani--Frisch (mSF) formalism read
\begin{eqnarray}
\rho_{\uparrow \uparrow, \downarrow \downarrow}^{\textrm{mSF}} & = & 
\frac{1}{2} \left[ \rho_{\textrm{p}} \pm \rho_{\textrm{s}} \right] \\
\rho_{\textrm{s}} & = & \frac{1}{3} \left[ \rho_{\textrm{m}}^x + \rho_{\textrm{m}}^y + \rho_{\textrm{m}}^z \right] \\
\gamma_{\uparrow \uparrow, \downarrow \downarrow}^{\textrm{mSF}} & = &
\frac{1}{4} \left[ \vec{\nabla} \rho_{\textrm{p}} \cdot \vec{\nabla} \rho_{\textrm{p}}
+ \vec{\nabla} \vec{\rho}_{\textrm{m}} \odot \vec{\nabla} \vec{\rho}_{\textrm{m}} \right]
\pm \frac{f_{\nabla}}{2} \Gamma_{\text{s}} \\
\Gamma_{\text{s}} & = & \vec{\nabla} \rho_{\textrm{p}} \cdot \vec{\nabla} \rho_{\textrm{s}} \\
\gamma_{\uparrow \downarrow}^{\textrm{mSF}} & = & \frac{1}{4} \left[ \vec{\nabla} \rho_{\textrm{p}} \cdot \vec{\nabla} \rho_{\textrm{p}}
- \vec{\nabla} \vec{\rho}_{\textrm{m}} \odot \vec{\nabla} \vec{\rho}_{\textrm{m}} \right], \\
\tau_{\uparrow,\downarrow}^{\textrm{mSF}} & = & \frac{1}{2} \tau_{\textrm{p}} \pm \frac{f_{\tau}}{2} \tau_{\textrm{s}}, \\
\tau_{\textrm{s}} & = & \frac{1}{3} \left[  \tau_{\textrm{m}}^x +  \tau_{\textrm{m}}^y + \tau_{\textrm{m}}^z \right].
\end{eqnarray}
This leads to the semilocal XC potential matrices
\begin{equation}
\label{eq:Vxc0-msf}
{\small
\begin{split}
X^{\textrm{mSF}, 0, \vec{L}}_{\mu \nu} = & \phantom{+} \int \frac{1}{2} \left[
\frac{\partial f^{\text{XC}}}{\partial \rho_{\uparrow}^{\textrm{mSF}}} +
\frac{\partial f^{\text{XC}}}{\partial \rho_{\downarrow}^{\textrm{mSF}}} \right]
\xi_{\mu}^{\vec{0}} (\vec{r}) ~ \xi_{\nu}^{\vec{L}} (\vec{r}) ~ \textrm{d}^3 r \\
 - \int & \frac{1}{2}
\left[ 
\frac{\partial f^{\text{XC}}}{\partial \gamma_{\uparrow \uparrow}^{\textrm{mSF}}} +
\frac{\partial f^{\text{XC}}}{\partial \gamma_{\downarrow \downarrow}^{\textrm{mSF}}} +
\frac{\partial f^{\text{XC}}}{\partial \gamma_{\uparrow \downarrow}^{\textrm{mSF}}} \right] \vec{\nabla} \rho_{\textrm{p}} \cdot \\
 & \left[ \left\{ \vec{\nabla} \xi_{\mu}^{\vec{0}} (\vec{r}) \right\} \xi_{\nu}^{\vec{L}} (\vec{r})
+ \xi_{\mu}^{\vec{0}} (\vec{r}) \left\{ \vec{\nabla} \xi_{\nu}^{\vec{L}} (\vec{r}) \right\}
\right]  \textrm{d}^3 r \\
 - \int &
\left[ 
\frac{\partial f^{\text{XC}}}{\partial \gamma_{\uparrow \uparrow}^{\textrm{mSF}}} -
\frac{\partial f^{\text{XC}}}{\partial \gamma_{\downarrow \downarrow}^{\textrm{mSF}}} \right]
\frac{f_{\nabla}}{2} \vec{\nabla} \rho_{\textrm{s}} \cdot
 \\
 & \left[ \left\{ \vec{\nabla} \xi_{\mu}^{\vec{0}} (\vec{r}) \right\} \xi_{\nu}^{\vec{L}} (\vec{r})
+ \xi_{\mu}^{\vec{0}} (\vec{r}) \left\{ \vec{\nabla} \xi_{\nu}^{\vec{L}} (\vec{r}) \right\}
\right]  \textrm{d}^3 r \\
 + \int & \frac{1}{2} \left[ 
\frac{\partial f^{\text{XC}}}{\partial \tau_{\uparrow}^{\textrm{mSF}}} + 
\frac{\partial f^{\text{XC}}}{\partial \tau_{\downarrow}^{\textrm{mSF}}}
\right]
\left[ \vec{\nabla} \xi_{\mu}^{\vec{0}} (\vec{r})  \right] \cdot
\left[ \vec{\nabla} \xi_{\nu}^{\vec{L}} (\vec{r})  \right] \textrm{d}^3 r
\end{split}
}
\end{equation}
and
\begin{equation}
\label{eq:Vxcu-msf}
{\small
\begin{split}
X^{\textrm{mSF}, u, \vec{L}}_{\mu \nu} = & \phantom{+} \int \frac{1}{6} \left[
\frac{\partial f^{\text{XC}}}{\partial \rho_{\uparrow}^{\textrm{mSF}}} - 
\frac{\partial f^{\text{XC}}}{\partial \rho_{\downarrow}^{\textrm{mSF}}} \right]
\xi_{\mu}^{\vec{0}} (\vec{r}) ~ \xi_{\nu}^{\vec{L}} (\vec{r}) ~ \textrm{d}^3 r \\
 - \int & \frac{1}{2}
\left[ 
\frac{\partial f^{\text{XC}}}{\partial \gamma_{\uparrow \uparrow}^{\textrm{mSF}}} +
\frac{\partial f^{\text{XC}}}{\partial \gamma_{\downarrow \downarrow}^{\textrm{mSF}}} -
\frac{\partial f^{\text{XC}}}{\partial \gamma_{\uparrow \downarrow}^{\textrm{mSF}}} \right] \vec{\nabla} \rho_{\textrm{m}}^u \cdot \\
 & \left[ \left\{ \vec{\nabla} \xi_{\mu}^{\vec{0}} (\vec{r}) \right\} \xi_{\nu}^{\vec{L}} (\vec{r})
+ \xi_{\mu}^{\vec{0}} (\vec{r}) \left\{ \vec{\nabla} \xi_{\nu}^{\vec{L}} (\vec{r}) \right\}
\right]  \textrm{d}^3 r \\
 - \int &
\left[ 
\frac{\partial f^{\text{XC}}}{\partial \gamma_{\uparrow \uparrow}^{\textrm{mSF}}} -
\frac{\partial f^{\text{XC}}}{\partial \gamma_{\downarrow \downarrow}^{\textrm{mSF}}} \right]
\frac{f_{\nabla}}{6} \vec{\nabla} \rho_{\textrm{p}} \cdot \\
 & \left[ \left\{ \vec{\nabla} \xi_{\mu}^{\vec{0}} (\vec{r}) \right\} \xi_{\nu}^{\vec{L}} (\vec{r})
+ \xi_{\mu}^{\vec{0}} (\vec{r}) \left\{ \vec{\nabla} \xi_{\nu}^{\vec{L}} (\vec{r}) \right\}
\right]  \textrm{d}^3 r \\
+ \int \Big[ &
\frac{\partial f^{\text{XC}}}{\partial \tau_{\uparrow}^{\textrm{mSF}}} -
\frac{\partial f^{\text{XC}}}{\partial \tau_{\downarrow}^{\textrm{mSF}}}
\Big]
\frac{f_{\tau}}{6}
\left[ \vec{\nabla} \xi_{\mu}^{\vec{0}} (\vec{r})  \right] \cdot
\left[ \vec{\nabla} \xi_{\nu}^{\vec{L}} (\vec{r})  \right] \textrm{d}^3 r.
\end{split}
}
\end{equation}

The non-collinear formalism is needed for open-shell systems.
For closed-shell systems, the real and symmetric part of the spin-vector density matrix
vanishes due to Kramers' theorem, i.e., 
$\Re(D_{\nu \mu}^{\alpha \beta,\vec{L}}) + \Re(D_{\nu \mu}^{\beta \alpha,\vec{L}})$,
$\Im(D_{\nu \mu}^{\alpha \beta,\vec{L}}) - \Im(D_{\nu \mu}^{\beta \alpha,\vec{L}})$, and
$\Re(D_{\nu \mu}^{\alpha \alpha,\vec{L}}) - \Re(D_{\nu \mu}^{\beta \beta,\vec{L}})$ are zero.
Further, time-reversal symmetry holds.
Therefore, the general or unrestricted 2c formalism can be reduced to a Kramers-restricted
(KR) framework and the semilocal XC potential can be evaluated as done in the 1c formalism.

Hybrid density functionals \cite{Csonka.Perdew.ea:Global.2010, Becke:Perspective.2014,
Mardirossian.Head-Gordon:Thirty.2017} include a portion of non-local Fock exchange.
Thus, the exchange-correlation matrix becomes
\begin{equation}
\boldsymbol{X}^{\sigma \sigma'} = \boldsymbol{X}^{\sigma \sigma'}_{\textrm{C}} + (1- a)
\boldsymbol{X}^{\sigma \sigma'}_{\textrm{X}} + a \boldsymbol{K}^{\sigma \sigma'}
\end{equation}
with $\boldsymbol{X}^{\sigma \sigma'}_{\textrm{C}}$ and $\boldsymbol{X}^{\sigma \sigma'}_{\textrm{X}}$
denoting the semilocal DFT correlation and exchange contributions discussed so far, and $a$ is
the mixing parameter. The Fock exchange matrix $\boldsymbol{K}^{\sigma \sigma'}$ reads
\begin{equation}
\begin{split}
K^{\sigma \sigma', \vec{L} \vec{L}\,'}_{\mu \nu} = &
\sum_{\lambda \kappa} \sum_{\vec{M} \vec{N}} D_{\lambda \kappa}^{\sigma \sigma', \vec{M} \vec{N}}
\int \int \left[ \xi_{\mu}^{\vec{L}} (\vec{r}) ~ \xi_{\kappa}^{\vec{N}} (\vec{r}\,') \right. \\
& \left. \frac{1}{|\vec{r} - \vec{r}\,'|} ~ \xi_{\lambda}^{\vec{M}}  (\vec{r}) ~ \xi_{\nu}^{\vec{L}\,'}  (\vec{r}\,') \right]
\textrm{d}^3 r ~ \textrm{d}^3 r'.
\end{split}
\end{equation}
with $\vec{L}$, $\vec{L}\,'$, $\vec{M}$, $\vec{N}$ referring to lattice vectors.
For range-separated hybrid functionals, the two-electron interaction operator
is replaced with an effective operator
\cite{Gill.Adamson.ea:Coulomb-attenuated.1996, Leininger.Stoll.ea:Combining.1997,
Yanai.Tew.ea:new.2004, Henderson.Janesko.ea:Range.2008}.
The respective 2c complex form reads \cite{Bulik.Scalmani.ea:Noncollinear.2013}
\begin{equation}
\boldsymbol{K}^{\vec{L} \vec{L}\,'} = 
\begin{pmatrix}
\boldsymbol{K}^{\alpha \alpha} & \boldsymbol{K}^{\alpha \beta} \\
\boldsymbol{K}^{\beta \alpha}  & \boldsymbol{K}^{\beta \beta} 
\end{pmatrix}^{\vec{L} \vec{L}\,'}
\end{equation}
using all blocks of the 2c complex AO density matrix, i.e.\ the particle and spin density
contributions $\Re(D_{\nu \mu}^{\alpha \alpha,  \vec{M} \vec{N}}) + \Re(D_{\nu \mu}^{\beta \beta, \vec{M} \vec{N}})$,
$\Re(D_{\nu \mu}^{\alpha \beta, \vec{M} \vec{N}}) + \Re(D_{\nu \mu}^{\beta \alpha, \vec{M} \vec{N}})$,
$\Im(D_{\nu \mu}^{\alpha \beta, \vec{M} \vec{N}}) - \Im(D_{\nu \mu}^{\beta \alpha, \vec{M} \vec{N}})$,
$\Re(D_{\nu \mu}^{\alpha \alpha, \vec{M} \vec{N}}) - \Re(D_{\nu \mu}^{\beta \beta, \vec{M} \vec{N}})$,
as well as
$\Im(D_{\nu \mu}^{\alpha \alpha, \vec{M} \vec{N}}) + \Im(D_{\nu \mu}^{\beta \beta, \vec{M} \vec{N}})$,
$\Im(D_{\nu \mu}^{\alpha \beta, \vec{M} \vec{N}}) + \Im(D_{\nu \mu}^{\beta \alpha, \vec{M} \vec{N}})$,
$\Re(D_{\nu \mu}^{\alpha \beta, \vec{M} \vec{N}}) - \Re(D_{\nu \mu}^{\beta \alpha, \vec{M} \vec{N}})$, and
$\Im(D_{\nu \mu}^{\alpha \alpha, \vec{M} \vec{N}}) - \Im(D_{\nu \mu}^{\beta \beta, \vec{M} \vec{N}})$.
The latter four linear combinations are related to the particle current density and the three spin-current
densities \cite{Desmarais.Flament.ea:Spin-orbit.2020}. Thus, Fock exchange naturally includes
the current density in the formalism and also features a local magnetic torque as discussed
in Ref.~\cite{Desmarais.Flament.ea:Spin-orbit.2020}. However, the semilocal DFT exchange and
correlation parts still do not explicitly depend on the current density.

Based on the eight linear combinations above, the 2c Fock exchange can be evaluated
straightforwardly as discussed in Ref.~\cite{Bulik.Scalmani.ea:Noncollinear.2013}.
In this spirit, we extended the implementation of Ref.~\cite{Irmler.Burow.ea:Robust.2018},
making use of a minimum image convention or the truncated Coulomb interaction to the 2c
formalism.

For closed-shell Kramers-restricted systems, the spin density and particle current density
contributions vanish also for the Fock exchange. However, the spin-current density
contributions, i.e.\ the three linear combinations
$\Im(D_{\nu \mu}^{\alpha \beta, \vec{M} \vec{N}}) + \Im(D_{\nu \mu}^{\beta \alpha, \vec{M} \vec{N}})$,
$\Re(D_{\nu \mu}^{\alpha \beta, \vec{M} \vec{N}}) - \Re(D_{\nu \mu}^{\beta \alpha, \vec{M} \vec{N}})$, and
$\Im(D_{\nu \mu}^{\alpha \alpha, \vec{M} \vec{N}}) - \Im(D_{\nu \mu}^{\beta \beta, \vec{M} \vec{N}})$,
are not necessarily zero. Thus, the 2c formulation of the Fock exchange still requires
changes of the underlying 1c code for Kramers-restricted calculations.

\subsection{Two-Component Energy Gradients}
\label{subsec:gradients}

Energy gradients are required to optimize coordinates of atoms in the unit cell
and to optimize the corresponding lattice vectors based on the stress tensor
\cite{Lazarski.Burow.ea:Density.2016, Becker.Sierka:Density.2019,
Kudin.Scuseria:Analytic.2000, Bucko.Hafner.ea:Geometry.2005,
Doll:Analytical.2010, Knuth.Carbogno.ea:All-electron.2015}. Here, we
neglect Fock exchange and only consider semilocal or ``pure'' density
functional approximations.
The derivative of the SCF energy with respect to a nuclear displacement reads
\begin{equation}
E_{\text{SCF}}^{I, \lambda} = E_{\textrm{T}}^{I, \lambda}
+ E_{\textrm{J}}^{I, \lambda} + E_{\textrm{SO}}^{I, \lambda}
+ E_{\textrm{XC}}^{I, \lambda} + E_{\textrm{NN}}^{I, \lambda},
\label{eq:gradient}
\end{equation}
where the superscript $\{I, \lambda\}$ indicates that we move the nucleus $I$ in
the reference cell $\vec{L} = \vec{0}$ along the Cartesian direction $\lambda$.
For the positions $R_{I}^{\vec{L}} = R_{I}^{\vec{0}} + \vec{L}$ holds. 
Note that derivatives are generally formed
in the limit of a vanishing perturbation \cite{Schlegel:Geometry.2011, Pulay:Analytical.2014}.
The nuclear repulsion term $E_{\textrm{NN}}^{I, \lambda}$ associated with 
$V^{\vec{L}}_{\textrm{NN}}$ is trivial and the same as in the non-relativistic limit
and the 1c methodology.
According to Pulay, none of the derivatives includes a response of the density matrix
\cite{Pulay:Ab.1969}. This finding holds for both DFT and Hartree--Fock theory based
on converged 1c or 2c SCF procedures \cite{Versluis.Ziegler:determination.1988}.
Instead of such an explicit response, the energy-weighted density matrix $\boldsymbol{W}$ arises for
the first term given by
\begin{equation}
\begin{split}
E_{\textrm{T}}^{I, \lambda} = & {\phantom{-}}  \sum_{\mu \nu} \sum_{\vec{L}}
\left( T_{\mu \nu}^{\vec{L}} \right)^{I, \lambda}
\left[ \Re(D_{\nu \mu}^{\alpha \alpha, \vec{L}}) + \Re(D_{\nu \mu}^{\beta \beta, \vec{L}}) \right] \\
& - \sum_{\mu \nu} \sum_{\vec{L}} \left(S_{\mu \nu}^{\vec{L}} \right)^{I, \lambda}
\left[ \Re(W_{\nu \mu}^{\alpha \alpha, \vec{L}}) + \Re(W_{\nu \mu}^{\beta \beta, \vec{L}}) \right].
\end{split}
\end{equation}
Here, the energy-weighted density matrix is defined as
\begin{equation}
W_{\mu \nu}^{\sigma \sigma',\vec{L}} = \frac{1}{V_{\textrm{FBZ}}} \sum_{i=1}^n \int_{\textrm{FBZ}}^{\epsilon_i^{\vec{k}} < \epsilon_{\textrm{F}}} 
 e^{\text{i} \vec{k}\cdot\vec{L}} \left(c_{\mu i}^{\sigma,\vec{k}} ~ \epsilon_i^{\vec{k}} ~ c_{\nu i}^{*\sigma',\vec{k}} \right) \textrm{d}^3k,
\label{W}
\end{equation}
which only differs from the density matrix by the inclusion of the spinor energy $\epsilon_i^{\vec{k}}$ in
the integral, c.f.\ Eq.~(\ref{D0}).
Owing to the properties of the overlap matrix, we only need the real part of the diagonal
spin blocks, i.e.\ $\sigma = \sigma'$, which is the analog of the particle density matrix. 
Therefore, this term can be implemented straightforwardly based on the 1c routines.

The derivative of the Coulomb energy $E_{\textrm{J}}^{I, \lambda}$ does not depend
on the spin-vector density but only on the particle density. Therefore, its derivative is the
same as in the 1c formalism. Thus, it consists of a crystal near field and a crystal
far field contribution
\begin{equation}
E_{\textrm{J}}^{I, \lambda} = E_{\textrm{J}, \text{CNF}}^{I, \lambda}
+ E_{\textrm{J}, \text{CFF}}^{I, \lambda}.
\end{equation}
They are defined in Ref.~\cite{Lazarski.Burow.ea:Density.2016}, and we point to
this reference for details.

For the spin--orbit ECPs, the derivative $E_{\textrm{SO}}^{I, \lambda}$ follows similarly
to the kinetic energy contribution, and can be expressed as
\begin{equation}
E_{\textrm{SO}}^{I, \lambda} = \sum_{\mu \nu} \sum_{\vec{L}} \Tr \left[ 
\left(\boldsymbol{h}^{\textrm{SO},\vec{L}}_{\mu \nu} \right)^{I, \lambda}
\boldsymbol{D}^{\vec{L}}_{\nu \mu} \right].
\end{equation}
Consequently the one-electron SO-ECP integral derivatives are simply contracted
with the antisymmetric linear combinations of the complex AO density matrix. These
are the same as for the SCF energy calculation.

Finally, the derivative of the exchange-correlation energy $E_{\textrm{XC}}^{I, \lambda}$
is needed. Again, no derivative of the density matrix arises for first-order derivatives.
Therefore, we only need to form the derivative of the XC potential matrix
\begin{equation}
\left(\boldsymbol{X}_{\mu\nu}^{\vec{L}} \right)^{I, \lambda} = \phantom{+} \left( \int
\xi_\mu^{\vec{0}}\left(\vec{r}\right) ~ \boldsymbol{V}^{\textrm{XC}}(\vec{r}) ~ \xi_\nu^{\vec{L}}
\left(\vec{r}\right)\textrm{d}^3r \right)^{I, \lambda}.
\end{equation}
That is, the derivatives of $X^{0, \vec{L}}_{\mu \nu}$ and $X^{u, \vec{L}}_{\mu \nu}$ from Eqs.~(\ref{eq:Vxc0}) and
(\ref{eq:Vxcm}) (with $u \in \{x,y,z\}$) are calculated and contracted directly with the respective density
matrices to yield $E_{\textrm{XC}}^{I, \lambda}$. These derivatives are now essentially given by the
gradient of a product of Gaussian basis functions and the existing functional ingredients
for SCF energies, c.f.\ Refs.~\cite{Pople.Gill.ea:Kohn-Sham.1992,Johnson.Gill.ea:performance.1993}.
Weight derivatives for the numerical integration of the XC part are included based
on the 2c generalization of previous work \cite{Burow.Sierka:Linear.2011,Lazarski.Burow.ea:Density.2016,Becker.Sierka:Density.2019}.
For closed-shell systems, the Kramers restriction and time-reversal symmetry can be
exploited as done for SCF energies. This way, the 2c Kramers-restricted implementation
with semilocal functionals is almost completely available from an existing 1c implementation.

\subsection{Stress Tensor}
\label{subsec:stress-tensor}

Unit cell parameters are optimized with the stress tensor \cite{Becker.Sierka:Density.2019,
Doll:Analytical.2010, Knuth.Carbogno.ea:All-electron.2015} according to
\begin{equation}
\frac{\partial E_{\text{SCF}}}{\partial v_{n, p}} = \sum_{q = \{x,y,z\}} \left[ V_{\text{UC}} ~ \sigma_{pq}
- \sum_I \frac{\partial E_{\text{SCF}}}{\partial R_{I, p}} R_{I, q}
\right] \left( \boldsymbol{A}^{-1} \right)_{nq}
\end{equation}
where $\boldsymbol{A}$ is a $\left( 3 \times 3 \right)$ matrix consisting of the three vectors
$\vec{v}_n$ describing the unit cell of a three dimensional periodic system and
$p, q \in \{x, y, z\}$ are the Cartesian components.
The stress tensor components are given by
{\begin{equation}
\sigma_{pq} = \frac{1}{V_{\text{UC}}} \frac{\partial E_{\text{SCF}}}{\partial \epsilon_{pq}}
\end{equation}
with the volume of the unit cell $V_{\text{UC}}$ and the symmetric strain tensor $\epsilon_{pq}$.
The latter describes the change of an atomic position $\vec{R}_{I}$ in a direct
lattice cell $\vec{L}$ for an elastic deformation, i.e.
\begin{equation}
\left(R_{I, p}^{\vec{L}}\right)' = \sum_{q = \{x,y,z\}} \left(\delta_{pq} + \epsilon_{pq} \right) R_{I, q}^{\vec{L}}
\end{equation}
with the Kronecker delta $\delta_{pq}$.
The stress tensor follows as a sum over UC contributions
{\begin{equation}
\sigma_{pq} = \frac{1}{V_{\text{UC}}} \sum_{\vec{L}} \sum_I \frac{\partial E_{\text{SCF}}}{\partial R_{I, p}^{\vec{L}}} R_{I, q}^{\vec{L}}
\end{equation}
and the derivatives are obtained using Eq.~(\ref{eq:gradient}). This leads to
\begin{equation}
\begin{split}
\sigma_{pq} = & \phantom{ + } ~ \frac{1}{V_{\text{UC}}} \sum_{\vec{L}} \sum_I \left[
\frac{\partial E_{\textrm{T}}}{\partial R_{I, p}^{\vec{L}}}
+ \frac{\partial E_{\textrm{J}}}{\partial R_{I, p}^{\vec{L}}}
+ \frac{\partial E_{\textrm{NN}}}{\partial R_{I, p}^{\vec{L}}} \right] R_{I, q}^{\vec{L}} \\
& + \frac{1}{V_{\text{UC}}} \sum_{\vec{L}} \sum_I \left[
\frac{\partial E_{\textrm{SO}}}{\partial R_{I, p}^{\vec{L}}}
+ \frac{\partial E_{\textrm{XC}}}{\partial R_{I, p}^{\vec{L}}}
\right] R_{I, q}^{\vec{L}},
\label{eq:stress-contributions}
\end{split}
\end{equation}
where the kinetic energy term again includes the energy-weighted density matrix
\begin{equation}
\begin{split}
\frac{\partial E_{\textrm{T}}}{\partial R_{I, p}^{\vec{L}}} = & \phantom{-} ~ \sum_{\mu \nu} \sum_{\vec{L}\,'}
\frac{\partial T_{\mu \nu}^{\vec{L}\,'}}{\partial R_{I, p}^{\vec{L}}}
\left[ \Re(D_{\nu \mu}^{\alpha \alpha, \vec{L}\,'}) + \Re(D_{\nu \mu}^{\beta \beta, \vec{L}\,'}) \right] \\
& - \sum_{\mu \nu} \sum_{\vec{L}\,'} \frac{\partial S_{\mu \nu}^{\vec{L}\,'}}{\partial R_{I, p}^{\vec{L}}}
\left[ \Re(W_{\nu \mu}^{\alpha \alpha, \vec{L}\,'}) + \Re(W_{\nu \mu}^{\beta \beta, \vec{L}\,'}) \right].
\end{split}
\end{equation}
Hence, the calculation of the stress tensor contributions essentially reduces to the calculation
of derivatives with respect to nuclear displacements.

The Coulomb contribution is
evaluated like in the non-relativistic or scalar 1c limit \cite{Becker.Sierka:Density.2019},
whereas the calculation of the SO-ECP and XC terms in Eq.~(\ref{eq:stress-contributions}) requires the
additional AO density matrix linear combinations.

The derivatives of $E_{\textrm{SO}}$ use the antisymmetric linear
combinations $\Im(D_{\nu \mu}^{\alpha \beta,\vec{L}\,'}) + \Im(D_{\nu \mu}^{\beta \alpha,\vec{L}\,'})$,
$\Re(D_{\nu \mu}^{\alpha \beta,\vec{L}\,'}) - \Re(D_{\nu \mu}^{\beta \alpha,\vec{L}\,'})$, and
$\Im(D_{\nu \mu}^{\alpha \alpha,\vec{L}\,'}) - \Im(D_{\nu \mu}^{\beta \beta,\vec{L}\,'})$ for
the contraction with $\partial h^{\textrm{SO}, u, \vec{L}\,'}_{\mu \nu} / \partial R_{I, p}^{\vec{L}}$, with
$u$ denoting the spin--orbit components $x$, $y$, $z$.

For the XC contribution, the respective derivatives of
$X^{0, \vec{L}\,'}_{\mu \nu}$ and $X^{u, \vec{L}\,'}_{\mu \nu}$ are contracted with the symmetric
AO density matrices $\Re(D_{\nu \mu}^{\alpha \alpha,\vec{L}\,'}) + \Re(D_{\nu \mu}^{\beta \beta,\vec{L}\,'})$
for the scalar contributions, and 
$\Re(D_{\nu \mu}^{\alpha \beta,\vec{L}\,'}) + \Re(D_{\nu \mu}^{\beta \alpha,\vec{L}\,'})$,
$\Im(D_{\nu \mu}^{\alpha \beta,\vec{L}\,'}) - \Im(D_{\nu \mu}^{\beta \alpha,\vec{L}\,'})$, and
$\Re(D_{\nu \mu}^{\alpha \alpha,\vec{L}\,'}) - \Re(D_{\nu \mu}^{\beta \beta,\vec{L}\,'})$
for the non-collinear $x$, $y$, and $z$ components. The evaluation of the derivatives of the
XC potential ingredients (${\partial f^{\text{XC}}}/{\partial \rho}$, $\xi_{\mu}$, etc.)
is done as described in Refs.~\cite{Becker.Sierka:Density.2019,
Gill.Johnson.ea:standard.1993, Baker.Andzelm.ea:effect.1994, Stratmann.Scuseria.ea:Achieving.1996}.
We emphasize that weight derivatives of the DFT part are of great importance for the stress
tensor \cite{Baker.Andzelm.ea:effect.1994, Tobita.Hirata.ea:analytical.2003} and, therefore,
we always include them based on Refs.~\cite{Gill.Johnson.ea:standard.1993, Stratmann.Scuseria.ea:Achieving.1996}.

\subsection{Implementation}
\label{subsec:implementation}
Due to the integral evaluation in real space, our implementation is largely based on the
existing molecular implementations \cite{Armbruster.Weigend.ea:Self-consistent.2008,
Baldes.Weigend:Efficient.2013, Holzer.Franke.ea:Current.2022}. That is, the SO-ECP integrals
are evaluated with the McMurchie--Davidson scheme \cite{McMurchie.Davidson:Calculation.1981,
Pitzer.Winter:Spin-orbit.1991}. All one- and two-electron integral routines
are parallelized with the OpenMP paradigm \cite{OpenMP, Holzer.Franzke:2020}.
Algebraic operations make use of the Math Kernel Library (MKL). Depending on the
size of the basis set and the number of $k$ points in reciprocal space, parallelization
is either done over the $k$ points or for the algebraic operations and transformations
inside the loop over $k$ points. The first option is used with many $k$ points and
comparably small basis sets, whereas the second option is used for large basis sets
with more than, e.g., 5,000 functions in the sparse Cartesian AO basis.
The direct inversion in the iterative subspace (DIIS) is exploited to accelerate the
SCF convergence \cite{Pulay:Convergence.1980} in its \textGamma-point version.
Interfaces to \textsc{Libxc} \cite{Marques.Oliveira.ea:Libxc.2012,
Lehtola.Steigemann.ea:Recent.2018, LIBXC.2022} are provided to support (almost) all
semilocal density functional approximations. Both the canonical and the (modified)
Scalmani--Frisch non-collinear formalism were implemented. The latter was also
added for the molecular parts of the program suite, i.e.\ DFT energies
\cite{Armbruster.Weigend.ea:Self-consistent.2008, Baldes.Weigend:Efficient.2013,
Peng.Middendorf.ea:efficient.2013}, gradients \cite{Baldes.Weigend:Efficient.2013,
Franzke.Middendorf.ea:Efficient.2018}, electron paramagnetic resonance properties
\cite{Franzke.Yu:Hyperfine.2022, Franzke.Yu:Quasi-Relativistic.2022},
and the Green's function $GW$ approach \cite{Kehry.Franzke.ea:Quasirelativistic.2020,
Holzer.Klopper:Ionized.2019, Franzke.Holzer.ea:NMR.2022, Holzer:Practical.2023}.
Global and range-separated hybrid functionals are available for the SCF procedure.

Our implementation supports the SCF initial construction of the band structure based on
a discrete H\"uckel guess, core Hamiltonian guess, (non-collinear) superposition of atomic
densities, converged 1c molecular orbitals (MOs), 2c  molecular spinors, or 1c bands.
The Kramers-restricted framework can only be used on top of MOs or bands of an RKS calculation.
For the KU formalism, the starting wave function can be constructed as an eigenfunction
of the spin operators $\hat{S}_x$, $\hat{S}_y$, or $\hat{S}_z$.
For simplicity, we use the last option as default setting. By default, a threshold
of $10^{-6}$ is used for the eigenvalues of the overlap matrix in the transformation
to an orthogonal basis for periodic systems \cite{TURBOMOLE-manual}.

\begin{table*}[htbp]
\centering
\caption{Computation time in seconds for the SCF steps and number of iterations for a calculation of
a three-dimensional Pb crystal (PBE functional \cite{Perdew.Burke.ea:Generalized.1996},
dhf-TZVP-2c basis \cite{Weigend.Baldes:Segmented.2010})
with a single thread of an Intel Xeon Gold 6212U central processing unit @ 2.40 GHz.
The number of atoms per unit cell is one, the number of $k$ points is 32,768 and the chosen grid size for
the XC potential is 4 \cite{Treutler:Entwicklung.1995, Treutler.Ahlrichs:Efficient.1995}.
The code was compiled with the Intel Fortran Compiler 19.0.1.144 (no just-in-time flags).
The SCF initial guess is obtained by constructing bands
from H\"uckel theory. The overlap, kinetic energy, ECP, and SO-ECP integrals are evaluated
prior to entering the SCF iterations (``Pre-SCF'').
Timings for the overlap and kinetic energy matrix are omitted, as they
are negligible.
``$D_{\mu \nu}$ Build'' refers to the construction of the density matrices in real space,
``$J$'' to the evaluation of the Coulomb integrals, and ``XC'' to the numerical integration of the XC potential.
``Diag'' and ``Iter'' denote the diagonalization for the Kohn--Sham equations and the number of iterations.
``Time'' denotes the computation time per iteration of the SCF procedure.
Note that the energy contributions from the Coulomb term $J$ and the XC part are again calculated
after the last iteration. 
Bands are stored in binary format, while the MOs or spinors of the unit cell are
stored in ASCII format (Band/MO dump) after converging the SCF procedure (``Post-SCF'').
``Total'' refers to the complete computation time.
}
\begin{tabular}{@{\extracolsep{6pt}}l
S[table-format = 1.1]
S[table-format = 2.1]
S[table-format = 2.1]
S[table-format = 2.1]
S[table-format = 2.1]
S[table-format = 3.1]
S[table-format = 3.1]
S[table-format = 3.1]
S[table-format = 2.0]
S[table-format = 4.1]
@{}}
\toprule
 & \multicolumn{2}{c}{{\text{Pre-SCF}}} & \multicolumn{6}{c}{{\text{SCF}}} & \multicolumn{1}{c}{{\text{Post-SCF}}} \\
 \cmidrule{2-3} \cmidrule{4-9} \cmidrule{10-10}
 & {\text{ECP}} & {\text{SO-ECP}} & {\text{$D_{\mu \nu}$ Build}} &  {\text{$J$}} & {\text{XC}} & {\text{Diag}} &
 {\text{Time}} & {\text{Iter}} & {\text{Dump}} & {\text{Total}} \\
\hline
1c RKS    & 4.7 & {{--}} & 12.3 & 33.8 & 38.4 & 20.1  & 108.8 & 14 & 53.0  & 1656.4 \\
1c UKS    & 4.9 & {{--}} & 24.7 & 33.8 & 50.6 & 39.6  & 156.7 & 14 & 104.6 & 2488.2 \\
2c GKS KR & 4.7 & 38.0   & 49.4 & 33.5 & 37.9 & 82.7  & 214.3 & 14 & 150.2 & 3253.5 \\
2c GKS KU & 4.9 & 38.4   & 96.9 & 33.6 & 90.4 & 231.3 & 483.9 & 14 & 301.7 & 7050.8 \\
\bottomrule
\end{tabular}
\label{tab:CPU-Pb}
\end{table*}

The 2c geometry gradients and the related stress tensor were implemented based on the
1c routines \cite{Lazarski.Burow.ea:Density.2016, Becker.Sierka:Density.2019}.
The SO-ECP integral derivatives are currently evaluated numerically, c.f.\ the
molecular implementation in Ref.~\cite{Baldes.Weigend:Efficient.2013}.
The other contributions use analytical integral derivatives. All integral
derivatives are parallelized with the OpenMP scheme. Structures can be optimized
with Grimme's DFT dispersion correction D3 \cite{Grimme.Antony.ea:consistent.2010},
including Becke--Johnson (D3-BJ) damping \cite{Grimme.Ehrlich.ea:Effect.2011}.

\subsection{SCF Computation Times}
\label{subsec:computation-times}
Table~\ref{tab:CPU-Pb} shows the computation time for a three-dimensional
face-centered cubic (fcc) Pb crystal (primitive unit cell vectors' length 3.500\,\AA), employing 1c
restricted Kohn--Sham (RKS), 1c unrestricted Kohn--Sham (UKS), 2c GKS KR, and 2c GKS KU frameworks.
Here, we used the singlet state as initial guess, since this is the ground state in 2c calculations.
The number of atoms per unit cell is one, the number of $k$ points used is 32,768 and the chosen grid
size for the exchange-correlation potential is 4 \cite{Treutler:Entwicklung.1995, Treutler.Ahlrichs:Efficient.1995}.
Note that all frameworks except for 2c GKS KU use time-reversal symmetry for the $k$ points.
The PBE functional \cite{Perdew.Burke.ea:Generalized.1996} was combined with the dhf-TZVP-2c orbital
and auxiliary basis set \cite{Weigend.Baldes:Segmented.2010}. Thus, small-core Dirac--Fock ECPs
are applied (ECP-60) \cite{Metz.Stoll.ea:Small-core.2000}.
SCF thresholds of $10^{-8}$\,Hartree for the energy are chosen.
Gaussian smearing is used with a criterion of 0.005\,Hartree.
For comparison, the initial bands are obtained from a standard 1c H\"uckel guess.

The extension from a 1c to a 2c formalism has a crucial impact on the running time of the SCF procedure,
due to the complex nature of the SCF orbitals in real space. This covers the following major steps.
\begin{itemize}
\item \textbf{Construction of the density matrix.} For the 2c KU formalism, the particle density $\rho_\text{p}$
and the spin densities $\rho_\textrm{m}^x$, $\rho_\textrm{m}^y$, and $\rho_\textrm{m}^z$ have to be constructed
for both $+\vec{k}$ and $-\vec{k}$. Additionally, the densities for the contraction with the SO-ECP part
are needed. Formally, this leads to a factor of eight, however, several densities can be processed
simultaneously and symmetric/antisymmetric linear combinations are formed.
In the KR framework, time-reversal symmetry is used as done for the 1c approaches.
Thus, only half the number of $k$ points is needed and the computation time is reduced.
\item \textbf{Coulomb matrix.} There is no structural change and no increase of the computation time. 
\item \textbf{Exchange-correlation matrix.} For the 2c KU formalism, it is
necessary to evaluate four densities  $\rho_\text{p}$, $\rho_\text{m}^x$,
$\rho_\text{m}^y$, and $\rho_\text{m}^z$ on the DFT grid.
Compared to the 1c RKS procedure, this leads to a factor of four, or a factor of two compared
to the 1c UKS procedure. The 2c KR approach does not lead to any extra costs relative to the 1c RKS
ansatz.
\item\textbf{Diagonalization of the Kohn--Sham--Fock matrix.}
For periodic systems, a single diagonalization of the Kohn--Sham--Fock matrix is eight times more involved
compared to a 1c RKS calculation, as the dimensionality of matrices doubles due to the 2c construction.
As the computational time scales as cubically with the number of basis functions, $N_{\text{BF}}^3$,
this leads to the observed prefactor.
In UKS theory, the spin components of the Kohn--Sham--Fock matrix can be decoupled, which is not possible
in 2c GKS due to spin--orbit interaction.
\end{itemize}

The time investment needed for the diagonalization of the Kohn--Sham--Fock matrix depends on two factors, i.e.\ on
the number of basis functions $N_{\text{BF}}$ and the total number of $k$ points $N_k$, as one diagonalization
per $k$ point is executed. This means that the computational costs of the diagonalization are proportional to
$N_{\text{BF}}^3$ and to $N_k$. 
For smaller systems in terms of basis functions, the diagonalization time plays only a minor role for the total
computation time. Ultimately, it is obvious that the diagonalization time for the 2c calculations is substantially
larger than for the 1c calculations.

We conclude that for periodic systems, the 2c formalism provides an efficient way of dealing with physics based on
spin--orbit coupling. To reduce the computational costs, the unit cell size can be decreased to the minimum extent,
while simultaneously increasing the number of $k$ points in order to retain the same effective system size.
This way, a linear increase of the computational demands proportional to $N_k$ rather than the cubic proportionality
of $N_{\text{BF}}^3$ can be exploited.
That allows us to perform calculations on large systems in a reasonable amount of time with standard computer
hardware. Note, that this statement is not only true for 2c periodic calculations but for periodic calculations in general.

\section{Applications to Discrete and Periodic Systems}
\label{sec:application}
In this section, we present calculations of various systems to illustrate applications of the 2c GTO-based
approach and to show differences of the electronic structure by the inclusion of spin--orbit coupling.
First, we discuss discrete atoms and subsequently systems that are periodic in three to one dimensions.
Computational details are discussed in each subsection.

\subsection{Ionization Energies of Zero-Dimensional Heavy $p$-Block Atoms}
\label{subsec:atoms}
To begin with, we validate the new implementation in the \texttt{RIPER} module of TURBOMOLE
by comparison to the existing 2c molecular functionalities in the \texttt{RIDFT} module
\cite{Armbruster.Weigend.ea:Self-consistent.2008, Baldes.Weigend:Efficient.2013,
Eichkorn.Treutler.ea:Auxiliary.1995, Eichkorn.Weigend.ea:Auxiliary.1997, Weigend.Kattannek.ea:Approximated.2009}.
We study the ionization energies of heavy $p$ elements.
Compared to lighter elements, the ionization energies of heavy $5p$ and $6p$ elements follow a
different trend. While the light elements show an increasing ionization energy from left
to right in the periodic table with especially stable half-filled sub-shells \cite{Huheey.Keiter.ea:1994, Housecroft.Sharpe:2012},
this is not the case for the heavy elements as they feature a heavier core and, therefore,
stronger spin--orbit coupling. Strong spin--orbit coupling is accompanied by a larger splitting of $p$, $d$,
and $f$ orbitals. Here, the splitting of the $p$ shell is of particular importance.
The $p$ shells split into two-fold degenerated $p_{1/2}$ and four-fold degenerated $p_{3/2}$ orbitals,
indicating a stable $p^2$ configuration for lead. Thus, the ground state of lead is a triplet based
on 1c calculations, while it is a singlet in 2c approaches.

\begin{table}[t]
\centering
\caption{Ionization energies based on energy differences of the atom and its cation
for the heavy $p$-block elements In--I and Tl--At.
The PBE \cite{Perdew.Burke.ea:Generalized.1996} and PBE0 \cite{Adamo.Barone:Toward.1999}
exchange-correlation functionals combined with the dhf-SVP-2c Gaussian basis set
\cite{Weigend.Baldes:Segmented.2010} and small-core Dirac--Fock
ECPs \cite{Metz.Stoll.ea:Small-core.2000} are employed.
Experimentally determined ionization potentials (``Experiment'') are taken from Ref.~\cite{NIST}.
All values are given in eV.}
\begin{tabular}{@{\extracolsep{4pt}}
l
S[table-format = 1.2]
S[table-format = 1.2]
S[table-format = 1.2]
S[table-format = 1.2]
S[table-format = 2.2]
@{}}
\toprule
$5p$ block & {\text{In}} & {\text{Sn}} & {\text{Sb}} & {\text{Te}} & {\text{I}} \\
\midrule
1c \texttt{RIPER} PBE	     & 5.58 & 7.26 & 8.95 & 8.66  & 10.42 \\
2c \texttt{RIPER} PBE	     & 5.70 & 7.14 & 8.67 & 8.86  & 10.31 \\
2c \texttt{RIDFT} PBE        & 5.70 & 7.14 & 8.67 & 8.86  & 10.31 \\ 
1c \texttt{RIPER} PBE0	     & 5.61 & 7.31 & 9.05 & 8.68  & 10.55 \\
2c \texttt{RIPER} PBE0	     & 5.75 & 7.18 & 8.76 & 8.88  & 10.37 \\
2c \texttt{RIDFT} PBE0       & 5.75 & 7.18 & 8.76 & 8.88  & 10.37 \\ 
Experiment                   & 5.70 & 7.34 & 8.64 & 9.01  & 10.45 \\
\midrule
$6p$ block & {\text{Tl}} & {\text{Pb}} & {\text{Bi}} & {\text{Po}} & {\text{At}} \\
\midrule
1c \texttt{RIPER} PBE        & 5.41 & 6.97 & 8.55 & 8.26 & 9.99 \\
2c \texttt{RIPER} PBE        & 6.06 & 7.12 & 7.26 & 8.32 & 9.16 \\
2c \texttt{RIDFT} PBE        & 6.06 & 7.12 & 7.26 & 8.32 & 9.16 \\
1c \texttt{RIPER} PBE0       & 5.43 & 7.01 & 8.65 & 8.25 & 9.99 \\
2c \texttt{RIPER} PBE0       & 6.10 & 7.15 & 7.33 & 8.36 & 9.20 \\
2c \texttt{RIDFT} PBE0       & 6.10 & 7.15 & 7.33 & 8.36 & 9.20 \\
Experiment                   & 6.11 & 7.42 & 7.29 & 8.43 & 9.54 \\
\bottomrule
\end{tabular}
\label{tab:r0dion}
\end{table}

According to the results in Table~\ref{tab:r0dion}, the computed ionization energies validate this trend.
For all calculations, the PBE \cite{Perdew.Burke.ea:Generalized.1996} (grid size 4
\cite{Treutler.Ahlrichs:Efficient.1995, Treutler:Entwicklung.1995})
and PBE0 \cite{Adamo.Barone:Toward.1999} exchange-correlation functionals combined with
the dhf-SVP-2c GTO basis set \cite{Weigend.Baldes:Segmented.2010} with small-core
Dirac--Fock ECPs \cite{Metz.Stoll.ea:Small-core.2000} are employed.
The 2c procedure shows a good qualitative agreement with the experimental values, while the 1c calculations exhibit
partly large deviations. On quantitative considerations, the 2c PBE simulations yield an average deviation
of $0.14$\,eV from the experimental values, while we find an average deviation of $0.39$\,eV for the 1c procedure.
PBE0 yields similar overall errors, as it does not consistently improve the results.
Calculations with the \texttt{RIDFT} module lead to the same values as the \texttt{RIPER} module, demonstrating 
consistency.

\subsection{Band Structures of Three-Dimensional Gold and Lead Crystals} 
\label{subsec:au-pb}
Next, we demonstrate the consistency of our implementation with well-established codes
such as the plane-wave \textsc{Quantum Espresso} program \cite{QE-2020}. Therefore, we study the band structures
of three-dimensional gold and lead crystals. Figure~\ref{fig:r3dband} shows the simulated electronic band structure
for bulk gold and lead, forming a face-centered cubic (fcc) lattice with a lattice constant of $a = 4.0800$\,\AA \ and
$a = 4.9508$\,\AA, respectively \cite{Latdata}.

\begin{figure}
\centering
\includegraphics[width=0.99\columnwidth]{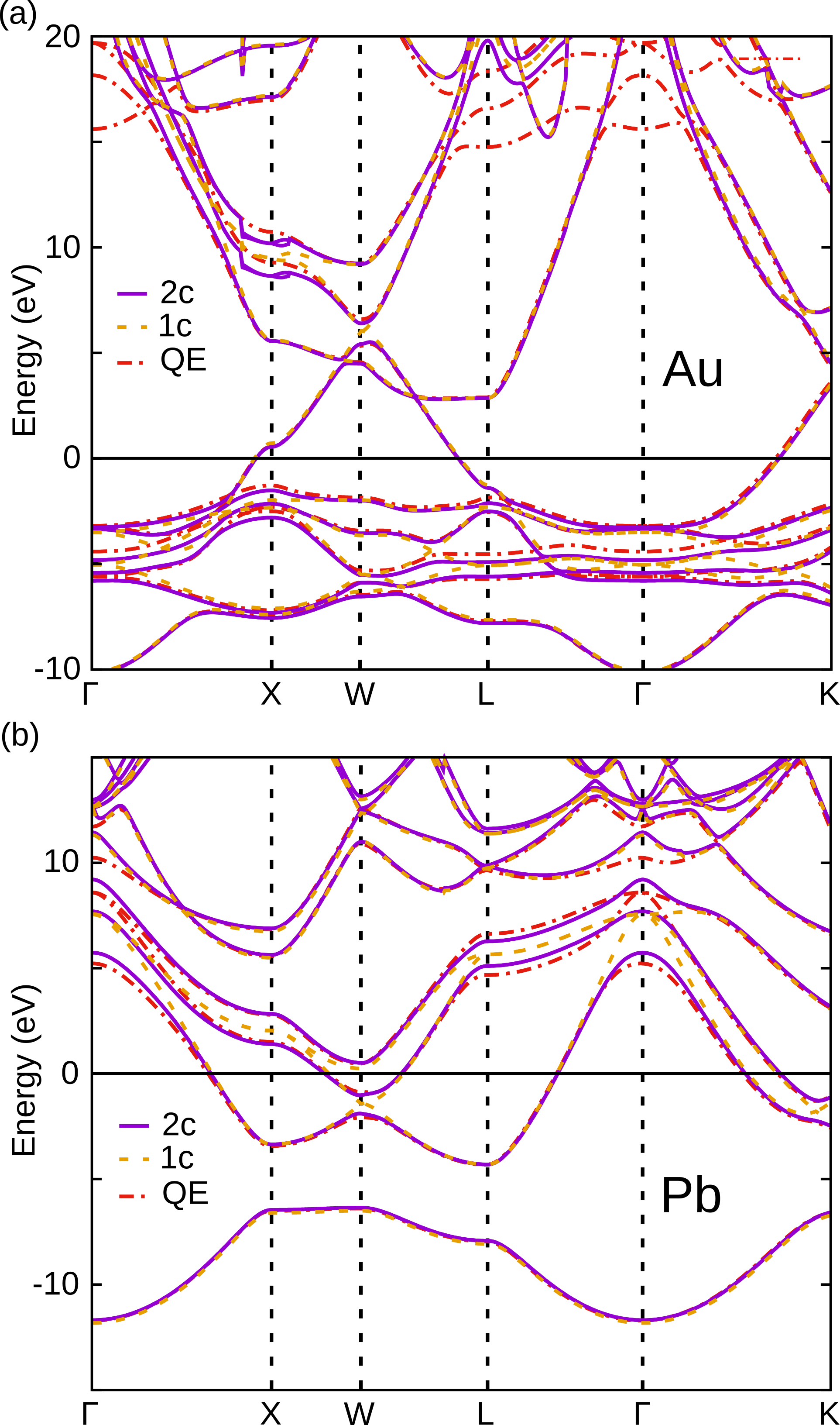}
\caption{Simulated band structure for (a) an fcc bulk gold crystal (lattice constant $a = 4.0800$\,\AA\ 
\cite{Latdata}) and (b) an fcc bulk lead crystal ($a = 4.9508$\,\AA\ \cite{Latdata}).
Results for the \texttt{RIPER} module with (2c) and without (1c) spin--orbit coupling are shown together
with relativistic 2c calculations using the \textsc{Quantum Espresso} (QE) code \cite{QE-2020}.
All computations employ the PBE exchange-correlation functional \cite{Perdew.Burke.ea:Generalized.1996}.
We use the dhf-TZVP-2c GTO basis set for \texttt{RIPER} and a plane-wave basis for \textsc{Quantum Espresso}.
Vertical dashed lines mark high symmetry points of the FBZ.}
\label{fig:r3dband}
\end{figure}

For the \texttt{RIPER} calculations (1c, 2c), the PBE functional \cite{Perdew.Burke.ea:Generalized.1996} (grid size 4) and the
dhf-TZVP-2c GTO basis set \cite{Weigend.Baldes:Segmented.2010}
with small-core Dirac--Fock ECPs \cite{Metz.Stoll.ea:Small-core.2000, Figgen.Peterson.ea:Energy-consistent.2009}
(ECP-60 for Au and Pb) are employed.
A $k$ point mesh of $32 \times 32 \times 32$ is used in combination with a Gaussian smearing of
$0.005$\,Hartree \cite{gsmear} to ensure convergence. Total SCF energies are converged up to $10^{-8}$\,Hartree.
For the \textsc{Quantum Espresso} calculations, we use the relativistic pseudopotentials \texttt{Au.rel-pbe-dn-rrkjus\_psl.0.1.UPF}
for gold and \texttt{Pb.rel-pbe-dn-rrkjus\_psl.0.2.2.UPF} for lead, respectively.
Cutoff radii of 100\,Hartree for the wave function and $1000$\,Hartree for the electron density are set,
while the other parameters such as the number of $k$ points, the Gaussian smearing, and the convergence
threshold are the same as for the \texttt{RIPER} calculations to ensure comparability.

The impact of spin--orbit interaction on the band structure of gold can be quantified by comparing 1c (orange) and 2c (purple) \texttt{RIPER} band structures
in Fig.~\ref{fig:r3dband}a. We find that the influence of spin--orbit interaction
is of minor relevance for bulk gold.
For large parts of the electronic band structure, both curves are in excellent agreement, especially close
to the Fermi level at $\epsilon_{\textrm{F}}=0$\,eV. The most prominent deviations between 1c and 2c calculations
occur for band splittings at about $5.5$\,eV for the W point and
at around $-5$\,eV for the $\Gamma$ point.
The comparison of the band structure of the 2c \texttt{RIPER} (purple) and the \textsc{Quantum Espresso} (red) calculations
reveals a good agreement of energies below and around the Fermi level $\epsilon_{\textrm{F}}=0$\,eV.
This indicates consistent results of the 2c \texttt{RIPER} implementation and the plane-wave
\textsc{Quantum Espresso} code, while providing the reduced calculation times of GTO over plane-wave basis sets.
However, the band structures exhibit larger deviations in the region above $7$\,eV.
This is not surprising as plane-wave basis sets are known to be superior in the description of energy
states in the continuum above the gold work function, which amounts to around $5$\,eV \cite{Sachtler1966work}. 

In contrast to gold, the band structure of lead in Fig.~\ref{fig:r3dband}b reveals a more pronounced impact of spin--orbit
coupling due to the heavier atoms.
Comparison of the 1c (orange) and the 2c (purple) \texttt{RIPER} band structures shows a band splitting of $0.87$\,eV at
the W point for the energetic region around $-1.5$\,eV. That is, the proper description of bulk electronic structures of heavy $p$
block systems generally necessitates the consideration of spin--orbit interaction. Figure~\ref{fig:r3dband}b
furthermore shows a similar behavior concerning the quality of our implementation for
bulk lead. For energies below and around the Fermi level $\epsilon_{\textrm{F}}=0$\,eV, the band structures of
both the relativistic \textsc{Quantum Espresso} (red) and the 2c \texttt{RIPER} (purple) calculations match almost perfectly.
In contrast, deviations are observed for energies above 5\,eV. Again, this is a consequence of the better
description of high energy states close to the continuum with plane-wave basis sets, c.f.\ the work function of
Pb amounts to around 4\,eV \cite{Anderson1956}.

\subsection{Band Gaps of Three-Dimensional Silver Halide Crystals}
\label{subsec:agx}

Silver halide crystals AgX (X = Cl, Br, I) are typical model systems to study relativistic
effects \cite{Peralta.Uddin.ea:Scalar.2005, Zhao.Zhang.ea:Exact.2016, Kadek.Repisky.ea:All-electron.2019,
Yeh.Shee.ea:Relativistic.2022}.
Studies have been performed with relativistic all-electron and ECP-based or quasi-relativistic
Hamiltonians. Due to the small gaps and the densely packed crystal structures, the systems
are a challenging case for computational simulations \cite{Kadek.Repisky.ea:All-electron.2019}.
In Table~\ref{tab:agx}, we compare results with our ECP-based implementation to previously
reported ones from relativistic all-electron approaches including all-electron exact two-component
(X2C) theory \cite{Zhao.Zhang.ea:Exact.2016} and the four-component (4c) Dirac--Kohn--Sham (DKS)
ansatz directly based on the many-electron Dirac--Coulomb equation \cite{Kadek.Repisky.ea:All-electron.2019}.
Additionally, we used a non-self-consistent second-variational-like (SV) approach for SO effects \cite{Huhn.Blum:One.2017}. 
Technically, the converged 1c bands are taken as initial guess for a 2c calculation and only one
2c diagonalization is carried out. For these calculations, the damping was turned off.
We note that in the terminology of Ref.~\cite{Huhn.Blum:One.2017} we use $N_{\text{states}} = N_{\text{basis}}$
and these authors would call our procedure a non-self-consistent first-variational approach.

\begin{table}
\begin{center}
\caption{Band gaps (in eV) of three-dimensional AgCl, AgBr, and AgI crystals
(lattice constants $a =5.612$\,\AA, $a = 5.843$\,\AA, $a = 6.169$\,\AA, all rocksalt structure)
at high symmetry points of the FBZ with the PBE functional \cite{Perdew.Burke.ea:Generalized.1996}
and grid size 4. Non-relativistic calculations (NR) are performed with the
TZVPalls2 (Ag) \cite{Ahlrichs.May:Contracted.2000},
def2-TZVP (Cl, Br) \cite{Weigend.Ahlrichs:Balanced.2005},
and the TZVPall (I) basis sets \cite{Ahlrichs.May:Contracted.2000},
whereas the ECP-based 1c and 2c calculations use the dhf-SVP bases \cite{Weigend.Baldes:Segmented.2010}
and small-core Dirac--Fock ECPs for Ag and I \cite{Figgen.Rauhut.ea:Energy-consistent.2005, Peterson.Shepler.ea:On.2006}.
Results with scalar relativistic (SR) and spin--orbit (SO) X2C employing (truncated) Slater-type
bases are taken from Ref.~\cite{Zhao.Zhang.ea:Exact.2016}.
Four-component DKS results are taken from Ref.~\cite{Kadek.Repisky.ea:All-electron.2019},
employing uncontracted double-$\zeta$ basis sets \cite{Peintinger.Oliveira.ea:Consistent.2013,
Dyall:Relativistic.2006,Dyall:Relativistic.2007}.
}
\label{tab:agx}
\begin{tabular}{@{\extracolsep{16pt}}l
S[table-format = -1.2]
S[table-format = -1.2]
S[table-format = -1.2]
S[table-format = -1.2]
@{}}
\toprule
AgCl & {\text{L--L}} & {\text{\textGamma--\textGamma}} & {\text{X--X}} & {\text{L--\textGamma}} \\
\midrule
NR     & 4.86 & 3.72 & 5.32 & 1.66 \\
1c ECP & 4.62 & 3.11 & 4.16 & 0.91 \\
SV ECP & 4.57 & 2.94 & 3.99 & 0.86 \\
2c ECP & 4.57 & 2.94 & 3.99 & 0.86 \\
SR X2C & 4.31 & 3.09 & 4.23 & 0.92 \\
SO X2C & 4.27 & 2.99 & 4.03 & 0.88 \\
4c DKS & 4.47 & 2.93 & 4.20 & 0.87 \\
\midrule
AgBr & {\text{L--L}} & {\text{\textGamma--\textGamma}} & {\text{X--X}} & {\text{L--\textGamma}} \\
\midrule
NR     & 4.28 & 3.19 & 4.77 & 1.56 \\
1c ECP & 4.07 & 2.64 & 3.70 & 0.86 \\
SV ECP & 4.02 & 2.64 & 3.54 & 0.82 \\
2c ECP & 4.02 & 2.64 & 3.54 & 0.82 \\
SR X2C & 3.87 & 2.43 & 3.87 & 0.68 \\
SO X2C & 3.77 & 2.25 & 3.67 & 0.60 \\
4c DKS & 3.82 & 2.24 & 3.68 & 0.61 \\
\midrule
AgI & {\text{L--L}} & {\text{\textGamma--\textGamma}} & {\text{X--X}} & {\text{L--X}} \\
\midrule
NR     & 3.89 & 3.42 & 3.71 & 1.48 \\
1c ECP & 3.49 & 2.16 & 2.98 & 0.65 \\
SV ECP & 3.25 & 1.82 & 2.69 & 0.41 \\
2c ECP & 3.25 & 1.82 & 2.69 & 0.41 \\
SR X2C & 3.42 & 2.27 & 3.07 & 0.74 \\
SO X2C & 3.17 & 1.90 & 2.76 & 0.49 \\
4c DKS & 3.25 & 1.88 & 2.74 & 0.49 \\
\bottomrule
\end{tabular}
\end{center}
\end{table}

Computational settings are chosen in accordance with previous studies
\cite{Peralta.Uddin.ea:Scalar.2005, Zhao.Zhang.ea:Exact.2016, Kadek.Repisky.ea:All-electron.2019}.
Lattice constants are taken from Ref.~\cite{Peralta.Uddin.ea:Scalar.2005}, i.e.\ $a =5.612$\,\AA
\ for AgCl, $a = 5.843$\,\AA \ for AgBr, and $a = 6.169$\,\AA \ for AgI. Calculations are carried
out with the primitive unit cell and a $k$ mesh of $7 \times 7 \times 7$ points. Increasing 
this to $12 \times 12 \times 12$ changed the energy of AgCl by less than $3\cdot10^{-6}$\,Hartree.
SCF convergence thresholds are set to $10^{-7}$\,Hartree. 
Non-relativistic calculations are performed with the TZVPalls2 (Ag),
def2-TZVP \cite{Weigend.Ahlrichs:Balanced.2005}
(Cl, Br), and the TZVPall (I) all-electron basis sets \cite{Ahlrichs.May:Contracted.2000}.
Scalar-relativistic and spin--orbit calculations employ the dhf-SVP basis
sets \cite{Weigend.Baldes:Segmented.2010} together with small-core Dirac--Fock ECPs
(ECP-28) \cite{Figgen.Rauhut.ea:Energy-consistent.2005, Peterson.Shepler.ea:On.2006}.
Note that we always transform the Kohn--Sham--Fock matrices into an orthogonal basis
by diagonalization of the overlap matrix (threshold $10^{-6}$,
c.f.\ Sec.~\ref{subsec:implementation}) \cite{Lazarski.Burow.ea:Density.2015}.
Thus, we have not removed diffuse functions, i.e.\ functions with exponents smaller than $0.1$\,Bohr$^{-1}$,
in contrast to related studies of silver halide crystals \cite{Peralta.Uddin.ea:Scalar.2005,
Zhao.Zhang.ea:Exact.2016}.

We find that the non-relativistic framework is clearly insufficient for all systems,
as it leads to a large deviation from the formally superior 4c DKS approach.
Relativistic effects are mainly captured by scalar ECPs, i.e.\ the spatial
contraction of the electron density is the leading relativistic
correction. However, some band gaps such as the \textGamma--\textGamma\ 
and the X--X gap of AgI require the inclusion of spin--orbit coupling,
as it lowers the gap by 0.2\,eV. Likewise, the \textGamma--\textGamma\ 
and X--X gaps of AgCl and AgBr are notably affected.
The SV and 2c-ECP approaches lead to the same band gaps for the
considered semilocal functionals. The same holds for total energies.
This finding is in line with the literature \cite{Huhn.Blum:One.2017}.
The SV approach is not sufficient for hybrid functionals which include
the current density in the two-electron part, see below and
Ref.~\cite{Desmarais.Ambrogio.ea:Generalized.2024}.

The 2c-ECP approach leads to a good agreement with relativistic
all-electron X2C and DKS schemes.
The qualitative trends of the band gaps are well described,
and the 2c approach generally reduces the deviation towards the 4c DKS results.
Exceptions in this regard are the X--X gaps of AgCl and AgBr.
The same holds for scalar-relativistic and spin--orbit X2C
calculations of the X--X gap of AgCl, see Refs.~\cite{Zhao.Zhang.ea:Exact.2016,
Kadek.Repisky.ea:All-electron.2019}.
For both ECPs and X2C, spin--orbit effects lower these band gaps by about
0.2\,eV. Therefore, considering these effects worsens the results,
which hints at subtle error cancellation.

\begin{table}[h!]
\begin{center}
\caption{Optimized lattice constants $a$ (in \AA, rocksalt structure)
of three-dimensional AgCl, AgBr, and AgI crystals and band gaps (in eV)
at high symmetry points of the FBZ with various density functional
approximations and the dhf-SVP basis sets \cite{Weigend.Baldes:Segmented.2010}.
Experimental data taken from Refs.~\cite{Berry:Physical.1955, Hidshaw.Lewis.ea:Elastic.1967,
Vogel.Kruger.ea:Ab.1998, Sommer:1986, Wang.Huang.ea:Highly.2009,
Bassani.Knox.ea:Band.1965, Piermarini.Weir:diamond.1962, Ves.Glotzel.ea:Pressure.1981},
as collected in Ref.~\cite{Zhao.Zhang.ea:Exact.2016}.
}
\label{tab:agx-dft}
\begin{tabular}{@{\extracolsep{4pt}}ll
S[table-format = 1.3]
S[table-format = -1.2]
S[table-format = -1.2]
S[table-format = -1.2]
S[table-format = -1.2]
@{}}
\toprule
AgCl & & $a$ & {\text{L--L}} & {\text{\textGamma--\textGamma}} & {\text{X--X}} & {\text{L--\textGamma}} \\
\midrule
S-VWN (V) & no D3  & 5.377 & 3.93 & 3.27 & 3.85 & 0.55 \\
PBE         & no D3  & 5.624 & 4.58 & 2.91 & 3.99 & 0.86 \\
PBEsol      & no D3  & 5.515 & 4.20 & 2.98 & 3.89 & 0.66 \\
TPSS        & no D3  & 5.586 & 4.55 & 3.04 & 4.13 & 0.97 \\
revTPSS     & no D3  & 5.561 & 4.38 & 3.02 & 4.13 & 0.89 \\
Tao--Mo     & no D3  & 5.541 & 4.24 & 3.10 & 4.24 & 0.97 \\
PKZB        & no D3  & 5.636 & 4.59 & 3.10 & 4.26 & 1.18 \\
r$^2$SCAN   & no D3  & 5.576 & 5.02 & 3.55 & 4.56 & 1.41 \\
PBE         & D3-BJ  & 5.537 & 4.47 & 3.08 & 4.03 & 0.85 \\
PBEsol      & D3-BJ  & 5.426 & 4.07 & 3.20 & 3.93 & 0.66 \\
TPSS        & D3-BJ  & 5.498 & 4.40 & 3.21 & 4.18 & 0.94 \\
revTPSS     & D3-BJ  & 5.476 & 4.23 & 3.20 & 4.17 & 0.87 \\
Tao--Mo     & D3-BJ  & 5.511 & 4.20 & 3.17 & 4.25 & 0.97 \\
r$^2$SCAN   & D3-BJ  & 5.517 & 4.94 & 3.68 & 4.59 & 1.42 \\
Experiment  &        & 5.550 & {\text{--}} & 5.2  & {\text{--}} & 3.0  \\
\midrule
AgBr & & $a$ & {\text{L--L}} & {\text{\textGamma--\textGamma}} & {\text{X--X}} & {\text{L--\textGamma}} \\
\midrule
S-VWN (V)   & no D3  & 5.604 & 3.42 & 2.82 & 3.39 & 0.54 \\
PBE         & no D3  & 5.849 & 4.03 & 2.63 & 3.54 & 0.82 \\
PBEsol      & no D3  & 5.692 & 3.60 & 2.77 & 3.45 & 0.65 \\
TPSS        & no D3  & 5.812 & 3.99 & 2.90 & 3.68 & 0.95 \\
revTPSS     & no D3  & 5.784 & 3.87 & 2.96 & 3.67 & 0.94 \\
Tao--Mo     & no D3  & 5.738 & 3.78 & 3.14 & 3.79 & 1.08 \\
PKZB        & no D3  & 5.868 & 4.13 & 2.99 & 3.81 & 1.19 \\
r$^2$SCAN   & no D3  & 5.811 & 4.50 & 3.36 & 4.09 & 1.42 \\
PBE         & D3-BJ  & 5.747 & 3.89 & 2.82 & 3.57 & 0.83 \\
PBEsol      & D3-BJ  & 5.613 & 3.48 & 2.95 & 3.48 & 0.67 \\
TPSS        & D3-BJ  & 5.708 & 3.82 & 3.09 & 3.71 & 0.95 \\
revTPSS     & D3-BJ  & 5.690 & 3.72 & 3.14 & 3.70 & 0.94 \\
Tao--Mo     & D3-BJ  & 5.743 & 3.79 & 3.13 & 3.79 & 1.08 \\
r$^2$SCAN   & D3-BJ  & 5.772 & 4.45 & 3.44 & 4.10 & 1.42 \\
Experiment  &        & 5.774 & {\text{--}} & 4.3  & {\text{--}} & 2.5  \\
\midrule
AgI & & $a$ & {\text{L--L}} & {\text{\textGamma--\textGamma}} & {\text{X--X}} & {\text{L--X}} \\
\midrule
S-VWN (V) & no D3  & 5.937 & 2.73 & 1.96 & 2.70 & -0.17 \\
PBE         & no D3  & 6.187 & 3.27 & 1.79 & 2.67 & 0.44 \\
PBEsol      & no D3  & 6.023 & 2.88 & 1.93 & 2.74 & 0.09 \\
TPSS        & no D3  & 6.153 & 3.25 & 2.05 & 2.94 & 0.58 \\
revTPSS     & no D3  & 6.116 & 3.15 & 2.13 & 3.00 & 0.54 \\
Tao--Mo     & no D3  & 6.071 & 3.10 & 2.34 & 3.11 & 0.62 \\
PKZB        & no D3  & 6.200 & 3.41 & 2.15 & 2.99 & 0.83 \\
r$^2$SCAN   & no D3  & 6.159 & 3.78 & 2.50 & 3.24 & 0.91 \\
PBE         & D3-BJ  & 6.067 & 3.11 & 1.99 & 2.78 & 0.28 \\
PBEsol      & D3-BJ  & 5.927 & 2.74 & 2.12 & 2.84 & -0.06 \\
TPSS        & D3-BJ  & 5.982 & 2.99 & 2.35 & 3.10 & 0.34 \\
revTPSS     & D3-BJ  & 5.949 & 2.88 & 2.44 & 3.16 & 0.29 \\
Tao--Mo     & D3-BJ  & 6.068 & 3.10 & 2.35 & 3.11 & 0.61 \\
r$^2$SCAN   & D3-BJ  & 6.156 & 3.77 & 2.51 & 3.24 & 0.91 \\
Experiment  &        & 6.067 & {\text{--}} & {\text{--}} & {\text{--}} & {\text{--}} \\
\bottomrule
\end{tabular}
\end{center}
\end{table}

For comparison, we study the impact of the basis set and the density functional approximation.
For the DFT study, we consider the first three rungs of Jacob's
ladder \cite{Perdew.Schmidt:Jacobs.2001}.
The local spin density approximation (LSDA) is represented by the S-VWN (V) functional
\cite{Slater:Simplification.1951, Vosko.Wilk.ea:Accurate.1980}, whereas
PBE \cite{Perdew.Burke.ea:Generalized.1996} and
PBEsol \cite{ Perdew.Ruzsinszky.ea:Restoring.2008}
serve as examples for generalized gradient approximations (GGAs).
mGGAs are included with the TPSS \cite{Tao.Perdew.ea:Climbing.2003},
revTPSS \cite{Perdew.Ruzsinszky.ea:Workhorse.2009, Perdew.Ruzsinszky.ea:Erratum.2011},
Tao--Mo \cite{Tao.Mo:Accurate.2016},
PKZB \cite{Perdew.Kurth.ea:Accurate.1999},
and r$^2$SCAN \cite{Furness.Kaplan.ea:Accurate.2020,
Furness.Kaplan.ea:Correction.2020}
approximations. Note that we use \textsc{Libxc} \cite{Marques.Oliveira.ea:Libxc.2012,
Lehtola.Steigemann.ea:Recent.2018, LIBXC.2022} for the PBEsol, revTPSS,
Tao--Mo, PKZB, and r$^2$SCAN functionals. The dhf-SVP and dhf-TZVP basis sets are
employed with fixed crystal structures, and results are listed in the Supplemental Material.
In addition, cell structure optimizations with these exchange-correlation functionals were
carried out with the dhf-SVP basis \cite{Weigend.Baldes:Segmented.2010,geo-note},
including the D3 correction with Becke--Johnson damping if available
\cite{Grimme.Antony.ea:consistent.2010, Grimme.Ehrlich.ea:Effect.2011,
Goerigk.Hansen.ea:look.2017, Patra.Jana.ea:Efficient.2020, Ehlert.Huniar.ea:r2SCAN-D4.2021}.
All other computational parameters such as SCF thresholds and grids are unchanged,
compared to the Hamiltonian study in Table~\ref{tab:agx}.
The main results are listed in Table~\ref{tab:agx-dft}, see Supplemental Material
for all results.

Obviously, the impact of the density functional approximations is
much larger than the deviations between the 2c ECP and the all-electron X2C or DKS ansatz.
Especially for the small L--\textGamma\  and L--X band gaps, the choice of the
semilocal functional substantially affects the results. Here, the gaps increase
from LSDA to GGA functionals and tend to further rise for the mGGAs. This
finding also holds for the other band gaps of all AgX systems.
For AgI, the  L--X band gap is very small and negative gaps
are obtained at the S-VWN (V) and PBEsol-D3 level.
Negative L--X band gaps of AgI were already found by
the group of Liu with S-VWN \cite{Zhao.Zhang.ea:Exact.2016}.

Dispersion correction leads to notably decreased lattice constants and thus
indirectly affects the band gaps. For the lattice constants, the
PBE-D3 functional performs best for AgCl, whereas r$^2$SCAN-D3 and Tao--Mo-D3
perform best for AgBr and AgI, respectively.

According to the Supplemental Material, the larger tripe-$\zeta$ basis sets
consistently lower the gaps, which is in agreement with similar studies
at the DKS level \cite{Kadek.Repisky.ea:All-electron.2019}.

\begin{table}
\begin{center}
\caption{Band gaps (in eV) of three-dimensional AgCl, AgBr, and AgI crystals
(lattice constants $a =5.612$\,\AA, $a = 5.843$\,\AA, $a = 6.169$\,\AA, all rocksalt structure)
at high symmetry points of the FBZ with the PBE \cite{Perdew.Burke.ea:Generalized.1996},
HSE03 \cite{Heyd.Scuseria.ea:Hybrid.2003, Heyd.Scuseria.ea:Erratum.2006},
HSE06 \cite{Heyd.Scuseria.ea:Hybrid.2003, Heyd.Scuseria.ea:Erratum.2006,
Krukau.Vydrov.ea:Influence.2006}, HSEsol \cite{Schmika.Harl.ea:Improved.2011},
and the HSE12 \cite{Moussa.Schultz.ea:Analysis.2012} functionals.
``no $j$'' and ``$j$'' denote that the spin current density contributions
are neglected or included for the 2c Fock exchange. Only the latter option is
the complete 2c generalization of the Fock exchange.
}
\label{tab:agx-hfx}
\begin{tabular}{@{\extracolsep{14pt}}l
S[table-format = -1.2]
S[table-format = -1.2]
S[table-format = -1.2]
S[table-format = -1.2]
@{}}
\toprule
AgCl & {\text{L--L}} & {\text{\textGamma--\textGamma}} & {\text{X--X}} & {\text{L--\textGamma}} \\
\midrule
1c PBE & 4.62 & 3.11 & 4.16 & 0.91 \\
2c PBE & 4.57 & 2.94 & 3.99 & 0.86 \\
1c HSE03 & 6.06 & 4.54 & 6.27 & 2.44 \\
2c HSE03 (no $j$) & 6.02 & 4.54 & 6.10 & 2.39 \\
2c HSE03 ($j$) & 6.01 & 4.54 & 6.09 & 2.39 \\
1c HSE06 & 5.97& 4.49 & 6.23 & 2.37 \\
2c HSE06 (no $j$) & 5.93 & 4.48 & 6.05 & 2.33 \\
2c HSE06 ($j$) & 5.93 & 4.48 & 6.05 & 2.33 \\
1c HSEsol & 5.81 & 4.42 & 6.15 & 2.26  \\
2c HSEsol (no $j$) & 5.76 & 4.41 & 5.97 & 2.22 \\
2c HSEsol ($j$) &  5.76 & 4.41 & 5.97 & 2.22 \\
1c HSE12 & 6.42 & 4.95 & 6.87 & 2.86 \\
2c HSE12 (no $j$) & 6.37 & 4.94 & 6.70 & 2.81 \\
2c HSE12 ($j$) & 6.37 & 4.94 & 6.69 & 2.81 \\
Experiment & {\text{--}} & 5.2  & {\text{--}} & 3.0 \\
\midrule
AgBr & {\text{L--L}} & {\text{\textGamma--\textGamma}} & {\text{X--X}} & {\text{L--\textGamma}} \\
\midrule
1c PBE & 4.07 & 2.64 & 3.70 & 0.86 \\
2c PBE & 4.02 & 2.64 & 3.54 & 0.82 \\
1c HSE03 & 5.34 & 3.88 & 5.60 & 2.23 \\
2c HSE03 (no $j$) & 5.30 & 3.87 & 5.54 & 2.19 \\
2c HSE03 ($j$) & 5.30 & 3.87 & 5.54 & 2.19 \\
1c HSE06 & 5.25 & 3.83 & 5.58 & 2.17 \\
2c HSE06 (no $j$) & 5.21 & 3.82 & 5.50 & 2.13 \\
2c HSE06 ($j$) & 5.21 & 3.82 & 5.49 & 2.13 \\
1c HSEsol & 5.09 & 3.77 & 5.57 & 2.08 \\
2c HSEsol (no $j$) & 5.05 & 3.77 & 5.42 & 2.04 \\
2c HSEsol ($j$) & 5.05 & 3.76 & 5.41 & 2.04 \\
1c HSE12 & 5.65 & 4.24 & 6.04 & 2.61 \\
2c HSE12 (no $j$) & 5.61 & 4.23 & 6.02 & 2.58 \\
2c HSE12 ($j$) & 5.61 & 4.23 & 6.02 & 2.58 \\
Experiment  & {\text{--}} & 4.3  & {\text{--}} & 2.5  \\
\midrule
AgI & {\text{L--L}} & {\text{\textGamma--\textGamma}} & {\text{X--X}} & {\text{L--X}} \\
\midrule
1c PBE & 3.49 & 2.16 & 2.98 & 0.65 \\
2c PBE & 3.25 & 1.82 & 2.69 & 0.41 \\
1c HSE03 & 4.54 & 3.22 & 4.06 & 1.76 \\
2c HSE03 (no $j$) & 4.27 & 2.87 & 3.77 & 1.50 \\
2c HSE03 ($j$) & 4.24 & 2.84 & 3.75 & 1.47 \\
1c HSE06 & 4.45 & 3.18 & 4.04 & 1.73 \\
2c HSE06 (no $j$) & 4.19 & 2.83 & 3.75 & 1.47 \\
2c HSE06 ($j$) & 4.16 & 2.80 & 3.73 & 1.44 \\
1c HSEsol & 4.31 & 3.12 & 4.06 & 1.70 \\
2c HSEsol (no $j$) & 4.05 & 2.78 & 3.77 & 1.44 \\
2c HSEsol ($j$) & 4.02 & 2.74 & 3.74 & 1.41 \\
1c HSE12 & 4.79 & 3.54 & 4.43 & 2.11 \\
2c HSE12 (no $j$) & 4.52 & 3.19 & 4.14 & 1.85 \\
2c HSE12 ($j$) & 4.48 & 3.15 & 4.10 & 1.81 \\
Experiment  & {\text{--}} & {\text{--}} & {\text{--}} & {\text{--}} \\
\bottomrule
\end{tabular}
\end{center}
\end{table}

We note, however, that none of the semilocal density functionals applied
herein is able to accurately reproduce the experimental band gaps for AgCl
and AgBr. Therefore, we study the performance of the range-separated
hybrid functionals HSE03 \cite{Heyd.Scuseria.ea:Hybrid.2003, Heyd.Scuseria.ea:Erratum.2006},
HSE06 \cite{Heyd.Scuseria.ea:Hybrid.2003, Heyd.Scuseria.ea:Erratum.2006,
Krukau.Vydrov.ea:Influence.2006}, HSEsol \cite{Schmika.Harl.ea:Improved.2011},
and HSE12 \cite{Moussa.Schultz.ea:Analysis.2012} using \textsc{Libxc}
\cite{Marques.Oliveira.ea:Libxc.2012, Lehtola.Steigemann.ea:Recent.2018, LIBXC.2022}.
Computational settings are unchanged except that the threshold for the
canonical orthogonalization was raised from $10^{-6}$ to $5.5 \cdot 10^{-5}$
for AgCl with all hybrid functionals and for AgBr with HSE12 in order to
facilitate the SCF convergence. This removes three vectors for AgCl with
all hybrid functionals and also three vectors for AgBr with HSE12.
For AgI, the smallest eigenvalue of the overlap matrix is larger than
$10^{-3}$, ensuring a smooth convergence.
Results are listed in Table~\ref{tab:agx-hfx}.

Admixture of Fock exchange substantially increases the band gaps and hence
improves the agreement with experiment. Especially, the L--\textGamma\ 
gap rises and the silver halides become small-gap semiconductors in
line with the experimental findings. The impact of the proper 2c generalization
using the spin-current density for Fock exchange is comparably small for
these systems.

\begin{figure*}
\begin{center}
\includegraphics[width=0.99\textwidth]{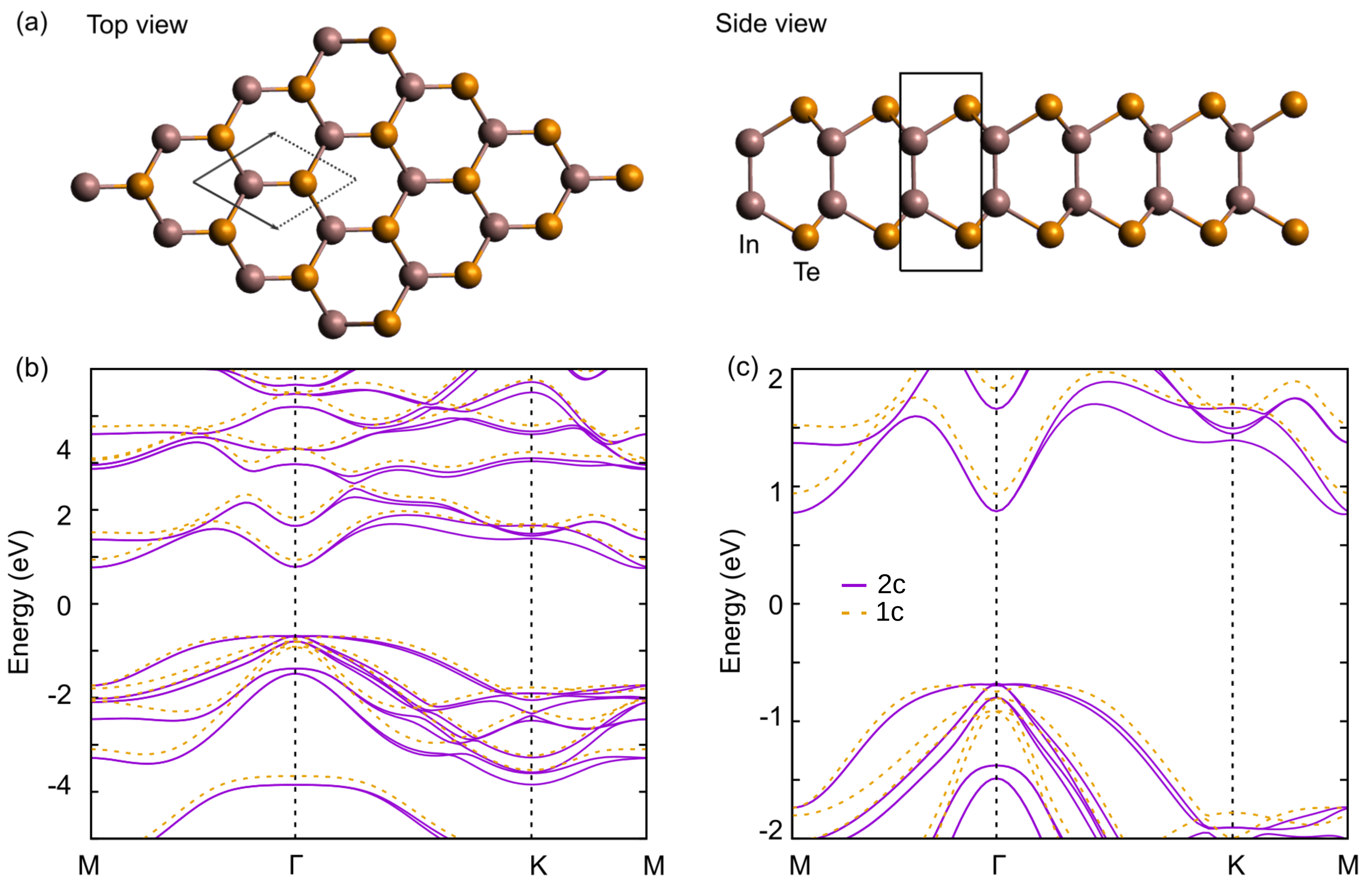}
\caption{(a) Top and side views of the two-dimensional InTe honeycomb system with indicated unit
cell. The unit cell consists of two In and two Te atoms.
(b, c) Electronic band structure of the FBZ. The orange dashed lines are calculated without
spin--orbit coupling, while the solid purple lines include spin--orbit interaction.
The black vertical dashed lines mark the $\Gamma$ and K points of the Brillouin zone.
Calculations are performed with the PBE functional \cite{Perdew.Burke.ea:Generalized.1996},
the D3-BJ dispersion correction \cite{Grimme.Antony.ea:consistent.2010, Grimme.Ehrlich.ea:Effect.2011},
and the dhf-TZVP-2c basis set \cite{Weigend.Baldes:Segmented.2010}. See Supplemental
Material for results with HSE06
\cite{Heyd.Scuseria.ea:Hybrid.2003, Heyd.Scuseria.ea:Erratum.2006, Krukau.Vydrov.ea:Influence.2006}.
}
\label{fig:r2dband}
\end{center}
\end{figure*}

In conclusion, an ECP-based implementation is sufficient for the band gaps
of crystals, as this property is driven by the valence region.
In contrast, all-electron approaches may be needed for other properties
or for a fully parameter-free description of relativistic effects.

\subsection{Indium(I,III)-Telluride Two-Dimensional Honeycomb System}
\label{subsec:honeycomb}

To illustrate the applicability of the implementation to two-dimensional systems, we
consider the indium(I,III)-telluride (InTe) honeycomb crystal displayed in Fig.~\ref{fig:r2dband}a.
For layered two-dimensional materials and their atomically thin layers dispersion interaction
may play an important role.
Here, we use the PBE functional \cite{Perdew.Burke.ea:Generalized.1996} (grid size 4)
combined with the D3-BJ correction \cite{Grimme.Antony.ea:consistent.2010,
Grimme.Ehrlich.ea:Effect.2011}. The dhf-TZVP-2c basis
set \cite{Weigend.Baldes:Segmented.2010} is applied, and a $k$ mesh of $32 \times 32$
is employed. A Gaussian smearing of 0.001\,Hartree \cite{gsmear} and an SCF threshold
of $10^{-8}$\,Hartree are chosen.
The cell structure is optimized with the 2c Hamiltonian, and band structures are
shown in Fig.~\ref{fig:r2dband}b and \ref{fig:r2dband}c. For comparison, the electronic band
structure based on the unit cell structure optimized without the D3-BJ correction and
spin--orbit interaction is displayed in the Supplemental Material.

We find that dispersion correction and weight derivatives are important for the unit cell
structure and consequently the band gap. Without D3-BJ and weight derivatives, a lattice
constant of $4.17$\,\AA, and band gaps of $1.56$\,eV (1c) and $1.12$\,eV (2c) are obtained.
Adding the D3-BJ correction and weight derivatives changes the lattice constant to $4.23$\,\AA.
Band gaps are increased to $1.63$\,eV (1c) and $1.44$\,eV (2c). This reveals a delicate
interplay of spin--orbit coupling and cell structure.

The most notable changes of the band structure induced by spin--orbit coupling
are found for the \textGamma\ point in the energetic region from 1 to 2\,eV below the Fermi
level of $\epsilon_{\textrm{F}}=0$\,eV. Here, the energies of the two occupied
bands are substantially decreased.
The shifts for these bands at about $-1.5$\,eV amount to more than 0.5\,eV.

The reduction of the bands gap from $1.63$ to $1.44$\,eV due to spin--orbit coupling
is mainly caused by the energetic decrease of the band $1$\,eV above the Fermi level.
Overall, our results are in reasonable agreement with those of Shang \textit{et al.}\ \cite{Shang2018},
who reported a band gap of $1.27$\,eV.

Application of Fock exchange using the HSE06 functional \cite{Heyd.Scuseria.ea:Hybrid.2003,
Heyd.Scuseria.ea:Erratum.2006, Krukau.Vydrov.ea:Influence.2006} with a $k$ mesh of
$17 \times 17$ and a threshold of $5.5 \cdot 10^{-5}$ for the orthogonal basis increases
the band gap to $2.29$\,eV. Inclusion of spin--orbit coupling changes this to $2.13$\,eV
without spin currents in the Fock exchange and to $2.09$\,eV with the complete 2c Fock exchange.
According to the band plots in the Supplemental Material, the increase of the band gap
for HSE06 compared to PBE is mainly due to the energetic shift of occupied bands.

\subsection{One-Dimensional Platinum Chains}
\label{subsec:pt}

Electron transport through atomically thin wires serves as a sensitive probe of their electronic structure.
Scalar-relativistic effects have been pointed out to be crucial for the chain formation in metallic
atomic contacts of Ir, Pt, and Au \cite{Smit2001}. From these three elements, Pt and Ir have been
suggested to show interesting magnetic effects based on the spin--orbit coupling, such as an
anisotropic magnetoresistance \cite{Garcia2009}. As chains are pulled, they transition from a
zigzag to a linear configuration \cite{Garcia2005}. While linear chains are thus not the ground-state
geometry for all interatomic distances, linear chains of Pt are by now a reference system for a
transition to a magnetic state \cite{Delin2003, Delin.Tosatti:Emerging.2004, Fernandez-Rossier2005,
Smogunov2008, Garcia2009}. We will reexamine this system with our new implementation here. 

\begin{figure*}
\begin{center}
\includegraphics[width=0.99\textwidth]{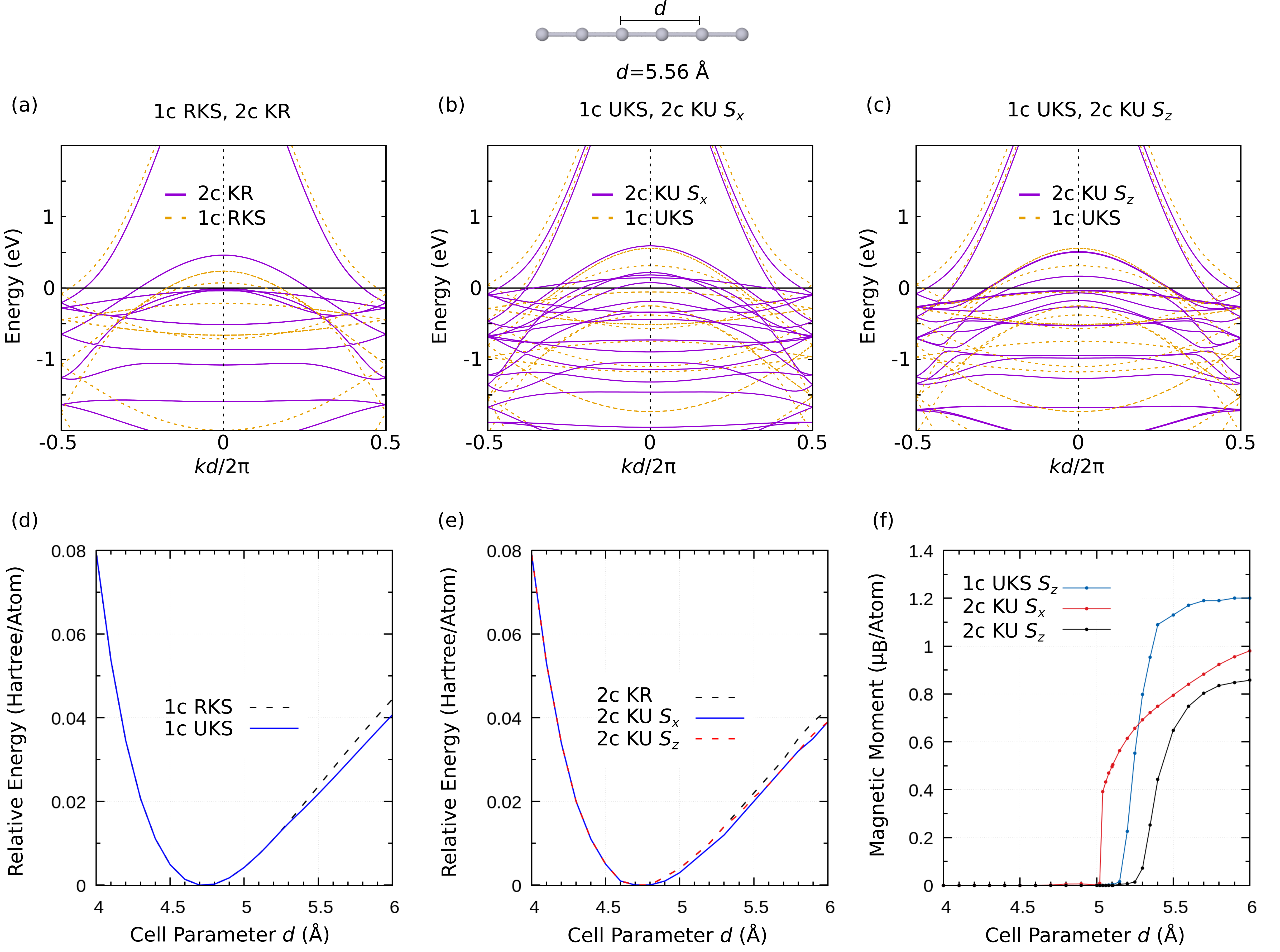}
\caption{(a)--(c) Electronic band structure in the FBZ for a linear platinum chain at $d=5.56$\,\AA.
The unit cell consists of two platinum atoms and is exemplarily shown above panel (b).
The black vertical dashed lines in each panel mark the \textGamma\ point of the FBZ.
Expectation values of the respective spin components are
1.16 [1c UKS, panels (b) and (c)], 0.82 [2c KU $S_x$, panel (b)],
and 0.71 [2c KU $S_z$, panel (c)], and the spin vector is fully aligned
along the $x$ and $z$ direction, respectively.
Note that we assume periodicity along the $x$ direction for one-dimensional
systems \cite{TURBOMOLE-manual}. The open-shell solutions are energetically favored
compared to the respective closed-shell solutions (1c RKS or 2c KR) by about
$4\cdot 10^{-3}$\,Hartree (1c UKS), $2\cdot 10^{-3}$\,Hartree (2c KU $S_x$ alignment),
and $10^{-3}$\,Hartree (2c KU $S_z$ alignment).
The magnetization parallel to the nanowire is thus energetically preferred
to the perpendicular one by about $10^{-3}$\,Hartree.
Calculations are performed with the PBE functional \cite{Perdew.Burke.ea:Generalized.1996}
and the dhf-TZVP-2c basis set \cite{Weigend.Baldes:Segmented.2010}.
(d) Dependence of the energy on the cell parameter in units of Hartree per atom for 1c
RKS and 1c UKS.
(e) Dependence of the energy on the cell parameter in units of Hartree per atom for 2c
KR, 2c KU $S_x$, and 2c KU $S_z$.
(f) Magnetic moment in units of Bohr's magneton $\mu_{\textrm{B}}$ per atom for the
spin contribution of 1c UKS, 2c KU $S_x$, and 2c KU $S_z$.
}
\label{fig:r1dband}
\end{center}
\end{figure*}

Our unit cell consists of two platinum atoms and is indicated in Fig.~\ref{fig:r3dband}(a).
The cell parameter $d$ was varied from 4.0\,\AA\ to 6.0\,\AA, with structures specified in the Supplemental
Material. For the calculations, the PBE exchange-correlation functional \cite{Perdew.Burke.ea:Generalized.1996}
(grid size 4 \cite{Treutler.Ahlrichs:Efficient.1995, Treutler:Entwicklung.1995}) and
the dhf-TZVP-2c GTO basis set \cite{Weigend.Baldes:Segmented.2010} with small-core Dirac--Fock
ECPs \cite{Figgen.Peterson.ea:Energy-consistent.2009} are employed. 32 $k$ points are used in combination with a
Gaussian smearing of $0.01$\,Hartree \cite{gsmear}.
SCF procedures are converged with a threshold of $10^{-8}$\,Hartree. We started the SCF calculations
both from a closed-shell initial guess and an open-shell initial guess based on four unpaired electrons.
In the 2c calculations, the initial wave function is chosen to be an eigenfunction of $S_x$ or $S_z$,
and the converged 2c wave function is thus aligned accordingly. We also confirmed the settings for the
2c calculations using a superposition of atomic densities with the magnetization aligned along the $x$
or $z$ axis for both Pt atoms as initial guess.
Note that we assume periodicity along the $x$ direction for one-dimensional systems \cite{TURBOMOLE-manual}.
Results with the Scalmani--Frisch approach are further listed in the Supplemental Material.

According to the restricted and unrestricted DFT calculations in Fig.~\ref{fig:r1dband}(d)--\ref{fig:r1dband}(f),
no notable spin polarization occurs for unit cells with $d$ smaller than 5.0\,\AA.
Here, the spin expectation values are essentially zero.
The most energetically favorable geometric structure is found for $d = 4.75$\,\AA\ in both scalar-relativistic and
spin--orbit calculations. Apparently the impact of spin--orbit coupling on the geometric structure is
small for this system. However, spin--orbit interaction has a large impact on the electronic band structure.
For the 2c Kramers-restricted solution in Fig.~\ref{fig:r1dband}(a), only two bands cross the Fermi level
at $\epsilon_{\textrm{F}}=0$\,eV. 
In contrast, seven and four bands cross the Fermi level for the Kramers-unrestricted solutions with the
$S_x$ and $S_z$ alignments, respectively, see Fig.~\ref{fig:r1dband}(b) and \ref{fig:r1dband}(c).
For the latter two cases, many bands are split due to spin--orbit effects. Thus, we find that
spin--orbit interaction substantially affects the band structure for all spin configurations.

Furthermore, spin--orbit coupling is of relevance for the magnetic moment and the spin expectation value.
The Pt chains transition into a magnetic state at $d = 5.25$\,\AA\ without taking spin--orbit interaction
into account, while spin--orbit effects change the transition geometry to $d = 5.04$\,\AA\ for the $S_x$
orientation and to $d = 5.35$\,\AA\ for the $S_z$ orientation.
For the absolute values of the spin expectation values, we find $\langle S_z \rangle = 1.2$ for
large $d = 6.0$\,\AA\ in the 1c UKS formalism, while the 2c calculations lead to $\langle S_x \rangle = 1.0$
and $\langle S_z \rangle = 0.85$. This corresponds to a large magnetic anisotropy as choosing the spin
parallel to the periodic direction ($x$ axis) notably affects the absolute spin expectation values and
consequently the magnetic moments. In the spin-only approximation, the latter are obtained in units
of Bohr's magneton $\mu_{\textrm{B}}$ by doubling the spin expectation value.
Since our unit cell contains two atoms, the spin expectation values listed above
directly correspond to the magnetic moments in $\mu_{\textrm{B}}$ per atom.

Changing the density functional approximation to the LSDA functional S-VWN (V) \cite{Slater:Simplification.1951,
Vosko.Wilk.ea:Accurate.1980} or to the mGGA functional TPSS \cite{Tao.Perdew.ea:Climbing.2003}
leads to a transition geometry of $ d= 5.3$\,\AA\ and $d = 5.2$\,\AA\ (1c).
Inclusion of spin--orbit coupling results in a jump of the magnetization at $d = 5.0$ \,\AA\ and
$4.9$\,\AA\ for the spin alignment parallel to the chain at the 2c level, respectively.
The spin expectation value for large cells is very similar to the PBE result. The 1c UKS formalism
predicts a jump of the magnetic moment at about $d = 5.1$\,\AA\ for both functionals. Notably,
the range-separated hybrid functional HSE06 \cite{Heyd.Scuseria.ea:Hybrid.2003, Heyd.Scuseria.ea:Erratum.2006,
Krukau.Vydrov.ea:Influence.2006} leads to a transition already at $d = 4.7$\,\AA\ (1c).
We refer to the Supplemental Material for the complete results with S-VWN (V), TPSS, and HSE06.

Our scalar-relativistic results are in excellent agreement with the study of Fern{\'a}ndez-Rossier
\textit{et al.}\ \cite{Fernandez-Rossier2005}, who predicted an equilibrium structure with
$d = 4.8$\,\AA.\ Additionally, they observed the magnetic transition at $d = 5.2$\,\AA\ and a
magnetic moment of 1.2\,$\mu_{\textrm{B}}$ per atom at $d=6.0$\,\AA\ based on 1c UKS PBE calculations
with the GTO-based code \textsc{crystal03}. Note that these authors used one platinum atom per unit
cell and hence the lattice spacing in their work has to be converted for comparison with our results.

Our results are also in qualitative agreement with the study of Smogunov \textit{et al.} using
plane-wave methods \cite{Smogunov2008}. They obtained an equilibrium distance of $d_0 \approx 4.8$\,\AA
(GGA), which is in close agreement with our prediction of $d_0 = 4.8$\,\AA.
Additionally, the transition to a magnetic state occurs at smaller $d$ values for a
magnetization parallel to the chain than for a perpendicular magnetization.
A jump of the magnetic moment is found at $d \approx 4.84$\,\AA\ for the first orientation with
GGA functionals and at slightly smaller $d$ for LSDA. For the perpendicular magnetization, the jump
is at $d \approx 5.2$\,\AA.
Smogunov \textit{et al.}\ \cite{Smogunov2008} also found that the parallel magnetization is
energetically favored, which is in agreement with our calculations.

\section{Summary and Outlook}
\label{sec:conclusion}
We presented an efficient two-component DFT procedure, which accounts for spin--orbit interaction
as well as for scalar-relativistic effects. Relativistic effects are introduced with effective
core potentials. Due to the use of atom-centered Gaussian-type orbitals, our implementation is
applicable to both molecular and periodic systems of any dimensionality. For ground-state
energy calculations hybrid functionals are available to account for the self-consistent
relaxation of the induced current density.

We demonstrated the validity of our approach by calculating the ionization energies
of heavy $p$-block atoms, the electronic bulk band structure of gold and lead, band gaps of
silver halide crystals, the geometry and band structure of the InTe honeycomb system,
as well as the spin polarizations of linear platinum chains.
In the process, we assessed the accuracy of the implementation by comparison with the
plane-wave-based \textsc{Quantum Espresso} program and other codes, showing excellent agreement.

Extension of this work is promising in multiple directions. In terms of density functional
approximations, this covers the extension of the mGGA framework to account
for the current density as done in Ref.~\cite{Holzer.Franke.ea:Current.2022}. 
An extension to local hybrid functionals \cite{Jaramillo.Scuseria.ea:Local.2003} may be useful
to allow for a more flexible admixture of Fock exchange \cite{Plessow.Weigend:Seminumerical.2012,
Maier.Arbuznikov.ea:Local.2018, Holzer.Franzke:Local.2022}.
Additionally, relativistic all-electron approaches are necessary to study
energetically low-lying states and the density in the vicinity of the nuclei
\cite{Peng.Middendorf.ea:efficient.2013, Franzke.Middendorf.ea:Efficient.2018}.

\section*{Supplemental Material}
Supplemental Material is available:
\begin{itemize}
\item Basis set study of silver halide crystals with non-relaxed structure,
\item density functional approximation study of silver halide crystals with relaxed structures:
scalar-relativistic and spin--orbit results,
\item further results for the indium(I,III)-telluride two-dimensional honeycomb system
(results with the geometry optimization without D3-BJ correction for PBE, results with HSE06),
\item energies and expectation values for the Pt chain with the canonical and
the Scalmani--Frisch non-collinear ansatz for PBE as well as results with S-VWN (V) and TPSS,
\item all structures (coordinate files) used in this work.
\end{itemize}

\section*{Data Availability Statement}
The data that support the findings of this study are available within the
article and its Supplemental Material.

\section*{Authors' Contributions}
Y.\ J.\ Franzke and W.\ M.\ Schosser contributed equally to this work.

\begin{acknowledgments}
We thank Christof Holzer (Karlsruhe), Marek Sierka (Jena), and
Florian Weigend (Marburg) for stimulating discussions.
Y.J.F.\ acknowledges TURBOMOLE GmbH for financial support.
Y.J.F.\ gratefully acknowledges support via the Walter--Benjamin
programme funded by the Deutsche Forschungsgemeinschaft (DFG, German
Research Foundation) --- 518707327.
\end{acknowledgments}

\bibliography{literature}

\begin{thebibliography}{217}
\bibitem [{geo()}]{geo-note}%
  \BibitemOpen
  \href@noop {} {}\bibinfo {note} {Minor deviations from the fcc structure were
  removed by symmetrization. This changed the energy by less than $3\cdot
  10^{-4}$\,Hartree or less than 1\,kJ/mol. Typically, changes are below
  $10^{-5}$\,Hartree.}\BibitemShut {Stop}%
\end{thebibliography}

\ifarXiv


\section*{Supplementary material}

## III.    INDIUM(I,III)-TELLURIDE TWO-DIMENSIONAL HONEYCOMB SYSTEM

Unit cell parameters and electronic band structures are calculated without the D3-BJ correction for comparison to the results in the main text. Additionally, the cell structure was optimized without taking spin–orbit coupling into account [22, 23]. All other computational parameters are unchanged compared to the study in the main text. In detail, we use the PBE functional [4] combined with the dhf-TZVP-2c basis sets [16]. A $k$ mesh of $32 \times 32$ is employed. A Gaussian smearing of 0.001 Hartree [24] and an SCF threshold of $10^{-8}$ Hartree are chosen. The electronic band structure is displayed in Figure 1.

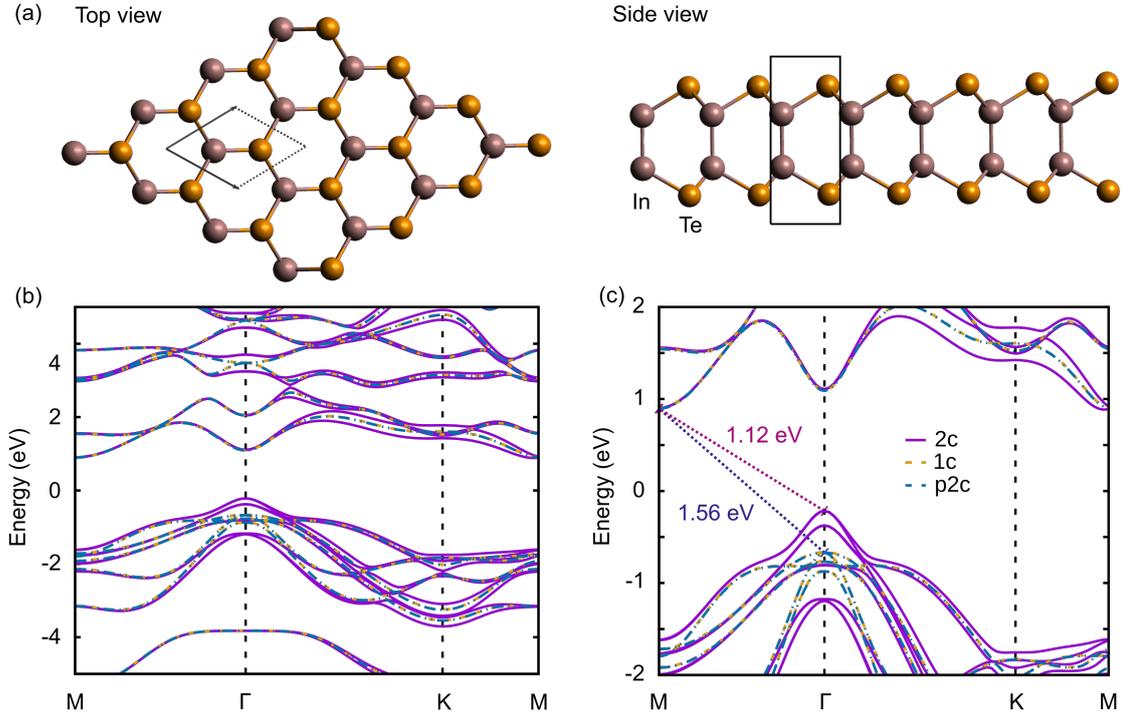

FIG. 1.  (a) Top and side views of the two-dimensional InTe honeycomb system, with indicated unit cell. The unit cell consists of two In and two Te atoms. (b,c) Electronic band structure of the FBZ. The orange dashed lines are calculated without spin–orbit coupling in a one-component (1c) formalism, while the solid purple lines include spin–orbit interaction in the two-component (2c) methodology. The powder blue lines mark the two-component calculations with spin–orbit ECPs scaled to zero, i.e. the pseudo two-component (p2c) version. The black vertical dashed lines mark the $\Gamma$ and K points of the Brillouin zone. Band gaps are indicated in blue (1c) and purple (2c) in panel (c).

Results with the HSE06 range-separated hybrid functional are shown in Figure 2. Here, unit cell parameters are optimized at the 2c PBE-D3-BJ level so that the structure is the same as that employed in the main text. The threshold of $5 \cdot 10^{-5}$ for the eigenvalues of the overlap matrix in the orthogonal basis transformation removes at most 4 linear combinations of basis functions (vectors). The total number of vectors is 232.

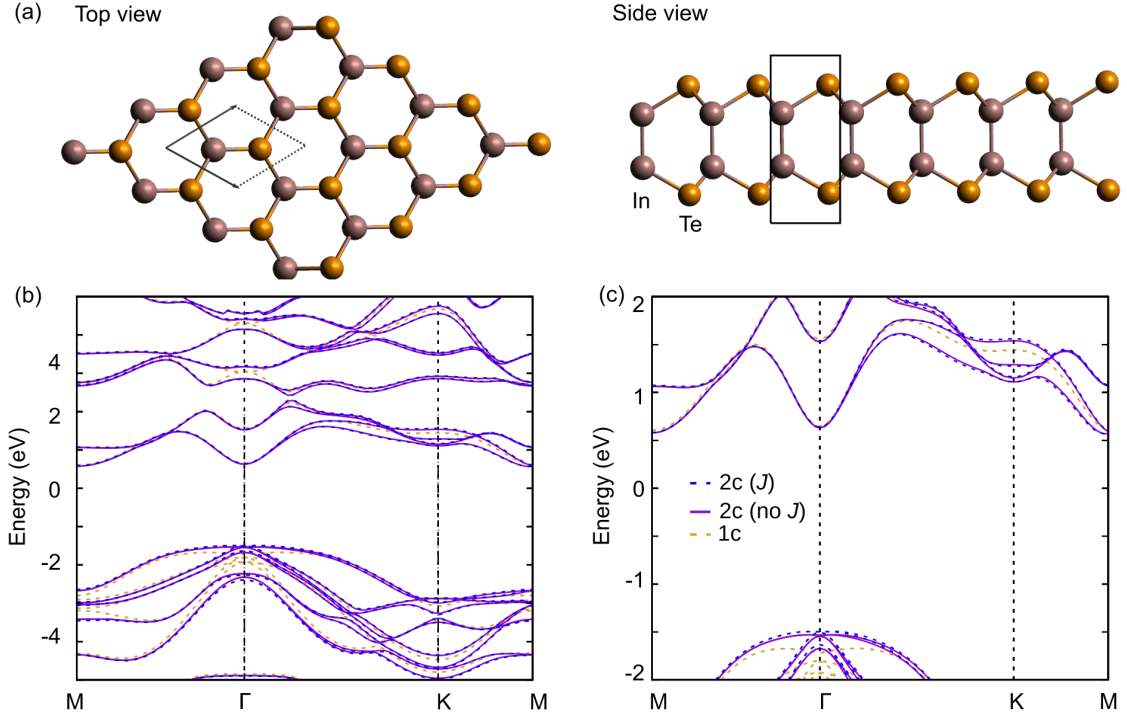

FIG. 2. (a) Top and side views of the two-dimensional InTe honeycomb system, with indicated unit cell. The unit cell consists of two In and two Te atoms. (b,c) Electronic band structure of the FBZ. The orange dashed lines are calculated without spin–orbit coupling in a one-component (1c) formalism, while the solid purple lines include spin–orbit interaction in the two-component (2c) methodology without the spin-current density contribution to the Fock exchange (2c no $J$). Dashed blue lines mark the two-component calculations with the complete 2c Fock exchange (2c $J$). The black vertical dashed lines mark the $\Gamma$ and K points of the Brillouin zone.

## IV. ONE-DIMENSIONAL PLATINUM CHAINS

TABLE VII. Total SCF energies in Hartree obtained with the 1c RKS and 1c UKS formalisms and energies for the limit of a vanishing Gaussian smearing (limit). For the UKS calculations, the numbers of $\alpha$ and $\beta$ electrons are denoted by $n_\alpha$ and $n_\beta$. The $S_z$ expectation value is calculated as $\langle S_z \rangle = (n_\alpha - n_\beta)/2$. The PBE functional [4] is employed.

| $d$ | RKS limit | RKS | UKS limit | UKS | $n_\alpha$ | $n_\beta$ | $\langle S_z \rangle$ |
|---|---|---|---|---|---|---|---|
| 4.000 | $-238.8565$ | $-238.8556$ | $-238.8565$ | $-238.8556$ | 18.0000 | 18.0000 | $1.03 \times 10^{-5}$ |
| 4.100 | $-238.9083$ | $-238.9074$ | $-238.9083$ | $-238.9074$ | 18.0000 | 18.0000 | $1.86 \times 10^{-5}$ |
| 4.200 | $-238.9467$ | $-238.9457$ | $-238.9467$ | $-238.9457$ | 18.0000 | 18.0000 | $2.56 \times 10^{-5}$ |
| 4.300 | $-238.9745$ | $-238.9733$ | $-238.9745$ | $-238.9733$ | 18.0000 | 18.0000 | $3.89 \times 10^{-5}$ |
| 4.400 | $-238.9936$ | $-238.9923$ | $-238.9936$ | $-238.9923$ | 18.0001 | 17.9999 | $8.50 \times 10^{-5}$ |
| 4.500 | $-239.0060$ | $-239.0045$ | $-239.0060$ | $-239.0045$ | 18.0001 | 17.9999 | $1.10 \times 10^{-4}$ |
| 4.600 | $-239.0131$ | $-239.0112$ | $-239.0131$ | $-239.0112$ | 18.0003 | 17.9998 | $2.50 \times 10^{-4}$ |
| 4.700 | $-239.0158$ | $-239.0137$ | $-239.0158$ | $-239.0137$ | 18.0007 | 17.9993 | $7.02 \times 10^{-4}$ |
| 4.800 | $-239.0153$ | $-239.0128$ | $-239.0153$ | $-239.0128$ | 18.0007 | 17.9993 | $7.42 \times 10^{-4}$ |
| 4.900 | $-239.0123$ | $-239.0093$ | $-239.0123$ | $-239.0093$ | 18.0007 | 17.9994 | $6.48 \times 10^{-4}$ |
| 5.000 | $-239.0073$ | $-239.0039$ | $-239.0073$ | $-239.0039$ | 18.0008 | 17.9992 | $7.65 \times 10^{-4}$ |
| 5.020 | $-239.0062$ | $-239.0026$ | $-239.0062$ | $-239.0026$ | 18.0003 | 17.9997 | $2.59 \times 10^{-4}$ |
| 5.040 | $-239.0049$ | $-239.0013$ | $-239.0049$ | $-239.0013$ | 18.0004 | 17.9996 | $3.65 \times 10^{-4}$ |
| 5.060 | $-239.0037$ | $-238.9999$ | $-239.0037$ | $-238.9999$ | 18.0005 | 17.9995 | $5.38 \times 10^{-4}$ |
| 5.080 | $-239.0023$ | $-238.9985$ | $-239.0023$ | $-238.9985$ | 18.0008 | 17.9992 | $8.32 \times 10^{-4}$ |
| 5.100 | $-239.0010$ | $-238.9970$ | $-239.0010$ | $-238.9970$ | 18.0029 | 17.9971 | $2.93 \times 10^{-3}$ |
| 5.150 | $-238.9974$ | $-238.9932$ | $-238.9974$ | $-238.9932$ | 18.0158 | 17.9842 | $1.58 \times 10^{-2}$ |
| 5.200 | $-238.9937$ | $-238.9892$ | $-238.9937$ | $-238.9892$ | 18.2246 | 17.7754 | $2.25 \times 10^{-1}$ |
| 5.250 | $-238.9897$ | $-238.9850$ | $-238.9899$ | $-238.9851$ | 18.5527 | 17.4473 | $5.53 \times 10^{-1}$ |
| 5.300 | $-238.9856$ | $-238.9806$ | $-238.9862$ | $-238.9814$ | 18.7969 | 17.2031 | $7.97 \times 10^{-1}$ |
| 5.350 | $-238.9815$ | $-238.9761$ | $-238.9828$ | $-238.9780$ | 18.9539 | 17.0461 | $9.54 \times 10^{-1}$ |
| 5.400 | $-238.9772$ | $-238.9716$ | $-238.9794$ | $-238.9747$ | 19.0855 | 16.9145 | 1.09 |
| 5.500 | $-238.9686$ | $-238.9625$ | $-238.9723$ | $-238.9674$ | 19.1328 | 16.8672 | 1.13 |
| 5.600 | $-238.9600$ | $-238.9533$ | $-238.9649$ | $-238.9598$ | 19.1699 | 16.8302 | 1.17 |
| 5.700 | $-238.9514$ | $-238.9442$ | $-238.9573$ | $-238.9519$ | 19.1857 | 16.8143 | 1.19 |
| 5.800 | $-238.9431$ | $-238.9353$ | $-238.9496$ | $-238.9439$ | 19.1936 | 16.8064 | 1.19 |
| 5.900 | $-238.9349$ | $-238.9267$ | $-238.9420$ | $-238.9359$ | 19.1976 | 16.8024 | 1.20 |
| 6.000 | $-238.9270$ | $-238.9184$ | $-238.9345$ | $-238.9281$ | 19.1997 | 16.8003 | 1.20 |

TABLE VIII. Total SCF energies in Hartree obtained with the 2c KR, 2c canonical KU $S_z$, and 2c canonical KU $S_x$ approaches and energies for the limit of a vanishing Gaussian smearing (limit). Expectation values of the spin $S_{z,x}$ are listed for the open-shell calculations. The PBE functional [4] is employed.

| $d$ | KR limit | KR | KU $S_z$ limit | KU $S_z$ | $\langle S_z \rangle$ | KU $S_x$ limit | KU $S_x$ | $\langle S_x \rangle$ |
|---|---|---|---|---|---|---|---|---|
| 4.000 | $-239.1439$ | $-239.1430$ | $-239.1439$ | $-239.1430$ | $-2.86 \times 10^{-6}$ | $-239.1439$ | $-239.1430$ | $-1.12 \times 10^{-5}$ |
| 4.100 | $-239.1956$ | $-239.1945$ | $-239.1956$ | $-239.1945$ | $1.43 \times 10^{-6}$ | $-239.1956$ | $-239.1945$ | $-1.41 \times 10^{-5}$ |
| 4.200 | $-239.2340$ | $-239.2325$ | $-239.2340$ | $-239.2325$ | $1.69 \times 10^{-6}$ | $-239.2340$ | $-239.2325$ | $3.33 \times 10^{-5}$ |
| 4.300 | $-239.2618$ | $-239.2601$ | $-239.2618$ | $-239.2601$ | $1.10 \times 10^{-5}$ | $-239.2618$ | $-239.2601$ | $2.89 \times 10^{-5}$ |
| 4.400 | $-239.2812$ | $-239.2792$ | $-239.2812$ | $-239.2792$ | $2.23 \times 10^{-5}$ | $-239.2812$ | $-239.2792$ | $1.05 \times 10^{-4}$ |
| 4.500 | $-239.2940$ | $-239.2916$ | $-239.2940$ | $-239.2916$ | $7.31 \times 10^{-5}$ | $-239.2940$ | $-239.2916$ | $9.33 \times 10^{-5}$ |
| 4.600 | $-239.3017$ | $-239.2988$ | $-239.3017$ | $-239.2988$ | $9.94 \times 10^{-5}$ | $-239.3017$ | $-239.2988$ | $5.48 \times 10^{-4}$ |
| 4.700 | $-239.3052$ | $-239.3019$ | $-239.3052$ | $-239.3019$ | $1.23 \times 10^{-4}$ | $-239.3052$ | $-239.3019$ | $1.31 \times 10^{-3}$ |
| 4.800 | $-239.3056$ | $-239.3019$ | $-239.3056$ | $-239.3019$ | $5.67 \times 10^{-4}$ | $-239.3056$ | $-239.3019$ | $5.43 \times 10^{-3}$ |
| 4.900 | $-239.3036$ | $-239.2994$ | $-239.3036$ | $-239.2994$ | $-9.93 \times 10^{-4}$ | $-239.3036$ | $-239.2994$ | $6.00 \times 10^{-3}$ |
| 5.000 | $-239.2998$ | $-239.2952$ | $-239.2998$ | $-239.2952$ | $6.06 \times 10^{-5}$ | $-239.2998$ | $-239.2952$ | $2.23 \times 10^{-3}$ |
| 5.020 | $-239.2989$ | $-239.2942$ | $-239.2989$ | $-239.2942$ | $2.88 \times 10^{-5}$ | $-239.2989$ | $-239.2942$ | $8.09 \times 10^{-3}$ |
| 5.040 | $-239.2979$ | $-239.2931$ | $-239.2979$ | $-239.2931$ | $5.00 \times 10^{-5}$ | $-239.2984$ | $-239.2939$ | $3.91 \times 10^{-1}$ |
| 5.060 | $-239.2969$ | $-239.2920$ | $-239.2969$ | $-239.2920$ | $7.79 \times 10^{-5}$ | $-239.2975$ | $-239.2930$ | $4.33 \times 10^{-1}$ |
| 5.080 | $-239.2958$ | $-239.2909$ | $-239.2958$ | $-239.2909$ | $1.30 \times 10^{-4}$ | $-239.2965$ | $-239.2921$ | $4.69 \times 10^{-1}$ |
| 5.100 | $-239.2947$ | $-239.2897$ | $-239.2947$ | $-239.2897$ | $5.17 \times 10^{-4}$ | $-239.2947$ | $-239.2911$ | $4.96 \times 10^{-1}$ |
| 5.150 | $-239.2918$ | $-239.2866$ | $-239.2918$ | $-239.2866$ | $4.19 \times 10^{-3}$ | $-239.2929$ | $-239.2884$ | $5.63 \times 10^{-1}$ |
| 5.200 | $-239.2887$ | $-239.2833$ | $-239.2887$ | $-239.2833$ | $6.97 \times 10^{-3}$ | $-239.2900$ | $-239.2855$ | $6.14 \times 10^{-1}$ |
| 5.250 | $-239.2855$ | $-239.2798$ | $-239.2855$ | $-239.2798$ | $1.42 \times 10^{-2}$ | $-239.2870$ | $-239.2824$ | $6.56 \times 10^{-1}$ |
| 5.300 | $-239.2821$ | $-239.2762$ | $-239.2821$ | $-239.2762$ | $7.18 \times 10^{-2}$ | $-239.2838$ | $-239.2790$ | $6.92 \times 10^{-1}$ |
| 5.350 | $-239.2785$ | $-239.2725$ | $-239.2786$ | $-239.2727$ | $2.52 \times 10^{-1}$ | $-239.2804$ | $-239.2755$ | $7.22 \times 10^{-1}$ |
| 5.400 | $-239.2749$ | $-239.2687$ | $-239.2753$ | $-239.2693$ | $4.43 \times 10^{-1}$ | $-239.2770$ | $-239.2719$ | $7.48 \times 10^{-1}$ |
| 5.500 | $-239.2676$ | $-239.2610$ | $-239.2684$ | $-239.2624$ | $6.48 \times 10^{-1}$ | $-239.2698$ | $-239.2645$ | $7.95 \times 10^{-1}$ |
| 5.600 | $-239.2600$ | $-239.2531$ | $-239.2613$ | $-239.2552$ | $7.48 \times 10^{-1}$ | $-239.2626$ | $-239.2569$ | $8.41 \times 10^{-1}$ |
| 5.700 | $-239.2525$ | $-239.2452$ | $-239.2541$ | $-239.2478$ | $8.03 \times 10^{-1}$ | $-239.2553$ | $-239.2493$ | $8.83 \times 10^{-1}$ |
| 5.800 | $-239.2450$ | $-239.2373$ | $-239.2469$ | $-239.2403$ | $8.35 \times 10^{-1}$ | $-239.2481$ | $-239.2418$ | $9.23 \times 10^{-1}$ |
| 5.900 | $-239.2376$ | $-239.2296$ | $-239.2398$ | $-239.2331$ | $8.48 \times 10^{-1}$ | $-239.2410$ | $-239.2345$ | $9.55 \times 10^{-1}$ |
| 6.000 | $-239.2303$ | $-239.2220$ | $-239.2329$ | $-239.2259$ | $8.57 \times 10^{-1}$ | $-239.2342$ | $-239.2274$ | $9.80 \times 10^{-1}$ |

TABLE IX. Total SCF energies in Hartree obtained with the 2c KR and the 2c Scalmani–Frisch KU $S_x$ approaches and energies for the limit of a vanishing Gaussian smearing (limit). Expectation values of the spin $S_x$ are listed for the open-shell calculations. The PBE functional [4] is employed.

| $d$ | KR limit | KR | KU $S_x$ limit | KU $S_x$ | $\langle S_x \rangle$ |
|---|---|---|---|---|---|
| 4.000 | −239.1439 | −239.1430 | −239.1439 | −239.1430 | $4.48 \times 10^{-6}$ |
| 4.100 | −239.1956 | −239.1945 | −239.1956 | −239.1945 | $5.69 \times 10^{-6}$ |
| 4.200 | −239.2340 | −239.2325 | −239.2340 | −239.2325 | $9.80 \times 10^{-6}$ |
| 4.300 | −239.2618 | −239.2601 | −239.2618 | −239.2601 | $-7.77 \times 10^{-5}$ |
| 4.400 | −239.2812 | −239.2792 | −239.2812 | −239.2792 | $-3.60 \times 10^{-5}$ |
| 4.500 | −239.2940 | −239.2916 | −239.2940 | −239.2916 | $-3.62 \times 10^{-4}$ |
| 4.600 | −239.3017 | −239.2988 | −239.3017 | −239.2988 | $-3.39 \times 10^{-5}$ |
| 4.700 | −239.3052 | −239.3019 | −239.3052 | −239.3019 | $-1.41 \times 10^{-3}$ |
| 4.800 | −239.3056 | −239.3019 | −239.3056 | −239.3019 | $-4.09 \times 10^{-3}$ |
| 4.900 | −239.3036 | −239.2994 | −239.3036 | −239.2994 | $1.22 \times 10^{-2}$ |
| 5.000 | −239.2998 | −239.2952 | −239.3000 | −239.2956 | $2.89 \times 10^{-1}$ |
| 5.020 | −239.2989 | −239.2942 | −239.2992 | −239.2948 | $3.46 \times 10^{-1}$ |
| 5.040 | −239.2979 | −239.2931 | −239.2984 | −239.2940 | $3.93 \times 10^{-1}$ |
| 5.060 | −239.2969 | −239.2920 | −239.2975 | −239.2931 | $4.34 \times 10^{-1}$ |
| 5.080 | −239.2958 | −239.2909 | −239.2965 | −239.2921 | $4.70 \times 10^{-1}$ |
| 5.100 | −239.2947 | −239.2897 | −239.2955 | −239.2911 | $5.01 \times 10^{-1}$ |
| 5.150 | −239.2918 | −239.2866 | −239.2929 | −239.2884 | $5.66 \times 10^{-1}$ |
| 5.200 | −239.2887 | −239.2833 | −239.2900 | −239.2855 | $6.13 \times 10^{-1}$ |
| 5.250 | −239.2855 | −239.2798 | −239.2870 | −239.2824 | $6.57 \times 10^{-1}$ |
| 5.300 | −239.2821 | −239.2762 | −239.2838 | −239.2790 | $6.91 \times 10^{-1}$ |
| 5.350 | −239.2785 | −239.2725 | −239.2804 | −239.2755 | $7.19 \times 10^{-1}$ |
| 5.400 | −239.2749 | −239.2687 | −239.2770 | −239.2719 | $7.46 \times 10^{-1}$ |
| 5.500 | −239.2676 | −239.2610 | −239.2698 | −239.2645 | $7.92 \times 10^{-1}$ |
| 5.600 | −239.2600 | −239.2531 | −239.2626 | −239.2569 | $8.36 \times 10^{-1}$ |
| 5.700 | −239.2525 | −239.2452 | −239.2553 | −239.2493 | $8.77 \times 10^{-1}$ |
| 5.800 | −239.2450 | −239.2373 | −239.2481 | −239.2418 | $9.15 \times 10^{-1}$ |
| 5.900 | −239.2376 | −239.2296 | −239.2410 | −239.2345 | $9.47 \times 10^{-1}$ |
| 6.000 | −239.2303 | −239.2220 | −239.2342 | −239.2274 | $9.72 \times 10^{-1}$ |

TABLE X. Total SCF energies in Hartree obtained with the 1c RKS and 1c UKS formalisms and energies for the limit of a vanishing Gaussian smearing (limit). For the UKS calculations, the numbers of $\alpha$ and $\beta$ electrons are denoted by $n_\alpha$ and $n_\beta$. The $S_z$ expectation value is calculated as $\langle S_z \rangle = (n_\alpha - n_\beta)/2$. The S-VWN (V) functional [2, 3] is employed.

| $d$ | RKS limit | RKS | UKS limit | UKS | $n_\alpha$ | $n_\beta$ | $\langle S_z \rangle$ |
|------|-----------|-----------|-----------|-----------|---------|---------|------------------------|
| 4.000 | −238.4885 | −238.4876 | −238.4885 | −238.4876 | 18.0000 | 18.0000 | 0.00 |
| 4.100 | −238.5363 | −238.5354 | −238.5363 | −238.5354 | 18.0000 | 18.0000 | $6.04 \times 10^{-14}$ |
| 4.200 | −238.5708 | −238.5698 | −238.5708 | −238.5698 | 18.0000 | 18.0000 | $1.86 \times 10^{-5}$ |
| 4.300 | −238.5949 | −238.5937 | −238.5949 | −238.5937 | 18.0000 | 18.0000 | $2.72 \times 10^{-5}$ |
| 4.400 | −238.6105 | −238.6091 | −238.6105 | −238.6091 | 18.0000 | 18.0000 | $4.46 \times 10^{-5}$ |
| 4.500 | −238.6195 | −238.6179 | −238.6195 | −238.6179 | 18.0001 | 17.9999 | $8.67 \times 10^{-5}$ |
| 4.600 | −238.6233 | −238.6214 | −238.6233 | −238.6214 | 18.0002 | 17.9998 | $2.02 \times 10^{-4}$ |
| 4.700 | −238.6230 | −238.6207 | −238.6230 | −238.6207 | 18.0006 | 17.9994 | $5.97 \times 10^{-4}$ |
| 4.800 | −238.6196 | −238.6169 | −238.6196 | −238.6169 | 18.0009 | 17.9991 | $9.02 \times 10^{-4}$ |
| 4.900 | −238.6138 | −238.6106 | −238.6138 | −238.6106 | 18.0011 | 17.9989 | $1.14 \times 10^{-3}$ |
| 5.000 | −238.6062 | −238.6026 | −238.6062 | −238.6026 | 18.0010 | 17.9990 | $1.03 \times 10^{-3}$ |
| 5.100 | −238.5973 | −238.5932 | −238.5973 | −238.5932 | 18.0006 | 17.9994 | $5.63 \times 10^{-4}$ |
| 5.200 | −238.5876 | −238.5829 | −238.5876 | −238.5829 | 18.0674 | 17.9326 | $6.74 \times 10^{-2}$ |
| 5.300 | −238.5773 | −238.5721 | −238.5775 | −238.5724 | 18.5460 | 17.4540 | $5.46 \times 10^{-1}$ |
| 5.400 | −238.5667 | −238.5609 | −238.5679 | −238.5627 | 18.9002 | 17.0998 | $9.00 \times 10^{-1}$ |
| 5.500 | −238.5561 | −238.5497 | −238.5587 | −238.5534 | 19.0715 | 16.9285 | 1.07 |
| 5.600 | −238.5455 | −238.5386 | −238.5494 | −238.5440 | 19.1478 | 16.8522 | 1.15 |
| 5.700 | −238.5351 | −238.5276 | −238.5400 | −238.5344 | 19.1842 | 16.8158 | 1.18 |
| 5.800 | −238.5249 | −238.5170 | −238.5307 | −238.5248 | 19.2022 | 16.7978 | 1.20 |
| 5.900 | −238.5151 | −238.5067 | −238.5215 | −238.5153 | 19.2111 | 16.7889 | 1.21 |
| 6.000 | −238.5056 | −238.4967 | −238.5125 | −238.5059 | 19.2159 | 16.7841 | 1.22 |

TABLE XI. Total SCF energies in Hartree obtained with the 2c KR, 2c canonical KU $S_z$, and 2c canonical KU $S_x$ approaches and energies for the limit of a vanishing Gaussian smearing (limit). Expectation values of the spin $S_{z,x}$ are listed for the open-shell calculations. The S-VWN (V) functional [2, 3] is employed.

| $d$ | KR limit | KR | KU $S_z$ limit | KU $S_z$ | $\langle S_z \rangle$ | KU $S_x$ limit | KU $S_x$ | $\langle S_x \rangle$ |
|---|---|---|---|---|---|---|---|---|
| 4.000 | −238.7779 | −238.7769 | −238.7779 | −238.7769 | $-1.60 \times 10^{-6}$ | −238.7779 | −238.7769 | $4.58 \times 10^{-5}$ |
| 4.100 | −238.8256 | −238.8243 | −238.8256 | −238.8243 | $1.82 \times 10^{-5}$ | −238.8256 | −238.8243 | $2.79 \times 10^{-5}$ |
| 4.200 | −238.8601 | −238.8586 | −238.8601 | −238.8586 | $-3.67 \times 10^{-9}$ | −238.8601 | −238.8586 | $2.74 \times 10^{-5}$ |
| 4.300 | −238.8844 | −238.8825 | −238.8844 | −238.8825 | $6.47 \times 10^{-8}$ | −238.8844 | −238.8825 | $2.01 \times 10^{-4}$ |
| 4.400 | −238.9004 | −238.8982 | −238.9004 | −238.8982 | $1.80 \times 10^{-7}$ | −238.9004 | −238.8982 | $-4.26 \times 10^{-4}$ |
| 4.500 | −238.9099 | −238.9074 | −238.9099 | −238.9074 | $-2.01 \times 10^{-6}$ | −238.9099 | −238.9074 | $6.26 \times 10^{-4}$ |
| 4.600 | −238.9144 | −238.9115 | −238.9144 | −238.9115 | $-1.79 \times 10^{-6}$ | −238.9144 | −238.9115 | $1.23 \times 10^{-3}$ |
| 4.700 | −238.9149 | −238.9116 | −238.9149 | −238.9116 | $1.38 \times 10^{-8}$ | −238.9149 | −238.9116 | $1.71 \times 10^{-3}$ |
| 4.800 | −238.9125 | −238.9087 | −238.9125 | −238.9087 | $2.03 \times 10^{-3}$ | −238.9125 | −238.9087 | $8.34 \times 10^{-4}$ |
| 4.900 | −238.9077 | −238.9035 | −238.9077 | −238.9035 | $1.12 \times 10^{-3}$ | −238.9077 | −238.9035 | $3.68 \times 10^{-3}$ |
| 5.000 | −238.9013 | −238.8967 | −238.9013 | −238.8967 | $1.09 \times 10^{-3}$ | −238.9014 | −238.8967 | $1.71 \times 10^{-2}$ |
| 5.100 | −238.8938 | −238.8887 | −238.8938 | −238.8887 | $1.01 \times 10^{-3}$ | −238.8939 | −238.8890 | $2.60 \times 10^{-1}$ |
| 5.200 | −238.8854 | −238.8799 | −238.8854 | −238.8799 | $2.94 \times 10^{-3}$ | −238.8861 | −238.8811 | $4.83 \times 10^{-1}$ |
| 5.300 | −238.8765 | −238.8706 | −238.8765 | −238.8706 | $5.99 \times 10^{-3}$ | −238.8776 | −238.8725 | $6.02 \times 10^{-1}$ |
| 5.400 | −238.8672 | −238.8609 | −238.8672 | −238.8609 | $3.43 \times 10^{-2}$ | −238.8686 | −238.8634 | $6.82 \times 10^{-1}$ |
| 5.500 | −238.8578 | −238.8511 | −238.8579 | −238.8513 | $2.58 \times 10^{-1}$ | −238.8595 | −238.8540 | $7.42 \times 10^{-1}$ |
| 5.600 | −238.8483 | −238.8413 | −238.8488 | −238.8422 | $5.32 \times 10^{-1}$ | −238.8503 | −238.8445 | $7.96 \times 10^{-1}$ |
| 5.700 | −238.8390 | −238.8316 | −238.8398 | −238.8330 | $6.56 \times 10^{-1}$ | −238.8412 | −238.8351 | $8.41 \times 10^{-1}$ |
| 5.800 | −238.8297 | −238.8220 | −238.8309 | −238.8239 | $7.27 \times 10^{-1}$ | −238.8322 | −238.8258 | $8.83 \times 10^{-1}$ |
| 5.900 | −238.8207 | −238.8126 | −238.8221 | −238.8150 | $7.72 \times 10^{-1}$ | −238.8235 | −238.8169 | $9.19 \times 10^{-1}$ |
| 6.000 | −238.8119 | −238.8035 | −238.8136 | −238.8062 | $8.03 \times 10^{-1}$ | −238.8151 | −238.8082 | $9.50 \times 10^{-1}$ |

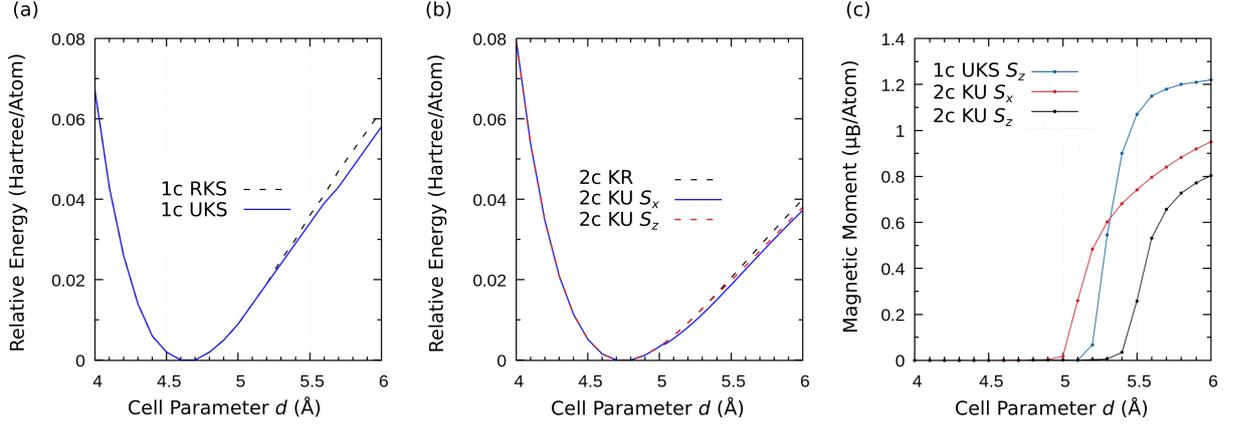

FIG. 3. (a) Dependence of the energy on the cell parameter in units of Hartree per atom for the 1c RKS and 1c UKS calculations with the S-VWN (V) functional [2, 3] and the dhf-SVP-2c basis set [16] combined with small-core Dirac–Fock ECPs [25]. (b) Dependence of the energy on the cell parameter in units of Hartree per atom for the 2c KR, 2c KU $S_x$, and 2c KU $S_z$ calculations. (c) Magnetic moment in units of Bohr's magneton $\mu_B$ per atom for the spin contribution of the 1c UKS, 2c KU $S_x$, and 2c KU $S_z$ calculations.

16

TABLE XII. Total SCF energies in Hartree obtained with the 1c RKS and 1c UKS formalisms and energies for the limit of a vanishing Gaussian smearing (limit). For the UKS calculations, the numbers of $\alpha$ and $\beta$ electrons are denoted by $n_\alpha$ and $n_\beta$. The $S_z$ expectation value is calculated as $\langle S_z \rangle = (n_\alpha - n_\beta)/2$. The TPSS functional [6] is employed.

| $d$ | RKS limit | RKS | UKS limit | UKS | $n_\alpha$ | $n_\beta$ | $\langle S_z \rangle$ |
|---|---|---|---|---|---|---|---|
| 4.000 | $-238.7223$ | $-238.7214$ | $-238.7223$ | $-238.7214$ | 18.0000 | 18.0000 | $1.30 \times 10^{-5}$ |
| 4.100 | $-238.7742$ | $-238.7732$ | $-238.7742$ | $-238.7732$ | 18.0000 | 18.0000 | $2.46 \times 10^{-5}$ |
| 4.200 | $-238.8125$ | $-238.8115$ | $-238.8125$ | $-238.8115$ | 18.0000 | 18.0000 | $3.56 \times 10^{-5}$ |
| 4.300 | $-238.8401$ | $-238.8389$ | $-238.8401$ | $-238.8389$ | 18.0001 | 17.9999 | $7.48 \times 10^{-5}$ |
| 4.400 | $-238.8590$ | $-238.8577$ | $-238.8590$ | $-238.8577$ | 18.0001 | 17.9999 | $1.27 \times 10^{-4}$ |
| 4.500 | $-238.8711$ | $-238.8696$ | $-238.8711$ | $-238.8696$ | 18.0002 | 17.9997 | $2.48 \times 10^{-4}$ |
| 4.600 | $-238.8778$ | $-238.8760$ | $-238.8778$ | $-238.8760$ | 18.0005 | 17.9995 | $5.20 \times 10^{-4}$ |
| 4.700 | $-238.8803$ | $-238.8782$ | $-238.8803$ | $-238.8782$ | 18.0014 | 17.9986 | $1.41 \times 10^{-3}$ |
| 4.800 | $-238.8794$ | $-238.8769$ | $-238.8794$ | $-238.8769$ | 18.0009 | 17.9991 | $9.09 \times 10^{-4}$ |
| 4.900 | $-238.8760$ | $-238.8731$ | $-238.8760$ | $-238.8731$ | 18.0006 | 17.9995 | $5.48 \times 10^{-4}$ |
| 5.000 | $-238.8706$ | $-238.8673$ | $-238.8706$ | $-238.8673$ | 18.0004 | 17.9996 | $3.93 \times 10^{-4}$ |
| 5.100 | $-238.8638$ | $-238.8600$ | $-238.8638$ | $-238.8600$ | 18.0209 | 17.9791 | $2.09 \times 10^{-2}$ |
| 5.200 | $-238.8561$ | $-238.8518$ | $-238.8562$ | $-238.8519$ | 18.5396 | 17.4604 | $5.40 \times 10^{-1}$ |
| 5.300 | $-238.8477$ | $-238.8428$ | $-238.8491$ | $-238.8446$ | 19.0238 | 16.9762 | 1.02 |
| 5.400 | $-238.8388$ | $-238.8334$ | $-238.8421$ | $-238.8377$ | 19.1479 | 16.8521 | 1.15 |
| 5.500 | $-238.8298$ | $-238.8239$ | $-238.8347$ | $-238.8302$ | 19.1888 | 16.8112 | 1.19 |
| 5.600 | $-238.8208$ | $-238.8143$ | $-238.8270$ | $-238.8222$ | 19.2055 | 16.7945 | 1.21 |
| 5.700 | $-238.8119$ | $-238.8048$ | $-238.8190$ | $-238.8139$ | 19.2133 | 16.7867 | 1.21 |
| 5.800 | $-238.8031$ | $-238.7955$ | $-238.8111$ | $-238.8057$ | 19.2192 | 16.7808 | 1.22 |
| 5.900 | $-238.7946$ | $-238.7866$ | $-238.8033$ | $-238.7976$ | 19.2244 | 16.7756 | 1.22 |
| 6.000 | $-238.7865$ | $-238.7780$ | $-238.7957$ | $-238.7896$ | 19.2293 | 16.7707 | 1.23 |

TABLE XIII. Total SCF energies in Hartree obtained with the 2c KR, 2c canonical KU $S_z$, and 2c canonical KU $S_x$ approaches and energies for the limit of a vanishing Gaussian smearing (limit). Expectation values of the spin $S_{z,x}$ are listed for the open-shell calculations. The TPSS functional [6] is employed.

| $d$ | KR limit | KR | KU $S_z$ limit | KU $S_z$ | $\langle S_z \rangle$ | KU $S_x$ limit | KU $S_x$ | $\langle S_x \rangle$ |
|---|---|---|---|---|---|---|---|---|
| 4.000 | $-239.0074$ | $-239.0065$ | $-239.0074$ | $-239.0065$ | $1.83 \times 10^{-5}$ | $-239.0074$ | $-239.0065$ | $3.62 \times 10^{-5}$ |
| 4.100 | $-239.0592$ | $-239.0580$ | $-239.0592$ | $-239.0580$ | $2.64 \times 10^{-5}$ | $-239.0592$ | $-239.0580$ | $2.35 \times 10^{-5}$ |
| 4.200 | $-239.0975$ | $-239.0961$ | $-239.0975$ | $-239.0961$ | $1.67 \times 10^{-5}$ | $-239.0975$ | $-239.0961$ | $4.72 \times 10^{-5}$ |
| 4.300 | $-239.1251$ | $-239.1234$ | $-239.1251$ | $-239.1234$ | $4.79 \times 10^{-5}$ | $-239.1251$ | $-239.1234$ | $3.41 \times 10^{-4}$ |
| 4.400 | $-239.1443$ | $-239.1423$ | $-239.1443$ | $-239.1423$ | $1.92 \times 10^{-4}$ | $-239.1443$ | $-239.1423$ | $4.77 \times 10^{-4}$ |
| 4.500 | $-239.1568$ | $-239.1544$ | $-239.1568$ | $-239.1544$ | $1.57 \times 10^{-4}$ | $-239.1568$ | $-239.1544$ | $5.10 \times 10^{-4}$ |
| 4.600 | $-239.1641$ | $-239.1613$ | $-239.1641$ | $-239.1613$ | $2.10 \times 10^{-4}$ | $-239.1641$ | $-239.1613$ | $5.05 \times 10^{-4}$ |
| 4.700 | $-239.1672$ | $-239.1640$ | $-239.1672$ | $-239.1640$ | $1.81 \times 10^{-4}$ | $-239.1672$ | $-239.1640$ | $3.94 \times 10^{-4}$ |
| 4.800 | $-239.1672$ | $-239.1636$ | $-239.1672$ | $-239.1636$ | $1.93 \times 10^{-4}$ | $-239.1672$ | $-239.1636$ | $3.97 \times 10^{-3}$ |
| 4.900 | $-239.1648$ | $-239.1607$ | $-239.1648$ | $-239.1607$ | $1.10 \times 10^{-4}$ | $-239.1650$ | $-239.1611$ | $2.54 \times 10^{-1}$ |
| 5.000 | $-239.1606$ | $-239.1560$ | $-239.1606$ | $-239.1560$ | $1.31 \times 10^{-3}$ | $-239.1614$ | $-239.1575$ | $4.91 \times 10^{-1}$ |
| 5.100 | $-239.1550$ | $-239.1501$ | $-239.1550$ | $-239.1501$ | $6.56 \times 10^{-3}$ | $-239.1565$ | $-239.1525$ | $6.27 \times 10^{-1}$ |
| 5.200 | $-239.1486$ | $-239.1432$ | $-239.1487$ | $-239.1434$ | $1.21 \times 10^{-1}$ | $-239.1506$ | $-239.1464$ | $7.13 \times 10^{-1}$ |
| 5.300 | $-239.1415$ | $-239.1357$ | $-239.1419$ | $-239.1365$ | $5.13 \times 10^{-1}$ | $-239.1439$ | $-239.1395$ | $7.76 \times 10^{-1}$ |
| 5.400 | $-239.1340$ | $-239.1278$ | $-239.1352$ | $-239.1299$ | $7.27 \times 10^{-1}$ | $-239.1367$ | $-239.1320$ | $8.30 \times 10^{-1}$ |
| 5.500 | $-239.1262$ | $-239.1197$ | $-239.1283$ | $-239.1231$ | $8.06 \times 10^{-1}$ | $-239.1293$ | $-239.1241$ | $8.84 \times 10^{-1}$ |
| 5.600 | $-239.1183$ | $-239.1115$ | $-239.1210$ | $-239.1157$ | $8.55 \times 10^{-1}$ | $-239.1217$ | $-239.1162$ | $9.27 \times 10^{-1}$ |
| 5.700 | $-239.1104$ | $-239.1032$ | $-239.1136$ | $-239.1081$ | $8.78 \times 10^{-1}$ | $-239.1141$ | $-239.1083$ | $9.69 \times 10^{-1}$ |
| 5.800 | $-239.1025$ | $-239.0950$ | $-239.1062$ | $-239.1005$ | $8.91 \times 10^{-1}$ | $-239.1067$ | $-239.1006$ | $1.00$ |
| 5.900 | $-239.0948$ | $-239.0869$ | $-239.0989$ | $-239.0929$ | $8.97 \times 10^{-1}$ | $-239.0994$ | $-239.0931$ | $1.03$ |
| 6.000 | $-239.0873$ | $-239.0791$ | $-239.0916$ | $-239.0853$ | $8.89 \times 10^{-1}$ | $-239.0924$ | $-239.0859$ | $1.04$ |

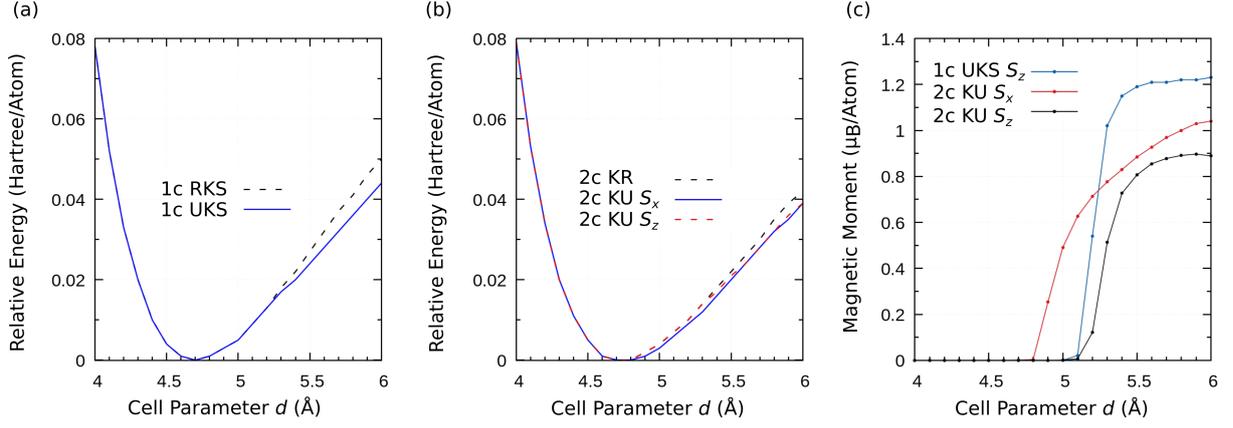

FIG. 4. (a) Dependence of the energy on the cell parameter in units of Hartree per atom for the 1c RKS and 1c UKS calculations with the TPSS functional [6] and the dhf-SVP-2c basis set [16] combined with small-core Dirac–Fock ECPs [25]. (b) Dependence of the energy on the cell parameter in units of Hartree per atom for the 2c KR, 2c KU $S_x$, and 2c KU $S_z$ calculations. (c) Magnetic moment in units of Bohr's magneton $\mu_B$ per atom for the spin contribution of the 1c UKS, 2c KU $S_x$, and 2c KU $S_z$ calculations.

TABLE XIV. Total SCF energies in Hartree with the 1c RKS and 1c UKS formalisms and energies for the limit of a vanishing Gaussian smearing (limit). For the UKS calculations, the numbers of $\alpha$ and $\beta$ electrons are denoted by $n_\alpha$ and $n_\beta$. The $S_z$ expectation value is calculated as $\langle S_z \rangle = (n_\alpha - n_\beta)/2$. The HSE06 functional [26–28] is used with a threshold of $5 \cdot 10^{-5}$ for the eigenvalues of the overlap matrix in the orthogonal basis transformation for $d = 4.0$ to $d = 4.7$Å. This removes at most 2 linear combinations of basis functions (vectors), total number of vectors is 86. For the other structures, the default threshold of $10^{-6}$ is applied. 33 $k$ points are employed throughout.

| $d$ | RKS limit | RKS | UKS limit | UKS | $n_\alpha$ | $n_\beta$ | $\langle S_z \rangle$ |
|---|---|---|---|---|---|---|---|
| 4.000 | $-238.7081$ | $-238.7073$ | $-238.7081$ | $-238.7073$ | 18.0000 | 18.0000 | $5.18 \times 10^{-8}$ |
| 4.100 | $-238.7576$ | $-238.7568$ | $-238.7576$ | $-238.7568$ | 18.0000 | 18.0000 | $1.82 \times 10^{-7}$ |
| 4.200 | $-238.7939$ | $-238.7930$ | $-238.7939$ | $-238.7930$ | 18.0000 | 18.0000 | $5.88 \times 10^{-7}$ |
| 4.300 | $-238.8197$ | $-238.8187$ | $-238.8197$ | $-238.8187$ | 18.0000 | 18.0000 | $2.54 \times 10^{-6}$ |
| 4.400 | $-238.8370$ | $-238.8358$ | $-238.8370$ | $-238.8358$ | 18.0000 | 18.0000 | $2.26 \times 10^{-5}$ |
| 4.500 | $-238.8476$ | $-238.8464$ | $-238.8476$ | $-238.8464$ | 18.0002 | 17.9998 | $1.76 \times 10^{-4}$ |
| 4.600 | $-238.8531$ | $-238.8517$ | $-238.8531$ | $-238.8517$ | 18.0048 | 17.9952 | $4.82 \times 10^{-3}$ |
| 4.700 | $-238.8545$ | $-238.8529$ | $-238.8566$ | $-238.8554$ | 18.3843 | 17.6157 | $3.84 \times 10^{-1}$ |
| 4.800 | $-238.8530$ | $-238.8512$ | $-238.8556$ | $-238.8543$ | 18.4481 | 17.5518 | $4.48 \times 10^{-1}$ |
| 4.900 | $-238.8488$ | $-238.8467$ | $-238.8550$ | $-238.8541$ | 19.1415 | 16.8585 | 1.14 |
| 5.000 | $-238.8430$ | $-238.8407$ | $-238.8532$ | $-238.8523$ | 19.1571 | 16.8429 | 1.16 |
| 5.100 | $-238.8361$ | $-238.8337$ | $-238.8494$ | $-238.8484$ | 19.1669 | 16.8331 | 1.17 |
| 5.200 | $-238.8286$ | $-238.8262$ | $-238.8470$ | $-238.8465$ | 19.2274 | 16.7726 | 1.23 |
| 5.300 | $-238.8205$ | $-238.8182$ | $-238.8422$ | $-238.8417$ | 19.2330 | 16.7670 | 1.23 |
| 5.400 | $-238.8120$ | $-238.8097$ | $-238.8331$ | $-238.8323$ | 19.2724 | 16.7276 | 1.27 |
| 5.500 | $-238.8033$ | $-238.8007$ | $-238.8311$ | $-238.8306$ | 19.2418 | 16.7582 | 1.24 |
| 5.600 | $-238.7944$ | $-238.7916$ | $-238.8253$ | $-238.8247$ | 19.2461 | 16.7539 | 1.25 |
| 5.700 | $-238.7856$ | $-238.7824$ | $-238.8195$ | $-238.8188$ | 19.2510 | 16.7490 | 1.25 |
| 5.800 | $-238.7851$ | $-238.7830$ | $-238.8137$ | $-238.8130$ | 19.2564 | 16.7436 | 1.26 |
| 5.900 | $-238.7788$ | $-238.7766$ | $-238.8080$ | $-238.8072$ | 19.2417 | 16.7583 | 1.24 |
| 6.000 | $-238.7724$ | $-238.7700$ | $-238.8060$ | $-238.8051$ | 19.1536 | 16.8464 | 1.15 |

# V. STRUCTURES

Geometric structures are given in TURBOMOLE format [29–32]. Cell parameters ($cell) are specified in atomic units unless stated differently, i.e. angs for Ångström. Coordinates inside the cell are given in atomic units ($coord) or with fractional and dimensionless units ($coord frac). Structures are listed under the subsection headlines of the main text, where they are used.

## A. SCF Computation Times

```
$cell
 6.61460836 6.61460836 6.61460836 60.0 60.0 60.0
$coord
    0.0000000000000        0.0000000000000        0.0000000000000       pb
$end
```

## B.   Ionization Energies of Zero-Dimensional Heavy *p*-Block Atoms

```
$coord
      0.00000000000000      0.00000000000000      0.00000000000000      in
$end
```

```
$coord
      0.00000000000000      0.00000000000000      0.00000000000000      sn
$end
```

```
$coord
      0.00000000000000      0.00000000000000      0.00000000000000      sb
$end
```

```
$coord
      0.00000000000000      0.00000000000000      0.00000000000000      te
$end
```

```
$coord
      0.00000000000000      0.00000000000000      0.00000000000000      i
$end
```

```
$coord
      0.00000000000000      0.00000000000000      0.00000000000000      tl
$end
```

```
$coord
      0.00000000000000      0.00000000000000      0.00000000000000      pb
$end
```

```
$coord
      0.00000000000000      0.00000000000000      0.00000000000000      bi
$end
```

```
$coord
      0.00000000000000      0.00000000000000      0.00000000000000      po
$end
```

```
$coord
      0.00000000000000      0.00000000000000      0.00000000000000      at
$end
```

## C. Band Structures of Three-Dimensional Gold and Lead Crystals

```
a) Three-dimensional gold fcc crystal
$cell
 5.45185 5.45185 5.45185 60.0 60.0 60.0
$coord
     0.00000000000000      0.00000000000000      0.00000000000000      au
$end


b) Three-dimensional lead fcc crystal
$cell
 6.57429 6.57429 6.57429 60.0 60.0 60.0
$coord
     0.00000000000000      0.00000000000000      0.00000000000000      pb
$end
```

## D. Band Gaps of Three-Dimensional Silver Halide Crystals

```
a) Non-relaxed structures


i) AgCl

$cell angs
   3.9683    3.9683    3.9683  60.0000    60.0000    60.0000
$coord frac
     0.00000000000000        0.00000000000000        0.00000000000000       ag
     0.50000000000000        0.50000000000000        0.50000000000000       cl
$end


ii) AgBr

$cell angs
   4.1316    4.1316    4.1316  60.0000    60.0000    60.0000
$coord frac
     0.00000000000000        0.00000000000000        0.00000000000000       ag
     0.50000000000000        0.50000000000000        0.50000000000000       br
$end


iii) AgI

$cell angs
   4.3621    4.3621    4.3621  60.0000    60.0000    60.0000
$coord frac
     0.00000000000000        0.00000000000000        0.00000000000000       ag
     0.50000000000000        0.50000000000000        0.50000000000000       i
$end
```

b) Relaxed structures without D3-BJ

i) AgCl

```
SVWN (V)
$cell  angs
  3.84825344503333333333   3.84825344503333333333   3.84825344503333333333   60.0   60.0   60.0
$coord frac
     0.00000000000000        0.00000000000000        0.00000000000000    ag
     0.50000000000000        0.50000000000000        0.50000000000000    cl
$end


PBE
$cell  angs
  3.97881790756666666666   3.97881790756666666666   3.97881790756666666666   60.0   60.0   60.0
$coord frac
     0.00000000000000        0.00000000000000        0.00000000000000    ag
     0.50000000000000        0.50000000000000        0.50000000000000    cl
$end


PBEsol
$cell  angs
   3.90080049883333333333   3.90080049883333333333   3.90080049883333333333   60.0   60.0   60.0
$coord frac
     0.00000000000000        0.00000000000000        0.00000000000000    ag
     0.50000000000000        0.50000000000000        0.50000000000000    cl
$end


TPSS
$cell  angs
   3.95192791410000000000   3.95192791410000000000   3.95192791410000000000   60.0   60.0   60.0
$coord frac
     0.00000000000000        0.00000000000000        0.00000000000000    ag
     0.50000000000000        0.50000000000000        0.50000000000000    cl
$end


revTPSS
$cell  angs
   3.93323644336667     3.93323644336667     3.93323644336667     60.0    60.0    60.0
$coord frac
     0.00000000000000        0.00000000000000        0.00000000000000    ag
     0.50000000000000        0.50000000000000        0.50000000000000    cl
$end
```

```
Tao-Mo
$cell  angs
   3.9184690781    3.9184690781    3.9184690781  60.0    60.0    60.0
$coord frac
     0.00000000000000       0.00000000000000       0.00000000000000  ag
     0.50000000000000       0.50000000000000       0.50000000000000  cl
$end


PKZB
$cell  angs
   3.9880650068    3.9880650068    3.9880650068   60.0    60.0    60.0
$coord frac
     0.00000000000000       0.00000000000000       0.00000000000000  ag
     0.50000000000000       0.50000000000000       0.50000000000000  cl
$end


r2SCAN
$cell  angs
   3.94398068143333333333    3.94398068143333333333    3.94398068143333333333    60.0    60.0    60.0
$coord frac
     0.00000000000000       0.00000000000000       0.00000000000000  ag
     0.50000000000000       0.50000000000000       0.50000000000000  cl
$end


2c-SVWN (V)
$cell  angs
     3.8017642716     3.8017642716     3.8017642716    60.0    60.0  60.0
$coord frac
     0.00000000000000       0.00000000000000       0.00000000000000  ag
     0.50000000000000       0.50000000000000       0.50000000000000  cl
$end


2c-PBE
$cell  angs
     3.97687054476666666666    3.97687054476666666666    3.97687054476666666666  60.0  60.0  60.0
$coord frac
     0.00000000000000       0.00000000000000       0.00000000000000  ag
     0.50000000000000       0.50000000000000       0.50000000000000  cl
$end


2c-PBEsol
$cell  angs
     3.89940465353333333333    3.89940465353333333333    3.89940465353333333333    60.0    60.0    60.0
$coord frac
     0.00000000000000       0.00000000000000       0.00000000000000  ag
     0.50000000000000       0.50000000000000       0.50000000000000  cl
$end
```

```
2c-TPSS
$cell  angs
     3.95007142833333333333   3.95007142833333333333   3.95007142833333333333   60.0   60.0  60.0
$coord frac
     0.00000000000000         0.00000000000000         0.00000000000000   ag
     0.50000000000000         0.50000000000000         0.50000000000000   cl
$end


2c-revTPSS
$cell  angs
     3.93227899606666666666   3.93227899606666666666     3.93227899606666666666   60.0  60.0  60.0
$coord frac
     0.00000000000000         0.00000000000000         0.00000000000000   ag
     0.50000000000000         0.50000000000000         0.50000000000000   cl
$end


2c-Tao-Mo
$cell  angs
    3.91842787876666666666   3.91842787876666666666   3.91842787876666666666  60.0  60.0  60.0
$coord frac
     0.00000000000000         0.00000000000000         0.00000000000000   ag
     0.50000000000000         0.50000000000000         0.50000000000000   cl
$end


2c-PKZB
$cell  angs
     3.9852599056      3.9852599056      3.9852599056      60.0   60.0   60.0
$coord frac
     0.00000000000000         0.00000000000000         0.00000000000000   ag
     0.50000000000000         0.50000000000000         0.50000000000000   cl
$end


2c-r2SCAN
$cell  angs
     3.9430444896      3.9430444896      3.9430444896      60.0   60.0   60.0
$coord frac
     0.00000000000000         0.00000000000000         0.00000000000000   ag
     0.50000000000000         0.50000000000000         0.50000000000000   cl
$end
```

ii) AgBr

SVWN (V)
```
$cell  angs
    3.9659229888    3.9659229888    3.9659229888  60.0    60.0  60.0
$coord frac
    0.00000000000000      0.00000000000000      0.00000000000000  ag
    0.50000000000000      0.50000000000000      0.50000000000000  br
$end
```

PBE
```
$cell  angs
    4.13768026056666666666  4.13768026056666666666  4.13768026056666666666  60.0  60.0  60.0
$coord frac
    0.00000000000000      0.00000000000000      0.00000000000000  ag
    0.50000000000000      0.50000000000000      0.50000000000000  br
$end
```

PBEsol
```
$cell  angs
    4.02721811653333333333  4.02721811653333333333  4.02721811653333333333  60.0  60.0  60.0
$coord frac
    0.00000000000000      0.00000000000000      0.00000000000000  ag
    0.50000000000000      0.50000000000000      0.50000000000000  br
$end
```

TPSS
```
$cell  angs
    4.11184283506666666666  4.11184283506666666666  4.11184283506666666666  60.0  60.0  60.0
$coord frac
    0.00000000000000      0.00000000000000      0.00000000000000  ag
    0.50000000000000      0.50000000000000      0.50000000000000  br
$end
```

revTPSS
```
$cell  angs
    4.09371116283333333333  4.09371116283333333333  4.09371116283333333333  60.0  60.0  60.0
$coord frac
    0.00000000000000      0.00000000000000      0.00000000000000  ag
    0.50000000000000      0.50000000000000      0.50000000000000  br
$end
```

```
Tao-Mo
$cell  angs
  4.08238349393333333333  4.08238349393333333333  4.08238349393333333333  60.0  60.0  60.0
$coord frac
    0.00000000000000      0.00000000000000      0.00000000000000    ag
    0.50000000000000      0.50000000000000      0.50000000000000    br
$end


PKZB
$cell  angs
  4.14890441196666666666  4.14890441196666666666  4.14890441196666666666  60.0 60.0 60.0
$coord frac
    0.00000000000000      0.00000000000000      0.00000000000000    ag
    0.50000000000000      0.50000000000000      0.50000000000000    br
$end


r2SCAN
$cell  angs
  4.11077895733333333333  4.11077895733333333333  4.11077895733333333333  60.0  60.0  60.0
$coord frac
    0.00000000000000      0.00000000000000      0.00000000000000    ag
    0.50000000000000      0.50000000000000      0.50000000000000    br
$end


2c-SVWN (V)
$cell  angs
    3.9626048018      3.9626048018      3.9626048018      60.0      60.0      60.0
$coord frac
    0.00000000000000      0.00000000000000      0.00000000000000    ag
    0.50000000000000      0.50000000000000      0.50000000000000    br
$end


2c-PBE
$cell  angs
    4.13617434303333333333    4.13617434303333333333    4.13617434303333333333  60.0  60.0  60.0
$coord frac
    0.00000000000000      0.00000000000000      0.00000000000000    ag
    0.50000000000000      0.50000000000000      0.50000000000000    br
$end


2c-PBEsol
$cell  angs
  4.02480791976666666666  4.02480791976666666666  4.02480791976666666666    60.0    60.0    60.0
$coord frac
    0.00000000000000      0.00000000000000      0.00000000000000    ag
    0.50000000000000      0.50000000000000      0.50000000000000    br
$end
```

```
2c-TPSS
$cell  angs
    4.11004402016666666666     4.11004402016666666666     4.11004402016666666666    60.0    60.0    60.0
$coord frac
    0.00000000000000          0.00000000000000          0.00000000000000       ag
    0.50000000000000          0.50000000000000          0.50000000000000       br
$end


2c-revTPSS
$cell  angs
    4.0897251878    4.0897251878    4.0897251878    60.0   60.0   60.0
$coord frac
    0.00000000000000          0.00000000000000          0.00000000000000       ag
    0.50000000000000          0.50000000000000          0.50000000000000       br
$end


2c-Tao-Mo
$cell  angs
    4.05717500276666666666     4.05717500276666666666     4.05717500276666666666    60.0    60.0    60.0
$coord frac
    0.00000000000000          0.00000000000000          0.00000000000000       ag
    0.50000000000000          0.50000000000000          0.50000000000000       br
$end


2c-PKZB
$cell  angs
    4.14933946790000000000     4.14933946790000000000     4.14933946790000000000    60.0    60.0    60.0
$coord frac
    0.00000000000000          0.00000000000000          0.00000000000000       ag
    0.50000000000000          0.50000000000000          0.50000000000000       br
$end


2c-r2SCAN
$cell  angs
    4.10922457123333333333     4.10922457123333333333     4.10922457123333333333    60.0    60.0    60.0
$coord frac
    0.00000000000000          0.00000000000000          0.00000000000000       ag
    0.50000000000000          0.50000000000000          0.50000000000000       br
$end
```

iii) AgI

SVWN (V)
```
$cell  angs
    4.19916806426666666666    4.19916806426666666666    4.19916806426666666666  60.0  60.0    60.0
$coord frac
      0.00000000000000        0.00000000000000        0.00000000000000  ag
      0.50000000000000        0.50000000000000        0.50000000000000  i
$end
```

PBE
```
$cell  angs
   4.375962455  4.375962455  4.375962455  60.0  60.0  60.0
$coord frac
      0.00000000000000        0.00000000000000        0.00000000000000  ag
      0.50000000000000        0.50000000000000        0.50000000000000  i
$end
```

PBEsol
```
$cell  angs
   4.2604273368  4.2604273368  4.2604273368  60.0  60.0  60.0
$coord frac
      0.00000000000000        0.00000000000000        0.00000000000000  ag
      0.50000000000000        0.50000000000000        0.50000000000000  i
$end
```

TPSS
```
$cell  angs
    4.35287278606666666666    4.35287278606666666666    4.35287278606666666666  60.0    60.0  60.0
$coord frac
      0.00000000000000        0.00000000000000        0.00000000000000  ag
      0.50000000000000        0.50000000000000        0.50000000000000  i
$end
```

revTPSS
```
$cell  angs
    4.326282372  4.326282372  4.326282372  60.0   60.0    60.0
$coord frac
      0.00000000000000        0.00000000000000        0.00000000000000  ag
      0.50000000000000        0.50000000000000        0.50000000000000  i
$end
```

```
Tao-Mo
$cell  angs
   4.31120483906666666666   4.31120483906666666666   4.31120483906666666666   60.0   60.0  60.0
$coord frac
    0.00000000000000       0.00000000000000       0.00000000000000  ag
    0.50000000000000       0.50000000000000       0.50000000000000  i
$end


PKZB
$cell  angs
   4.38623378086666666666   4.38623378086666666666   4.38623378086666666666  60.0  60.0  60.0
$coord frac
    0.00000000000000       0.00000000000000       0.00000000000000  ag
    0.50000000000000       0.50000000000000       0.50000000000000  i
$end


r2SCAN
$cell  angs
  4.35615418616666666666   4.35615418616666666666   4.35615418616666666666  60.0  60.0  60.0
$coord frac
    0.00000000000000       0.00000000000000       0.00000000000000  ag
    0.50000000000000       0.50000000000000       0.50000000000000  i
$end


2c-SVWN (V)
$cell  angs
   4.19790823766666666666   4.19790823766666666666   4.19790823766666666666  60.0 60.0 60.0
$coord frac
    0.00000000000000       0.00000000000000       0.00000000000000  ag
    0.50000000000000       0.50000000000000       0.50000000000000  i
$end


2c-PBE
$cell  angs
   4.374539102    4.374539102    4.374539102    60.0    60.0    60.0
$coord frac
    0.00000000000000       0.00000000000000       0.00000000000000  ag
    0.50000000000000       0.50000000000000       0.50000000000000  i
$end


2c-PBEsol
$cell  angs
   4.25857171496666666666   4.25857171496666666666   4.25857171496666666666   60.0   60.0   60.0
$coord frac
    0.00000000000000       0.00000000000000       0.00000000000000  ag
    0.50000000000000       0.50000000000000       0.50000000000000  i
$end
```

```
2c-TPSS
$cell  angs
   4.35102446196666666666    4.35102446196666666666    4.35102446196666666666  60.0   60.0   60.0
$coord frac
     0.00000000000000       0.00000000000000       0.00000000000000  ag
     0.50000000000000       0.50000000000000       0.50000000000000  i
$end


2c-revTPSS
$cell  angs
   4.32457321323333333333    4.32457321323333333333    4.32457321323333333333  60.0   60.0   60.0
$coord frac
     0.00000000000000       0.00000000000000       0.00000000000000  ag
     0.50000000000000       0.50000000000000       0.50000000000000  i
$end


2c-Tao-Mo
$cell  angs
   4.2926534394    4.2926534394    4.2926534394  60.0   60.0   60.0
$coord frac
     0.00000000000000       0.00000000000000       0.00000000000000  ag
     0.50000000000000       0.50000000000000       0.50000000000000  i
$end


2c-PKZB
$cell  angs
   4.38391013183333333333    4.38391013183333333333    4.38391013183333333333  60.0   60.0   60.0
$coord frac
     0.00000000000000       0.00000000000000       0.00000000000000  ag
     0.50000000000000       0.50000000000000       0.50000000000000  i
$end


2c-r2SCAN
$cell  angs
   4.35527771656666666666    4.35527771656666666666    4.35527771656666666666   60.0    60.0    60.0
$coord frac
     0.00000000000000       0.00000000000000       0.00000000000000  ag
     0.50000000000000       0.50000000000000       0.50000000000000  i
$end
```

c) Relaxed structures with D3-BJ


i) AgCl


PBE-D3-BJ
$cell angs
  3.9287962214666666666  3.9287962214666666666  3.9287962214666666666  60.0  60.0 60.0
$coord frac
     0.00000000000000      0.00000000000000      0.00000000000000  ag
     0.50000000000000      0.50000000000000      0.50000000000000  cl
$end


PBEsol-D3-BJ
$cell  angs
  3.83841735613333333333  3.83841735613333333333  3.83841735613333333333  60.0 60.0 60.0
$coord frac
     0.00000000000000      0.00000000000000      0.00000000000000  ag
     0.50000000000000      0.50000000000000      0.50000000000000  cl
$end


TPSS-D3-BJ
$cell angs
  3.8891246083666666666  3.8891246083666666666  3.8891246083666666666  60.0 60.0 60.0
$coord frac
     0.00000000000000      0.00000000000000      0.00000000000000  ag
     0.50000000000000      0.50000000000000      0.50000000000000  cl
$end


revTPSS-D3-BJ
$cell angs
  3.8724607048666666666  3.8724607048666666666  3.8724607048666666666  60.0  60.0  60.0
$coord frac
     0.00000000000000      0.00000000000000      0.00000000000000  ag
     0.50000000000000      0.50000000000000      0.50000000000000  cl
$end


Tao-Mo-D3-BJ
$cell  angs
  3.8983235666  3.8983235666  3.8983235666  60.0   60.0    60.0
$coord frac
     0.00000000000000      0.00000000000000      0.00000000000000  ag
     0.50000000000000      0.50000000000000      0.50000000000000  cl
$end

```
r2SCAN-D3-BJ
$cell  angs
  3.913670709  3.913670709  3.913670709  60.0    60.0    60.0
$coord frac
    0.00000000000000      0.00000000000000      0.00000000000000  ag
    0.50000000000000      0.50000000000000      0.50000000000000  cl
$end


2c-PBE-D3-BJ
$cell  angs
  3.91521636993333333333  3.91521636993333333333  3.91521636993333333333  60.0  60.0  60.0
$coord frac
    0.00000000000000      0.00000000000000      0.00000000000000  ag
    0.50000000000000      0.50000000000000      0.50000000000000  cl
$end


2c-PBEsol-D3-BJ
$cell  angs
  3.83689838036666666666  3.83689838036666666666  3.83689838036666666666  60.0  60.0  60.0
$coord frac
    0.00000000000000      0.00000000000000      0.00000000000000  ag
    0.50000000000000      0.50000000000000      0.50000000000000  cl
$end


2c-TPSS-D3-BJ
$cell  angs
  3.8874918142  3.8874918142  3.8874918142  60.0  60.0  60.0
$coord frac
    0.00000000000000      0.00000000000000      0.00000000000000  ag
    0.50000000000000      0.50000000000000      0.50000000000000  cl
$end


2c-revTPSS-D3-BJ
$cell  angs
  3.8711454888  3.8711454888  3.8711454888  60.0  60.0  60.0
$coord frac
    0.00000000000000      0.00000000000000      0.00000000000000  ag
    0.50000000000000      0.50000000000000      0.50000000000000  cl
$end


2c-Tao-Mo-D3-BJ
$cell  angs
  3.8970339839  3.8970339839  3.8970339839  60.0  60.0  60.0
$coord frac
    0.00000000000000      0.00000000000000      0.00000000000000  ag
    0.50000000000000      0.50000000000000      0.50000000000000  cl
$end
```

```
2c-r2SCAN-D3-BJ
$cell  angs
  3.90136349976666666666   3.90136349976666666666   3.90136349976666666666  60.0  60.0  60.0
$coord frac
     0.00000000000000        0.00000000000000        0.00000000000000   ag
     0.50000000000000        0.50000000000000        0.50000000000000   cl
$end
```

ii) AgBr

PBE-D3-BJ
```
$cell  angs
   4.0656560669  4.0656560669  4.0656560669  60.0  60.0  60.0
$coord frac
     0.00000000000000        0.00000000000000        0.00000000000000   ag
     0.50000000000000        0.50000000000000        0.50000000000000   br
$end
```

PBEsol-D3-BJ
```
$cell  angs
   3.9711451685  3.9711451685  3.9711451685  60.0  60.0  60.0
$coord frac
     0.00000000000000        0.00000000000000        0.00000000000000   ag
     0.50000000000000        0.50000000000000        0.50000000000000   br
$end
```

TPSS-D3-BJ
```
$cell  angs
   4.0377456864  4.0377456864  4.0377456864  60.0  60.0  60.0
$coord frac
     0.00000000000000        0.00000000000000        0.00000000000000   ag
     0.50000000000000        0.50000000000000        0.50000000000000   br
$end
```

revTPSS-D3-BJ
```
$cell  angs
   4.02469085823333333333  4.02469085823333333333  4.02469085823333333333  60.0  60.0  60.0
$coord frac
     0.00000000000000        0.00000000000000        0.00000000000000   ag
     0.50000000000000        0.50000000000000        0.50000000000000   br
$end
```

Tao-Mo-D3-BJ
```
$cell  angs
   4.06252851366666666666  4.06252851366666666666  4.06252851366666666666  60.0  60.0  60.0
$coord frac
     0.00000000000000        0.00000000000000        0.00000000000000   ag
     0.50000000000000        0.50000000000000        0.50000000000000   br
$end
```

```
r2SCAN-D3-BJ
$cell  angs
  4.082712298  4.082712298  4.082712298  60.0  60.0  60.0
$coord frac
    0.00000000000000      0.00000000000000      0.00000000000000  ag
    0.50000000000000      0.50000000000000      0.50000000000000  br
$end


2c-PBE-D3-BJ
$cell  angs
  4.06356866316666666666  4.06356866316666666666  4.06356866316666666666  60.0  60.0  60.0
$coord frac
    0.00000000000000      0.00000000000000      0.00000000000000  ag
    0.50000000000000      0.50000000000000      0.50000000000000  br
$end


2c-PBEsol-D3-BJ
$cell  angs
  3.96911415466666666666  3.96911415466666666666  3.96911415466666666666  60.0  60.0  60.0
$coord frac
    0.00000000000000      0.00000000000000      0.00000000000000  ag
    0.50000000000000      0.50000000000000      0.50000000000000  br
$end


2c-TPSS-D3-BJ
$cell  angs
  4.03638224546666666666  4.03638224546666666666  4.03638224546666666666  60.0  60.0  60.0
$coord frac
    0.00000000000000      0.00000000000000      0.00000000000000  ag
    0.50000000000000      0.50000000000000      0.50000000000000  br
$end


2c-revTPSS-D3-BJ
$cell  angs
  4.0234614894  4.0234614894  4.0234614894  60.0  60.0  60.0
$coord frac
    0.00000000000000      0.00000000000000      0.00000000000000  ag
    0.50000000000000      0.50000000000000      0.50000000000000  br
$end


2c-Tao-Mo-D3-BJ
$cell  angs
  4.0611942705  4.0611942705  4.0611942705  60.0  60.0  60.0
$coord frac
    0.00000000000000      0.00000000000000      0.00000000000000  ag
    0.50000000000000      0.50000000000000      0.50000000000000  br
$end
```

```
2c-r2SCAN-D3-BJ
$cell  angs
  4.0811353733  4.0811353733  4.0811353733  60.0  60.0  60.0
$coord frac
     0.00000000000000      0.00000000000000      0.00000000000000  ag
     0.50000000000000      0.50000000000000      0.50000000000000  br
$end
```

iii) AgI

PBE-D3-BJ
```
$cell  angs
   4.29762263753333333333   4.29762263753333333333   4.29762263753333333333   60.0   60.0   60.0
$coord frac
     0.00000000000000        0.00000000000000        0.00000000000000  ag
     0.50000000000000        0.50000000000000        0.50000000000000  i
$end
```

PBEsol-D3-BJ
```
$cell  angs
   4.19283430466666666666   4.19283430466666666666   4.19283430466666666666   60.0   60.0   60.0
$coord frac
     0.00000000000000        0.00000000000000        0.00000000000000  ag
     0.50000000000000        0.50000000000000        0.50000000000000  i
$end
```

TPSS-D3-BJ
```
$cell  angs
   4.23056622496666666666   4.23056622496666666666   4.23056622496666666666    60.0   60.0   60.0
$coord frac
     0.00000000000000        0.00000000000000        0.00000000000000  ag
     0.50000000000000        0.50000000000000        0.50000000000000  i
$end
```

revTPSS-D3-BJ
```
$cell  angs
   4.2109555586 4.2109555586 4.2109555586 60.0 60.0   60.0
$coord frac
     0.00000000000000        0.00000000000000        0.00000000000000  ag
     0.50000000000000        0.50000000000000        0.50000000000000  i
$end
```

Tao-Mo-D3-BJ
```
$cell  angs
   4.29789857256666666666   4.29789857256666666666   4.29789857256666666666   60.0   60.0   60.0
$coord frac
     0.00000000000000        0.00000000000000        0.00000000000000  ag
     0.50000000000000        0.50000000000000        0.50000000000000  i
$end
```

```
r2SCAN-D3-BJ
$cell  angs
  4.3548993639  4.3548993639  4.3548993639  60.0  60.0 60.0
$coord frac
     0.00000000000000        0.00000000000000        0.00000000000000  ag
     0.50000000000000        0.50000000000000        0.50000000000000  i
$end


2c-PBE-D3-BJ
$cell  angs
  4.2899779228  4.2899779228  4.2899779228  60.0  60.0  60.0
$coord frac
     0.00000000000000        0.00000000000000        0.00000000000000  ag
     0.50000000000000        0.50000000000000        0.50000000000000  i
$end


2c-PBEsol-D3-BJ
$cell  angs
  4.19121877706666666666  4.19121877706666666666  4.19121877706666666666  60.0  60.0  60.0
$coord frac
     0.00000000000000        0.00000000000000        0.00000000000000  ag
     0.50000000000000        0.50000000000000        0.50000000000000  i
$end


2c-TPSS-D3-BJ
$cell  angs
  4.22982934023333333333  4.22982934023333333333 4.22982934023333333333  60.0  60.0  60.0
$coord frac
     0.00000000000000        0.00000000000000        0.00000000000000  ag
     0.50000000000000        0.50000000000000        0.50000000000000  i
$end


2c-revTPSS-D3-BJ
$cell  angs
  4.20627833826666666666  4.20627833826666666666  4.20627833826666666666  60.0  60.0  60.0
$coord frac
     0.00000000000000        0.00000000000000        0.00000000000000  ag
     0.50000000000000        0.50000000000000        0.50000000000000  i
$end


2c-Tao-Mo-D3-BJ
$cell  angs
  4.29099278693333333333  4.29099278693333333333  4.29099278693333333333  60.0  60.0  60.0
$coord frac
     0.00000000000000        0.00000000000000        0.00000000000000  ag
     0.50000000000000        0.50000000000000        0.50000000000000  i
$end
```

```
2c-r2SCAN-D3-BJ
$cell  angs
  4.3530927047  4.3530927047  4.3530927047  60.0  60.0  60.0
$coord frac
    0.00000000000000      0.00000000000000      0.00000000000000  ag
    0.50000000000000      0.50000000000000      0.50000000000000  i
$end
```

## E.  Indium(I,III)-Telluride Two-Dimensional Honeycomb System

```
a) 1c optimization without D3
$cell
 8.0124390 8.0124390 60.0
$coord
     0.00000000000000      0.00000000000000      0.00000000000000      in
     0.00125810178926      0.00074629401429      5.30945552703617      in
     4.00628439107714      2.31300606929462     -2.75618224113402      te
     4.00743098931212      2.31368907685671      8.06634157724272      te
$end


b) 1c optimization with D3
$cell
  8.1284346 8.1284346 60.0
$coord
     0.00000000000000      0.00000000000000      0.00000000000000   in
     0.00000000000000      0.00000000000000      5.26273509093640   in
     4.05997252052396      2.34549030506129     -2.67054225966576   te
     4.05997252052396      2.34549030506129      7.93471996528309   te
$end


c) 2c optimization with D3
$cell
  8.1284346 8.1284346 60.0
$coord
     0.00000000000000      0.00000000000000      0.00000000000000   in
     0.00000000000000      0.00000000000000      5.26273509093640   in
     4.05997252052396      2.34549030506129     -2.67054225966576   te
     4.05997252052396      2.34549030506129      7.93471996528309   te
$end
```

## F. One-Dimensional Platinum Chains

```
$cell angs
4.0
$coord
    0.00000000000000    0.00000000000000    0.00000000000000    pt
    3.77945197720000    0.00000000000000    0.00000000000000    pt
$end


$cell angs
4.1
$coord
    0.00000000000000    0.00000000000000    0.00000000000000    pt
    3.87393827660000    0.00000000000000    0.00000000000000    pt
$end


$cell angs
4.2
$coord
    0.00000000000000    0.00000000000000    0.00000000000000    pt
    3.96842457600000    0.00000000000000    0.00000000000000    pt
$end


$cell angs
4.3
$coord
    0.00000000000000    0.00000000000000    0.00000000000000    pt
    4.06291087540000    0.00000000000000    0.00000000000000    pt
$end


$cell angs
4.4
$coord
    0.00000000000000    0.00000000000000    0.00000000000000    pt
    4.15739717490000    0.00000000000000    0.00000000000000    pt
$end


$cell angs
4.5
$coord
    0.00000000000000    0.00000000000000    0.00000000000000    pt
    4.25188347430000    0.00000000000000    0.00000000000000    pt
$end
```

```
$cell angs
4.6
$coord
    0.00000000000000        0.00000000000000        0.00000000000000        pt
    4.34636977370000        0.00000000000000        0.00000000000000        pt
$end


$cell angs
4.7
$coord
    0.00000000000000        0.00000000000000        0.00000000000000        pt
    4.44085607320000        0.00000000000000        0.00000000000000        pt
$end


$cell angs
4.8
$coord
    0.00000000000000        0.00000000000000        0.00000000000000        pt
    4.53534237260000        0.00000000000000        0.00000000000000        pt
$end


$cell angs
4.9
$coord
    0.00000000000000        0.00000000000000        0.00000000000000        pt
    4.62982867200000        0.00000000000000        0.00000000000000        pt
$end


$cell angs
5.0
$coord
    0.00000000000000        0.00000000000000        0.00000000000000        pt
    4.72431497140000        0.00000000000000        0.00000000000000        pt
$end


$cell ang
5.02
$coord
    0.00000000000000        0.00000000000000        0.00000000000000        pt
    4.74321300000000        0.00000000000000        0.00000000000000        pt
$end


$cell ang
5.04
$coord
    0.00000000000000        0.00000000000000        0.00000000000000        pt
    4.76211000000000        0.00000000000000        0.00000000000000        pt
$end
```

```
$cell ang
5.06
$coord
    0.00000000000000         0.00000000000000         0.00000000000000         pt
    4.78100700000000         0.00000000000000         0.00000000000000         pt
$end


$cell ang
5.08
$coord
    0.00000000000000         0.00000000000000         0.00000000000000         pt
    4.79990400000000         0.00000000000000         0.00000000000000         pt


$cell angs
5.1
$coord
    0.00000000000000         0.00000000000000         0.00000000000000         pt
    4.81880127090000         0.00000000000000         0.00000000000000         pt
$end


$cell angs
5.15
$coord
    0.00000000000000         0.00000000000000         0.00000000000000         pt
    4.86604500000000         0.00000000000000         0.00000000000000         pt
$end


$cell angs
5.2
$coord
    0.00000000000000         0.00000000000000         0.00000000000000         pt
    4.91328757030000         0.00000000000000         0.00000000000000         pt
$end


$cell angs
5.25
$coord
    0.00000000000000         0.00000000000000         0.00000000000000         pt
    4.96053100000000         0.00000000000000         0.00000000000000         pt
$end


$cell angs
5.3
$coord
    0.00000000000000         0.00000000000000         0.00000000000000         pt
    5.00777386970000         0.00000000000000         0.00000000000000         pt
$end
```

```
$cell angs
5.35
$coord
      0.00000000000000         0.00000000000000         0.00000000000000       pt
      5.05501500000000         0.00000000000000         0.00000000000000       pt
$end


$cell angs
5.4
$coord
      0.00000000000000         0.00000000000000         0.00000000000000       pt
      5.10226016920000         0.00000000000000         0.00000000000000       pt
$end


$cell angs
5.5
$coord
      0.00000000000000         0.00000000000000         0.00000000000000       pt
      5.19674646860000         0.00000000000000         0.00000000000000       pt
$end


$cell angs
5.6
$coord
      0.00000000000000         0.00000000000000         0.00000000000000       pt
      5.29123276800000         0.00000000000000         0.00000000000000       pt
$end


$cell angs
5.7
$coord
      0.00000000000000         0.00000000000000         0.00000000000000       pt
      5.38571906740000         0.00000000000000         0.00000000000000       pt
$end


$cell angs
5.8
$coord
      0.00000000000000         0.00000000000000         0.00000000000000       pt
      5.48020536690000         0.00000000000000         0.00000000000000       pt
$end
```

```
$cell angs
5.9
$coord
     0.00000000000000       0.00000000000000       0.00000000000000       pt
     5.57469166630000       0.00000000000000       0.00000000000000       pt
$end


$cell angs
6.0
$coord
     0.00000000000000       0.00000000000000       0.00000000000000       pt
     5.66917796570000       0.00000000000000       0.00000000000000       pt
$end
```


\fi

\end{document}